\documentclass[5p,times]{elsarticle}

\usepackage{amssymb}

\journal{TBA}
\usepackage[T1]{fontenc}
\usepackage[utf8]{inputenc}
\usepackage{acro}
\usepackage{adjustbox}
\usepackage{booktabs}
\usepackage{comment}
\usepackage{todonotes}
\usepackage{colortbl}
\usepackage{hyperref}
\usepackage[justification=centering]{subcaption} 
\usepackage{silence}
\usepackage{listings}

\newcolumntype{a}{>{\columncolor{gray!10!white}}c}
\newcolumntype{x}{>{\columncolor{green!10!white}}c}
\newcolumntype{y}{>{\columncolor{blue!10!white}}c}
\newcolumntype{z}{>{\columncolor{yellow!10!white}}c}
\newcolumntype{v}{>{\columncolor{red!10!white}}c}
\definecolor{OliveGreen}{rgb}{0,0.6,0}
\definecolor{ForestGreen}{RGB}{34,139,34}
\definecolor{myblue}{RGB}{37,165,203}
\definecolor{FAUblue}{rgb}{0.000, 0.2196, 0.3961}
\definecolor{myred}{RGB}{175,32,67}

\newcommand{\bq}{\begin{equation}}
\newcommand{\eq}{\end{equation}}
\newcommand{\bytes}{\mbox{B}}
\newcommand{\byte}{\mbox{byte}}
\newcommand{\second}{\mbox{s}}
\newcommand{\MS}{\mbox{ms}}

\newcommand{\NS}{\mbox{ns}}
\newcommand{\seconds}{\mbox{s}}
\newcommand{\flop}{\mbox{flop}}

\newcommand{\bit}{\mbox{bit}}

\newcommand{\GBPS}{\mbox{G\bit/\second}}

\newcommand{\GBS}{\mbox{GB/\second}}

\newcommand{\MBS}{\mbox{M\byte/\second}}

\newcommand{\GFS}{\mbox{G\flop/\second}}

\newcommand{\lup}{\mbox{LUP}}

\newcommand{\MLUPS}{\mbox{M\lup/\second}}
\newcommand{\GLUPS}{\mbox{G\lup/\second}}
\newcommand{\GHZ}{\mbox{GHz}}

\newcommand{\BF}{\mbox{\byte/\flop}}
\newcommand{\FB}{\mbox{\flop/\byte}}

\newcommand{\GB}{\mbox{GB}}
\newcommand{\TB}{\mbox{TB}}
\newcommand{\KB}{\mbox{kB}}
\newcommand{\MB}{\mbox{MB}}
\newcommand{\GiB}{\mbox{GiB}}
\newcommand{\MiB}{\mbox{MiB}}

\newcommand{\RFO}{read-for-ownership}

\newcommand{\CPP}{C\nolinebreak[4]\hspace{-.05em}\raisebox{.23ex}{\relsize{-1}{++}}}
\newcommand{\CER}{communication-to-execution ratio\xspace}
\newcommand{\PCA}{Principal Component analysis\xspace}

\newcommand{\ML}{machine learning\xspace}

\newcommand{\LLC}{last-level cache\xspace}

\lstdefinestyle{small}{
  language=C++,
  basicstyle=\footnotesize\tt,
  showspaces=false,
  showstringspaces=false,
  breakindent=0pt,
  breaklines,
  captionpos=b,
  breakatwhitespace=true,
  numbers=left,
  numberstyle=\tiny,
  stepnumber=2,
  numbersep=7pt,
  linewidth=0.47\textwidth,
  xleftmargin=.02\textwidth, 
  xrightmargin=.02\textwidth,
  keywordstyle=\color{blue},
  tabsize=2,
  backgroundcolor=\color{white!97!black},
}
\lstdefinestyle{smallWithColor}{
  basicstyle=\small\tt,
  language=C++,
  showspaces=false,
  showstringspaces=false,
  breakindent=0pt,
  breaklines,
  captionpos=b,
  breakatwhitespace=true,
  numbers=left,
  numberstyle=\tiny,
  stepnumber=1,
  numbersep=7pt,
  keywordstyle=\color{blue},
  tabsize=2,
  moredelim=**[is][\color{red}]{@}{@},
  moredelim=**[is][\color{green}]{|}{|},
  backgroundcolor=\color{white!97!black},
}

\lstdefinestyle{footnotesizeStyle}{
  basicstyle=\footnotesize\tt,
  language=C++,
  showspaces=false,
  showstringspaces=false,
  breakindent=0pt,
  breaklines,
  captionpos=b,
  breakatwhitespace=true,
  numbersep=5pt,
  keywordstyle=\color{blue},
  tabsize=2,
  moredelim=**[is][\color{red}]{@}{@},
  moredelim=**[is][\color{green}]{|}{|},
  backgroundcolor=\color{white!97!black},
}

\usepackage[htt]{hyphenat} 
  
\newcommand{\enquote}[1]{``#1''}  
\newcommand{\code}[1]{\texttt{#1}}
\newcommand{\mypara}[1]{\paragraph{#1}}

\colorlet{backgroundcol}{cyan!10!white}
\newcommand{\highlight}[1]{%
	\par\noindent
	\fcolorbox{black}{backgroundcol}{%
		\parbox{\dimexpr\linewidth-2\fboxsep\relax}{%
			#1
		}%
}}

\newcommand{\acrodef}[2]{\DeclareAcronym{#1}{short={#1},long={#2}}}
\acrodef{CER}{commu\-ni\-cation-to-exe\-cu\-tion ratio}
\acrodef{ChebFD}{Chebyshev filter diagonalization}
\acrodef{EOS}{Equation of State}
\acrodef{HPCG}{High Performance Conjugate Gradient}
\acrodef{LBM}{Lattice Boltzmann Menthod}
\acrodef{LULESH}{Livermore Unstructured Lagrangian Explicit Shock Hydrodynamics}
\acrodef{MG}{multigrid}
\acrodef{MST}{MPI-augmented STREAM Triad}
\acrodef{SPEC}{Standard Performance Evaluation Corporation}
\acrodef{SoA}{structure of arrays}
\acrodef{SpMVM}{sparse matrix-vector multiplication}

\usepackage{pgfplots}
\usepackage{cancel}
\usepackage{pgfplotstable}
\usepackage{multirow}
\usepgfplotslibrary{colorbrewer}
\usetikzlibrary{pgfplots.statistics, pgfplots.colorbrewer} \usepackage{forest}
\usetikzlibrary{arrows.meta, shapes.geometric, calc, shadows}
\usetikzlibrary{fit,calc}
\usepackage{smartdiagram}
\usepackage[ruled]{algorithm2e}

\colorlet{mygreen}{green!75!black}
\colorlet{col1in}{red!30}
\colorlet{col1out}{red!40}
\colorlet{col2in}{mygreen!40}
\colorlet{col2out}{mygreen!50}
\colorlet{col3in}{blue!30}
\colorlet{col3out}{blue!40}
\colorlet{col4in}{mygreen!20}
\colorlet{col4out}{mygreen!30}
\colorlet{col5in}{blue!10}
\colorlet{col5out}{blue!20}
\colorlet{col6in}{blue!20}
\colorlet{col6out}{blue!30}
\colorlet{col7out}{orange}
\colorlet{col7in}{orange!50}
\colorlet{col8out}{orange!40}
\colorlet{col8in}{orange!20}
\colorlet{linecol}{blue!60}
\usepgfplotslibrary{colorbrewer}

\usetikzlibrary{decorations.pathmorphing}
\tikzset{discont/.style={decoration={zigzag,segment length=1.5pt, amplitude=4pt},decorate}}
\def\discontarrow(#1)(#2)(#3)(#4);{
	\draw[discont] (#2) -- (#3);
	\draw[] (#1) -- (#2) (#3) -- (#4);
}

\usepgflibrary{decorations.fractals}
\pgfplotstableset{
rowfont/.style={
	postproc cell content/.append code={
		\count0=\pgfplotstablerow
		\advance\count0 by1
		\ifnum\count0=#1
		\pgfkeysalso{@cell content/.add={\bfseries\boldmath\color{red}}{}}
		\fi
	},
},
colfont/.style={
	postproc cell content/.append code={
		\count0=\pgfplotstablecol
		\advance\count0 by1
		\ifnum\count0=#1
		\pgfkeysalso{@cell content/.add={\color{black}}{}}
		\fi
	},
},
colfontred/.style={
	postproc cell content/.append code={
		\count0=\pgfplotstablecol
		\advance\count0 by1
		\ifnum\count0=#1
		\pgfkeysalso{@cell content/.add={\bfseries\boldmath\color{red}}{}}
		\fi
	},
},
colfontwhite/.style={
	postproc cell content/.append code={
		\count0=\pgfplotstablecol
		\advance\count0 by1
		\ifnum\count0=#1
		\pgfkeysalso{@cell content/.add={\bfseries\boldmath\color{white}}{}}
		\fi
	},
},	
}

\usepackage{colortbl}
\usepgfplotslibrary{colormaps}
\pgfplotsset{compat=1.17}
\pgfplotsset{colormap/Set3} 
\usepackage{cellspace}
\newcolumntype{C}{}
\newlength{\cellspacelimit}
\pgfplotstableset{
	/color cells/min/.initial=0,
	/color cells/max/.initial=100,
	/color cells/textcolor/.initial=,
	color cells/.code={%
		\pgfqkeys{/color cells}{#1}%
		\pgfkeysalso{%
			postproc cell content/.code={%
				\begingroup
				\pgfkeysgetvalue{/pgfplots/table/@preprocessed
					cell content}\value
				\ifx\value\empty
				\endgroup
				\else
				\pgfmathfloatparsenumber{\value}%
				\pgfmathfloattofixed{\pgfmathresult}%
				\let\value=\pgfmathresult
				\pgfplotscolormapaccess
				[\pgfkeysvalueof{/color cells/min}:\pgfkeysvalueof{/color
					cells/max}]
				{\value}
				{\pgfkeysvalueof{/pgfplots/colormap name}}%
				\pgfkeysgetvalue{/pgfplots/table/@cell content}\typesetvalue
				\pgfkeysgetvalue{/color cells/textcolor}\textcolorvalue
				\toks0=\expandafter{\typesetvalue}%
				\xdef\temp{%
					\noexpand\pgfkeysalso{%
						@cell content={%
							\noexpand\cellcolor[rgb]{\pgfmathresult}%
							\noexpand\definecolor{mapped
								color}{rgb}{\pgfmathresult}%
							\ifx\textcolorvalue\empty
							\else
							\noexpand\color{\textcolorvalue}%
							\fi
							\the\toks0 %
						}%
					}%
				}%
				\endgroup
				\temp
				\fi
			}%
		}%
	},
	every column/.style={column type={Cc!{\color{white}\vrule width 1pt}}},
	every first column/.style={reset styles, column type={l}, string type},
	every head row/.style={before row=\toprule, after row=\midrule},
	after row={\noalign{\global\arrayrulewidth=1pt}\arrayrulecolor{white}\hline},
	every last row/.style={after row=\arrayrulecolor{black}\bottomrule}, 
	@content options for rows/.style 2 args={
		postproc cell content/.add code={%
			\def\isInInputList{0}%
			\pgfplotsforeachungrouped \II in {#1} {%
				\ifnum\II<0
				\count0=\II
				\advance\count0 by\pgfplotstablerows
				\edef\II{\the\count0 }%
				\fi
				\ifnum\II=\pgfplotstablerow\relax
				\def\isInInputList{1}%
				\fi
			}%
			\if1\isInInputList%
			\pgfkeysalso{#2}%
			\fi
		}{},
	},
}

\usetikzlibrary{patterns}
\begin{document}

\begin{frontmatter}



\title{Making Applications Faster by Asynchronous Execution: Slowing Down Processes or Relaxing MPI Collectives}


\author[label1,label2]{Ayesha Afzal}
\ead{ayesha.afzal@fau.de}

\author[label1]{Georg Hager}
\ead{georg.hager@fau.de}

\author[label3]{Stefano Markidis}
\ead{markidis@kth.se}

\author[label1,label2]{Gerhard Wellein}
\ead{gerhard.wellein@fau.de}

\affiliation[label1]{%
	organization={Erlangen National High Performance Computing Center (NHR@FAU)},
    postcode={91058},
	city={Erlangen},
	country={Germany},
}
\affiliation[label2]{%
	organization={Department of Computer Science, Friedrich-Alexander-Universität Erlangen-Nürnberg},
	postcode={91058},
	city={Erlangen},
	country={Germany},
}
\affiliation[label3]{%
	organization={Department of Computer Science, KTH Royal Institute of Technology},
	postcode={11428},
	city={Stockholm},
	country={Sweden},
}
\sloppy
\emergencystretch10pt
\begin{abstract}
Comprehending the performance bottlenecks at the core of the intricate hardware-software interactions exhibited by highly parallel programs on HPC clusters is crucial.
This paper sheds light on the issue of automatically asynchronous MPI communication in memory-bound parallel programs on multicore clusters and how it can be facilitated. For instance, slowing down MPI processes by deliberate injection of delays can improve performance if certain conditions are met. This leads to the counter-intuitive conclusion that noise, independent of its source, is not always detrimental but can be leveraged for performance improvements. 
We employ phase-space graphs as a new tool to visualize parallel program dynamics. They are useful in spotting certain patterns in parallel execution that will easily go unnoticed with traditional tracing tools. 
We investigate five different microbenchmarks and applications on different supercomputer platforms: an MPI-augmented STREAM Triad, two implementations of Lattice-Boltzmann fluid solvers, and the LULESH and HPCG proxy applications. 
\end{abstract}



\end{frontmatter}


\section{Introduction and related work}\label{sec:intro}

\mypara{Motivation}
On contemporary HPC clusters, which are typically hybrid shared distributed-memory systems, numerous factors affect the performance of highly parallel applications, making it challenging to predict analytically.
Especially when there are resource bottlenecks, such as memory bandwidth or network bandwidth, simply adding the analytically predicted communication and computation runtimes does not always produce an accurate estimate of the parallel runtime; typically, system or application noise is to blame for this.
However, the dynamics of large-scale parallel programs on modern hardware and the true role of disturbances are not well understood despite extensive research on the characterization of noise, the identification of its sources, and the pinpointing of its influence on collective operations.

\mypara{Better resource utilization via asynchronicity}
This paper investigates the favorable consequences of \emph{noise}, the {\emph{implementation of MPI collectives}}, and generally a spectrum of {\emph{code properties and parameters}} as an enabling factor to achieve higher hardware efficiency of various memory-bound benchmarks and applications on modern clusters.
We investigate their role as potential triggers for the effects of \emph{bottleneck evasion} and \emph{automatic asynchronicity}.
The former means that a resource bottleneck is used concurrently by fewer processes than the possible maximum; the latter describes how processes gradually move out of their initial bulk-synchronous mode, allowing for communication to overlap with computation. 
A boost in asymptotic performance occurs most prominently in applications that are limited by computation and communication bottlenecks, such as memory-bound programs with relevant communication overhead.
This paper investigates the interplay of desynchronization and noise and how it can influence parallel program performance in a positive way. Using microbenchmarks and different implementations of a Lattice-Boltzmann (LBM) flow solver, we demonstrate how noise can actually be advantageous and used on purpose to speed up the transition of a parallel program to a state where communication is at least partially hidden by computation. The \emph{phase space plot} is introduced as a useful tool to identify typical patterns of desynchronization. It can substitute more data-heavy visualizations like, e.g., timeline traces in the context covered here.
Using the LULESH and HPCG proxy apps, we also show that the implementation of collective communication primitives in MPI and a spectrum of code properties can ease or hinder communication overlap.

\mypara{Related work}
The interaction of point-to-point communication with noise, which is frequent in distributed-memory parallel codes, is not covered in a significant portion of the literature~\cite{petrini2003case,nataraj2007ghost,ferreira2008characterizing}, which focuses only on the sources of noise and how it affects collective operations.
\emph{Idle waves} emerge when a disturbance (such as a delay or transient extra work) on an individual MPI process ripples through the other processes during each iteration at a speed that depends on the program's computational and communication characteristics~\cite{AfzalHW19,AfzalEuroMPI19Poster,markidis2015idle}.
Gamell et al.~\cite{Gamell:2015} observed the formation of idle waves in the context of local recovery and failure masking of stencil codes, while Böhme et al.~\cite{Boehme:2016} proposed a tool-based method to pinpoint the root causes of propagating wait states in MPI applications.
Afzal et al.~\cite{AfzalHW2021} explored how these idle waves interact nonlinearly with each other and gradually decay as a result of communication inhomogeneities, application noise, and system noise.
The strong positive correlation between a low propagation speed of idle waves and automatic communication-computation overlap was described in \cite{AfzalHW20}.
Using \ac{SpMVM} and \ac{ChebFD} benchmarks, Afzal et al.~\cite{AfzalHW:2022:4} could show that the smaller the minimum number of processes per memory domain required to saturate the memory bandwidth, the stronger the tendency towards asynchronicity.
The actual speedup that can be observed in such a scenario depends on a spectrum of code properties, such as decomposition strategies, sparse matrix structures, block vector sizes, communication concurrency, and the performance characteristics of back-to-back loops, which can all influence resource utilization~\cite{AfzalHWcpe22,AfzalHW:2022:3}. 
These prior studies show that \emph{bottleneck evasion via asynchronicity} can be regarded as a performance optimization technique, complementing traditional techniques such as explicitly asynchronous communication, noise mitigation, MPI process placement, dynamic load balancing, synchronization of operating kernel (OS) influence, lightweight OS kernels, etc.~\cite{petrini2003case,Bhatele2013,Leon2016,Weisbach:2018}.

\begin{table*}[tbh!]
    \centering
	\caption{Benchmark programs and applied analyses, parameter spaces, and communication properties; CER == communication-execution-ratio, P2P == point-to-point, CB == compute-bound, MB == memory-bound.
    } 
	\label{tab:app}
	\begin{adjustbox}{width=\textwidth}
            \setlength\extrarowheight{-0.7pt}
            \setlength\tabcolsep{2pt}
            \arrayrulecolor{blue}
            \begin{tabular}[fragile]{zzvxy}
            	\toprule
            	\rowcolor[gray]{0.9}
                Case & Parallel codes  & Research analyses & Parameter spaces & Communications\\
            	\midrule
                \cellcolor[gray]{0.9}1 & MST    & faster code with sparingly injected extra workload & noise & P2P\\
            	\cellcolor[gray]{0.9}2a & LBM (D3Q19)    & better measured performance than predicted synchronized performance  &  collectives occurrence and CER & P2P/collective\\
            	\cellcolor[gray]{0.9}2b & LBM (SPEC D2Q37)   & comparing asynchronicity-performance-interaction in CB and MB codes &  MB vs. CB implementation & P2P/collective\\
            	\cellcolor[gray]{0.9}3 & LULESH & slower code with significant imbalanced load & load imbalance & P2P/collective\\
            	\cellcolor[gray]{0.9}4 & HPCG &  overall faster with comparatively slow collective-only performance & collective algorithms and CER & P2P+collective\\
            	\bottomrule
            \end{tabular}
	\end{adjustbox}
\end{table*}
\section[Section title sans citation]{Prior contributions in~\cite{AfzalHW:2022:2}}\label{sec:PPAM-content}
\mypara{Research techniques and metrics}
This paper is a follow-up of \cite{AfzalHW:2022:2}, where we investigated techniques and metrics for quantifying asynchronicity by observing the behavior of MPI waiting times.
Five data analytics techniques (timelines, histograms, compact timelines, correlation coefficients~\cite{vetterling1992numerical}, and phase-space plots) and two \ML techniques (\PCA~\cite{jolliffe2016principal} and K-means clustering\footnote{The cluster center initialization was performed using the k-means++ algorithm~\cite{vassilvitskii2006k}, which heavily depends on the chosen distance metric type (\textit{squared Euclidean}, \textit{city-block}, \textit{cosine} and \textit{correlation}), while the quality of the clustering was quantified using a Silhouette analysis~\cite{kaufman2009finding}.}) were covered.
For the asynchronous execution of large-scale applications, these metrics and techniques were assessed based on their capacity to explore the difference in behavior between compute-bound and weak or strong memory-bound scenarios.

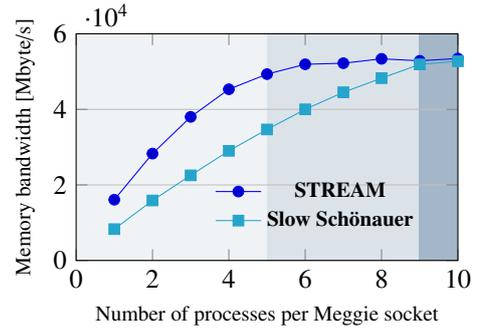
\begin{figure}
    \begin{minipage}[c]{0.5\textwidth}
    \centering
	\begin{tikzpicture}
		\begin{axis}[
		width=0.72\textwidth,height=0.5\textwidth,
		xlabel = {{Number of processes per Meggie socket}},
		ylabel = {{Memory bandwidth [\MBS]}},
		xmin=0,
		xmax=10,
		ymin=0,
		ymax=60000.00,
		ymajorgrids,
        label style={font=\footnotesize},
		xtick={0,2,4,6,8,10},
 		ytick={0,20000,40000,60000}, 
		legend columns = 1, 
		legend style = {
			draw=none,
			fill=none,
			font=\footnotesize,
			cells={align=left},
			anchor=east,
			at={(0.92,0.24)},
		},
		set layers, 
		]
        
        \begin{pgfonlayer}{axis background}
			\fill[shade, left color=FAUblue!6, right color=FAUblue!6]
			(rel axis cs:0,0)--(rel axis cs:0.5,0)--
			(rel axis cs:0.5,1)--(rel axis cs:0,1)--cycle;
			\fill[shade, left color=FAUblue!14, right color=FAUblue!14]
			(rel axis cs:0.5,0)--(rel axis cs:0.9,0)--
			(rel axis cs:0.9,1)--(rel axis cs:0.5,1)--cycle;
			\fill[shade, right color=FAUblue!36, left color=FAUblue!36]
			(rel axis cs:0.9,0)--(rel axis cs:1,0)--
			(rel axis cs:1,1)--(rel axis cs:0.9,1)--cycle;
		\end{pgfonlayer} 
		
		\addplot+[error bars/.cd, y dir=both, y explicit]
		table
		[
		x expr=\thisrow{Cores}, 
		y error minus expr=\thisrow{Median}-\thisrow{Min},
		y error plus expr=\thisrow{Max}-\thisrow{Median},
		row sep=crcr]{
		Cores	Median		Min		Max\\
1	16081.349	16081.349	16081.349\\
2	28282.9129	28282.9129	28282.9129\\
3	38011.239	38011.239	38011.239\\
4	45320.711	45320.711	45320.711\\
5	49324.4899	49324.4899	49324.4899\\
6	51933.2305	51933.2305	51933.2305\\
7	52217.5162	52217.5162	52217.5162\\
8	53376.9859	53376.9859	53376.9859\\
9	52863.7581	52863.7581	52863.7581\\
10	53463.242	53463.242	53463.242\\
		};
		\addlegendentry{\textbf{STREAM}}		
		\addplot+[myblue,mark options={fill=myblue},error bars/.cd, y dir=both, y explicit,]
		table
		[
		x expr=\thisrow{Cores}, 
		y error minus expr=\thisrow{Median}-\thisrow{Min},
		y error plus expr=\thisrow{Max}-\thisrow{Median},
		row sep=crcr
		]{
		Cores	Median		Min		Max\\
1	8296.5011	8296.5011	8296.5011\\
2	15875.1178	15875.1178	15875.1178\\
3	22530.8013	22530.8013	22530.8013\\
4	29022.0039	29022.0039	29022.0039\\
5	34677.882	34677.882	34677.882\\
6	40004.6952	40004.6952	40004.6952\\
7	44520.1861	44520.1861	44520.1861\\
8	48264.4511	48264.4511	48264.4511\\
9	51905.5896	51905.5896	51905.5896\\
10	52738.2617	52738.2617	52738.2617\\
        };
		\addlegendentry{\textbf{Slow Sch\"onauer}}
 		\end{axis}
        \end{tikzpicture}
        \end{minipage}
        \caption{Saturation behavior across the cores of a CPU socket of different MPI-parallel microbenchmarks (weakly or strongly memory bound).}
        \label{fig:saturation}
    \end{figure} 

One crucial trait which influences the desynchronization behavior is the memory boundedness of an application, which can be quantified by its saturation behavior across the cores of a ccNUMA domain. 
Figure~\ref{fig:saturation} illustrates the performance scaling behavior of MPI-parallel mmicrobenchmark codes with different characteristics across the cores of a multicore chip (a ccNUMA domain). The ``Slow Schönauer Triad'' \texttt{A(:) = B(:) + cos(C(:)/D(:))} is memory bound but weakly saturating because of the compu\-tation-heavy cosine and floating-point divide, and the STREAM Triad \texttt{A(:)=B(:)+s*C(:))} is strongly saturating because of its very low computational intensity. 
In~\cite{AfzalHW:2022:2}) we also employed PISOLVER, which numerically evaluates $\int_0^14/(1+x^2)\,\mathrm dx$ using the midpoint rule. This is a purely compute-bound workload dominated by floating-point divides and scales perfectly across cores. In all these micro\-bench\-marks, we added MPI communication to introduce inter-process dependencies, but no global MPI operations were done.

\mypara{Research objective}
Our objective of our prior work was to identify, classify, and characterize aspects of the dynamics of large-scale MPI parallel programs using a compact data representation extracted from tracing data without conducting a comprehensive analysis of the applications. We focused on the particular issue of desynchronized execution and how it may or may not influence the performance.

\mypara{Research method}
In order to bridge the gap between detailed timeline analysis and high-level performance analysis, we explored suitable techniques and metrics that serve as a halfway point. The time that MPI processes spend within the library (MPI waiting time) was chosen as a suitable metric that correlates with MPI asynchronicity.
        \begin{center}
            \includegraphics[scale=0.55]{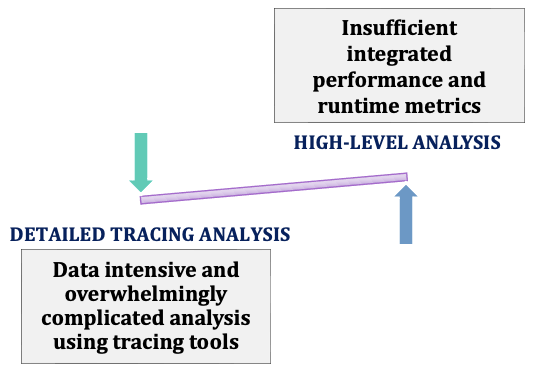}
        \end{center}
    
\mypara{Proposed future work}
The investigation of the behavior of \emph{real-world complex} parallel programs using an appropriate technique was acknowledged as a necessary future task.
By tracking the optimization potential, one can adjust performance modeling and optimization tactics to the dynamics of MPI parallel programs.

\section{Contributions}\label{sec:content}
\mypara{Research analyses}
The applications that were taken into consideration for this work, along with the corresponding analyses, parameter spaces, and communication properties, are listed in Table~\ref{tab:app}.
The selection aims to shed light on how performance across a wide range of program properties is impacted by MPI asynchronicity.
The existence of MPI collective routines is a crucial component.
The \ac{MST} is a microbenchmark with point-to-point communication that is flexible enough to study diverse scenarios.  Collectives are absolutely necessary for the \ac{HPCG} proxy app.
The other codes have avoidable collectives that can be adjusted in frequency and are only necessary to enforce time constraints (\ac{LULESH}) or check for correctness (\ac{LBM}).
Furthermore, both MST and LBM are two-phase programs that exhibit consecutive, identical com\-pute-communi\-cate cycles.
On the other hand, LULESH and HPCG are multi-phase applications whose various kernels exhibit different traits in terms of computational and communication resources.
Each program will be covered in greater detail in Sections~\ref{sec:mst}, \ref{sec:lbm},  \ref{sec:lulesh}, and \ref{sec:hpcg}.
We only use pure MPI programs in this work; in~\cite{AfzalHW20} we covered the fundamentals of hybrid MPI+OpenMP codes in terms of asynchronicity.

\mypara{Research techniques and metrics}
For the MST and LBM cases we concentrate on timelines and phase-space plots, which can be regarded as more explorative data analysis techniques. The phase-space plot was introduced by us in~\cite{AfzalHW:2022:2}.
Furthermore, two metrics are examined: performance per process and MPI time per process. The MPI time is the amount of time spent in the MPI library, i.e., when no computations are being done.
For the LULESH and HPCG analyses we mainly look at their native performance metrics to study the influence of load imbalance (LULESH) and implementation variants of collectives (HPCG). 

\mypara{Research method}
We present two different approaches to performance assessment of desynchronized applications.
First, a side-by-side comparison of performance (memory bandwidth utilization) and MPI times is presented for analysis.
Second, the \emph{composite} or \emph{synchronized} performance is compared with the measured performance. The composite or synchronized performance is the performance resulting from the summation of the individual times for communication and computation, assuming that all processes are in lock-step.

	\begin{table*}[tbh!]
		\centering
		\caption{Key hardware and software characteristics of systems.} 
		\label{tab:systems}
		\begin{adjustbox}{width=0.7\textwidth}
				\setlength\extrarowheight{-0.7pt}
                \setlength\tabcolsep{2pt}
                \arrayrulecolor{blue}
                \Huge
                \begin{tabular}[fragile]{c>{~}lvxy}
                	\toprule
                	\rowcolor[gray]{0.9}
                	\cellcolor[gray]{0.9}&Systems  & \href{https://hpc.fau.de/systems-services/documentation-instructions/clusters/meggie-cluster}{Meggie} (M)   & \href{https://doku.lrz.de/display/PUBLIC/SuperMUC-NG}{SuperMUC-NG} (S) & \href{https://hpc.fau.de/systems-services/documentation-instructions/clusters/fritz-cluster}{Fritz} (F) \\
                	\midrule
                	\cellcolor[gray]{0.9}&Processor  & Intel Xeon Broadwell EP   & Intel Xeon Skylake SP  & Intel Xeon Ice Lake \\    
                	\cellcolor[gray]{0.9}&Processor Model      & E5-2630 v4   & Platinum 8174  & Platinum 8360Y   \\
                	\cellcolor[gray]{0.9}&Base clock speed & $2.2$~\GHZ\    &  $3.10$~\GHZ\ ($2.3$~\GHZ\ used under power cap)   &  $2.4$~\GHZ\ \\
                	\cellcolor[gray]{0.9}&Physical cores per node    & 20   & 48    & 72      \\
                	\cellcolor[gray]{0.9}&Numa domains per node  &   2  & 2  & 4   \\
                	\cellcolor[gray]{0.9}&Last-level cache (LLC) size & $25$~\MiB\ (L3)  &  $24$~\MiB\ (L2) + $33$~\MiB\ (L3)  & $1.25$~\MiB\ (L2) + $54$~\MiB\ (L3) \\
                	\cellcolor[gray]{0.9}&Memory per node (type)& $64$~\GiB\ (DDR4)  & $96$~\GiB\ (DDR4)  & $256$~\GiB\ (DDR4) \\
                	\multirow{-8}{*}{\rotatebox{90}{\cellcolor[gray]{0.9} Micro-architecture}}&Theor. socket memory bandwidth & $68.3$~\GBS\ &  $128$~\GBS\ & $2\times 102.4$~\GBS\ \\
                	
                	\midrule
                	\cellcolor[gray]{0.9}&Node interconnect    & Omni-Path  & Omni-Path  & HDR100 Infiniband    \\
                	\cellcolor[gray]{0.9}&Interconnect topology & Fat-tree & Fat-tree  & Fat-tree \\
                	\multirow{-3}{*}{\rotatebox{90}{\cellcolor[gray]{0.9} Network}}&Raw bandwidth p. lnk n. dir & $100$~\GBPS\  &  $100$~\GBPS\ & $100$~\GBPS\   \\
                	
                	\midrule
                	\cellcolor[gray]{0.9}&Compiler    & Intel \CPP{} v2019.5.281     & Intel \CPP{} v2019.4.243   & Intel \CPP{} v2022.1 \\
                	\cellcolor[gray]{0.9}&Optimization flags & -O3 -xHost  & -O3 -qopt-zmm-usage=high & -O3 -qopt-zmm-usage=high \\
                	\cellcolor[gray]{0.9}&SIMD & -xAVX &-xCORE-AVX512 &  -xCORE-AVX512\\
                	\cellcolor[gray]{0.9}&Message passing library & Intel \verb.MPI. v2019u5     & Intel \verb.MPI. v2019u4    &  Intel \verb.MPI. v2021u7 \\
                	\multirow{-5}{*}{\rotatebox{90}{\cellcolor[gray]{0.9} Software}}&Operating system    & CentOS AlmaLinux v8.7  &  SUSE Linux ENT. Server 12 SP3  &  CentOS AlmaLinux v8.7 \\
                	
                	\midrule
                	\cellcolor[gray]{0.9}&\code{ITAC}    & v2019u5      & v2019  & v2021u6  \\
                   \cellcolor[gray]{0.9}&\code{ClusterCockpit}    &  2023   &  n/a  &  2023  \\           
                	\multirow{-3}{*}{\rotatebox{90}{\cellcolor[gray]{0.9} Tools}}&\code{LIKWID}   & 5.2.1         & 5.2.1     & 5.2.2 \\
                	\bottomrule
                \end{tabular}
		\end{adjustbox}
	\end{table*}
\mypara{Research contributions}
The focus of our previous publication~\cite{AfzalHW:2022:2} was on the exploration of various data analysis techniques; here we concentrate on the analysis of applications, particularly employing the new technique of phase-space analysis for temporal evolution.
The impact of MPI asynchronous execution on performance is explored, particularly to distinguish between parallel codes that are compute bound and those that are memory bandwidth limited.
Experiments were performed on more systems than in~\cite{AfzalHW:2022:2}.
While two of the applications (i.e., MST and 3DQ19 LBM) were expanded from our prior contribution, three new ones (LULESH, HPCG, and a 2DQ37 LBM code from SPEChpc 2021) were investigated.
The following significant contributions are made by this paper:
\begin{enumerate} 
    \item In \emph{\acf{MST}}, a deliberate injection of noise (extra workload) can accelerate spontaneous asynchronous execution, leading to better asymptotic performance if communication overhead is relevant.
    \item In \emph{\acf{LBM}}, we compare two variants:
    a memory-bound D3Q19 implementation and a compute-bound D2Q37 implementation from the SPEChpc 2021 suite. 
    Only the D3Q19 case exhibits a performance improvement with asynchronous execution.
    In order to leverage this advantage, it is also advisable to keep the frequency of required collectives low, not because of reduced overhead but because of relaxed resynchronization that allows processes to stay out of sync longer and thus benefit from better bottleneck utilization.
    \item In \emph{\acf{LULESH}}, any potential benefit of automatic asynchronicity is swamped by the significant artificial load imbalance even though the application fits the bill in terms of memory boundedness.
    \item In \emph{\acf{HPCG}}, we compare the performance of the overall application with that of the collective-only in an isolated benchmark.
    The actual implementation of \code{MPI\_Allreduce} is instrumental for enabling asynchronous execution and communication overlap.
    For certain problem sizes, the synchronizing quality of the collective is more important than its bare overhead.
    The collectives that support asynchronous execution allow for better application performance, even though they are not the fastest according to the micro-benchmark.
    \item In the presence of \emph{frequent synchronizing collectives}, any noise between two successive collectives causes significant loss because all other processes must wait for a delayed process.
    Petrini et al.\ resolved this in 2003~\cite{petrini2003case} by synchronizing the system noise, which concentrated the noise on all nodes in one time step and made subsequent time steps noise-free.
    In contrast to this synchronized noise effect, where bottleneck structure was irrelevant~\cite{petrini2003case}, we clearly make a point that any performance benefit of noise is dependent on the presence of a bottleneck.
\end{enumerate}

\mypara{Overview}
The organization of this paper is as follows:
We first go into detail about our experimental setup and methodology in Sect.~\ref{sec:setup}.
We then discuss the performance implications of asynchronicity, specifically focusing on \acs{MST} in Sect.~\ref{sec:mst}, \acs{LBM} in Sect.~\ref{sec:lbm}, \acs{LULESH} in Sect.~\ref{sec:lulesh}, and \acs{HPCG} Sect.~\ref{sec:hpcg}.
Finally, Section~\ref{sec:conclusion} concludes the paper and provides an outlook for future directions.

\begin{figure*}[tb]
	\centering
    \begin{minipage}{0.3\textwidth}
    \begin{tikzpicture}
            \put(-0.35,-0.01){\includegraphics[scale=0.14]{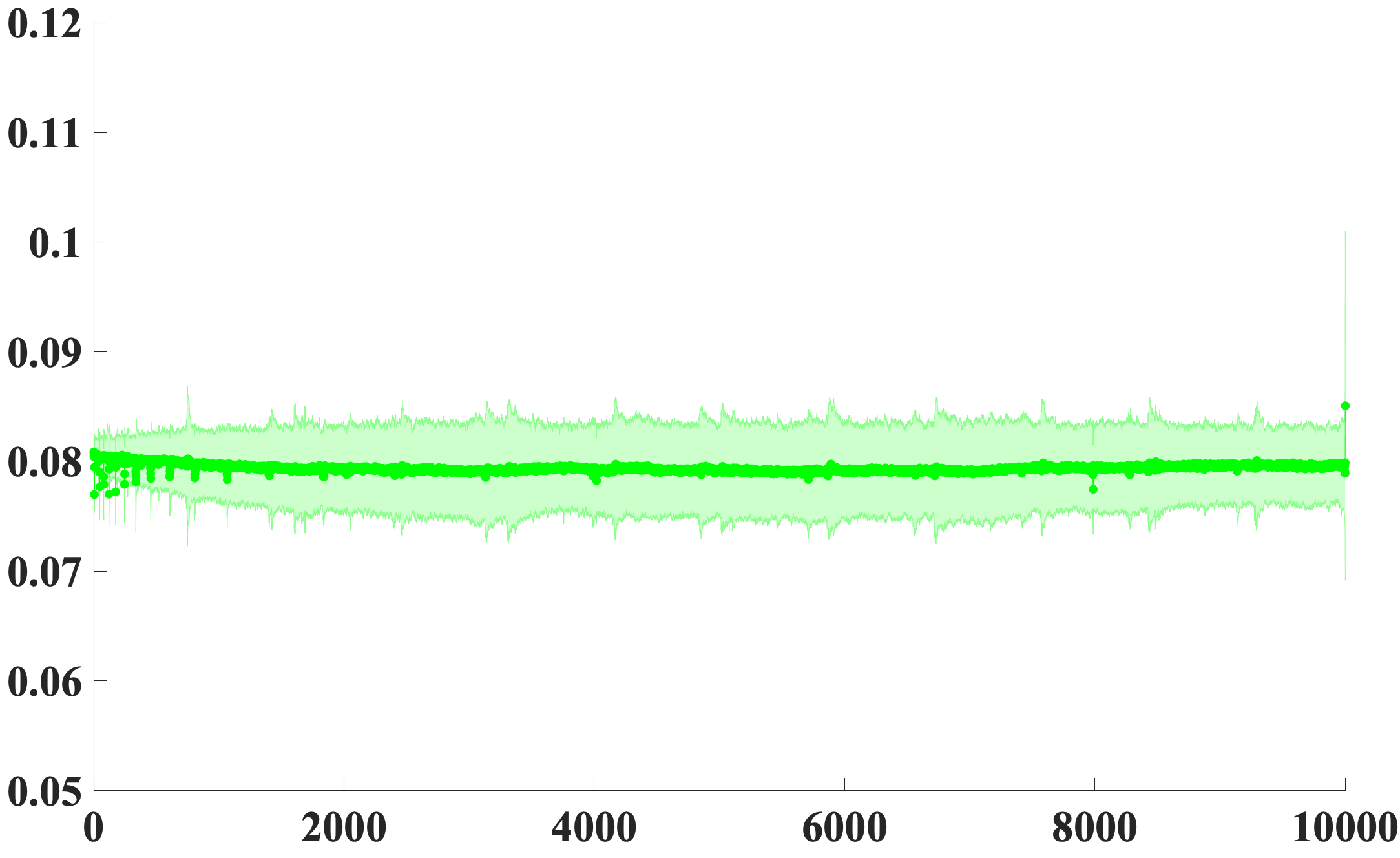}}
            \begin{axis}[
                trim axis left, trim axis right, scale only axis,
            	width = 0.24\textheight,
            	height = 0.16\textheight,
            	ylabel = {Performance [$\frac{iter}{s}$]},
            	xlabel = {Number of iterations},
                axis line style={draw=none},
            	ticks=none,
                x label style={font=\footnotesize},
                y label style={font=\footnotesize},
            ]
        \end{axis}
    \end{tikzpicture}
    
    \begin{tikzpicture}
        \put(-0.35,-0.01){\includegraphics[scale=0.14]{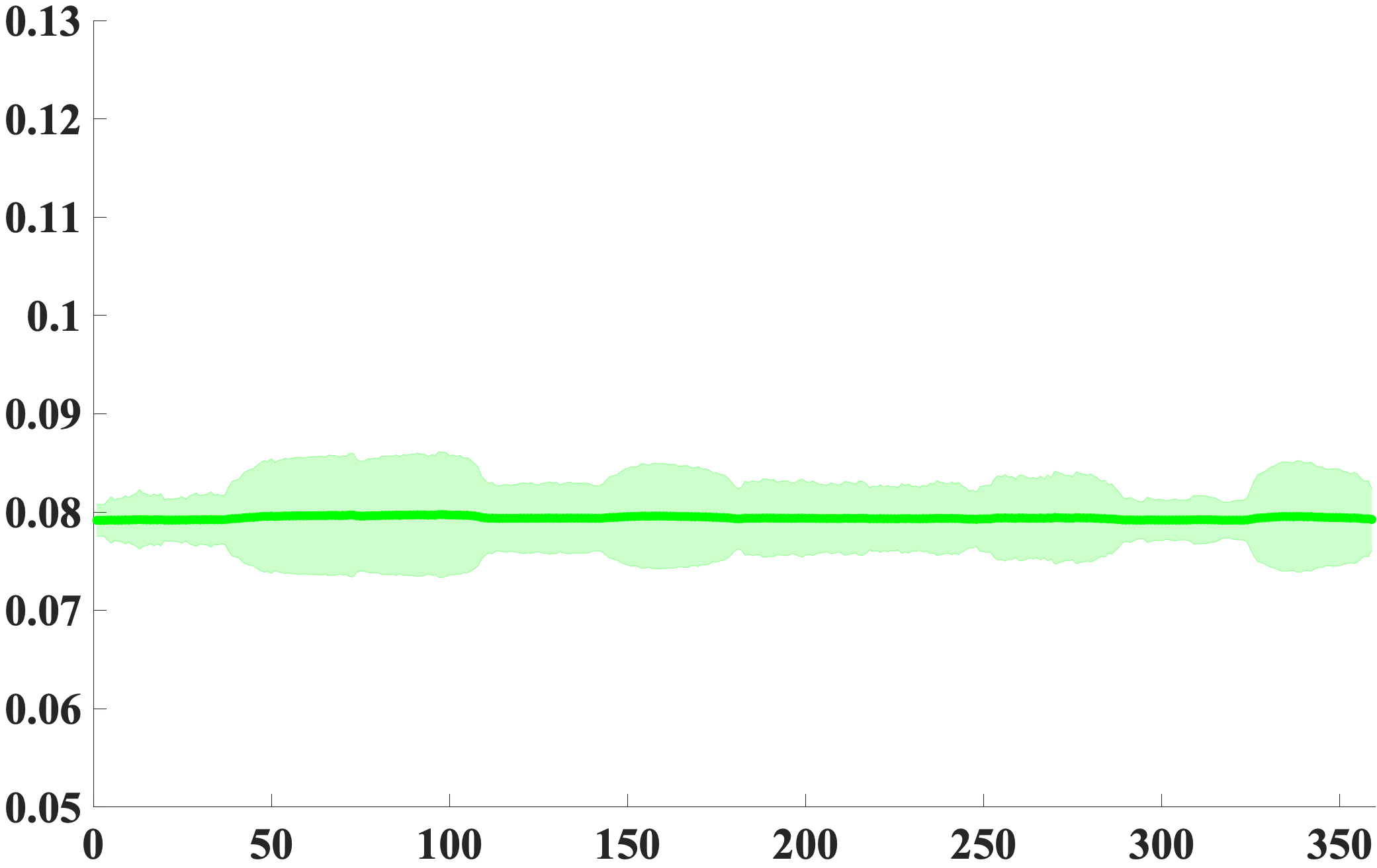}}
        \begin{axis}[
            trim axis left, trim axis right, scale only axis,
        	width = 0.22\textheight,
        	height = 0.16\textheight,
        	ylabel = {Performance [$\frac{iter}{s}$]},
        	xlabel = {Processes ID},
            axis line style={draw=none},
        	ticks=none,
            x label style={font=\footnotesize},
            y label style={font=\footnotesize},
        ]
        \end{axis}
    \end{tikzpicture}
    \caption*{(a) No extra work\\~\\~}
    \end{minipage}\quad
    \begin{minipage}{0.3\textwidth}
    \begin{tikzpicture}
        \put(-0.35,-0.01){\includegraphics[scale=0.115]{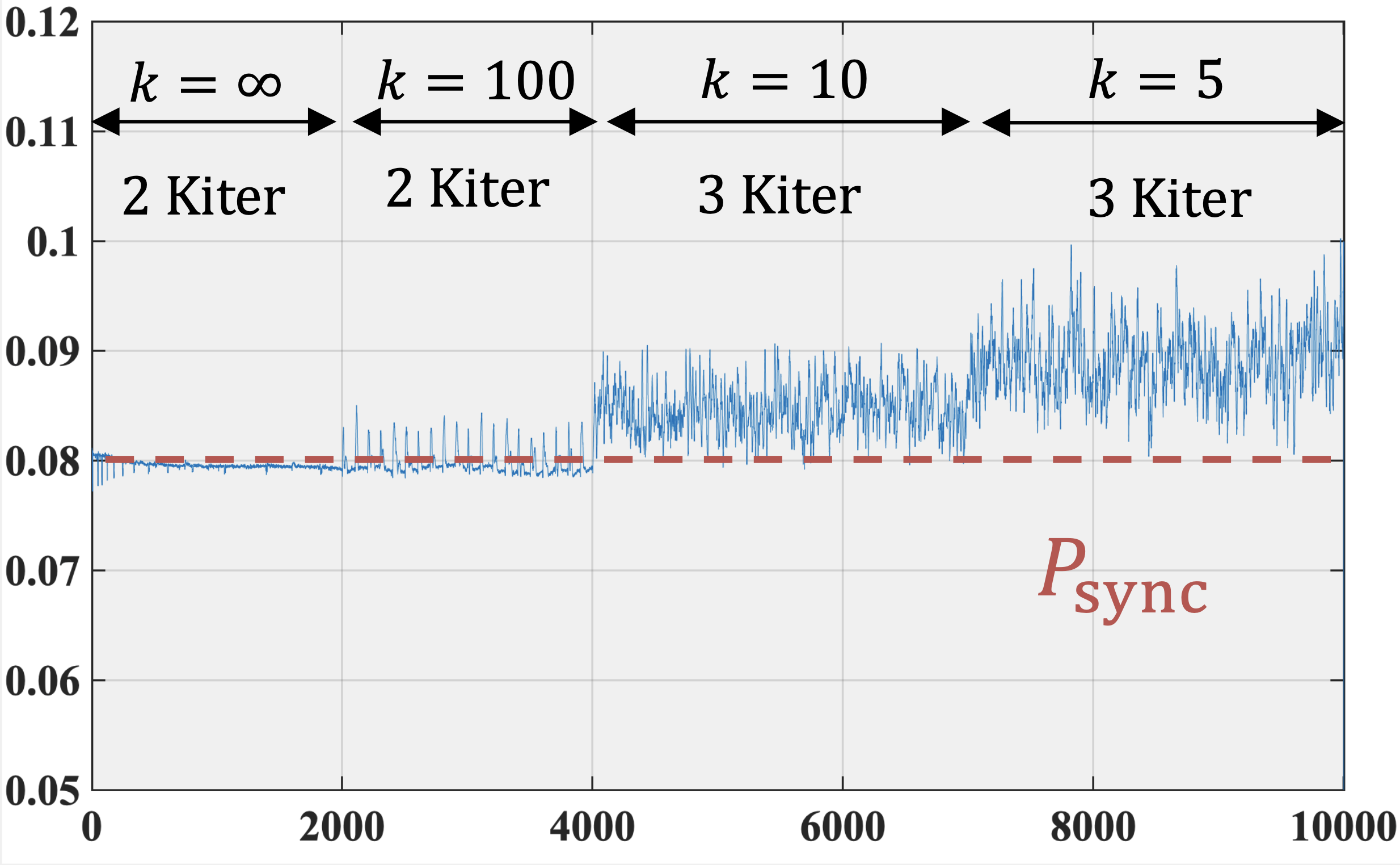}}
        \begin{axis}[
            trim axis left, trim axis right, scale only axis,
        	width = 0.22\textheight,
        	height = 0.16\textheight,
        	ylabel = {Performance [$\frac{iter}{s}$]},
        	xlabel = {Number of iterations},
            axis line style={draw=none},
        	ticks=none,
            x label style={font=\footnotesize},
            y label style={font=\footnotesize},
        ]
        \end{axis}
    \end{tikzpicture}
    
    \begin{tikzpicture}
        \put(-0.35,-0.01){\includegraphics[scale=0.14]{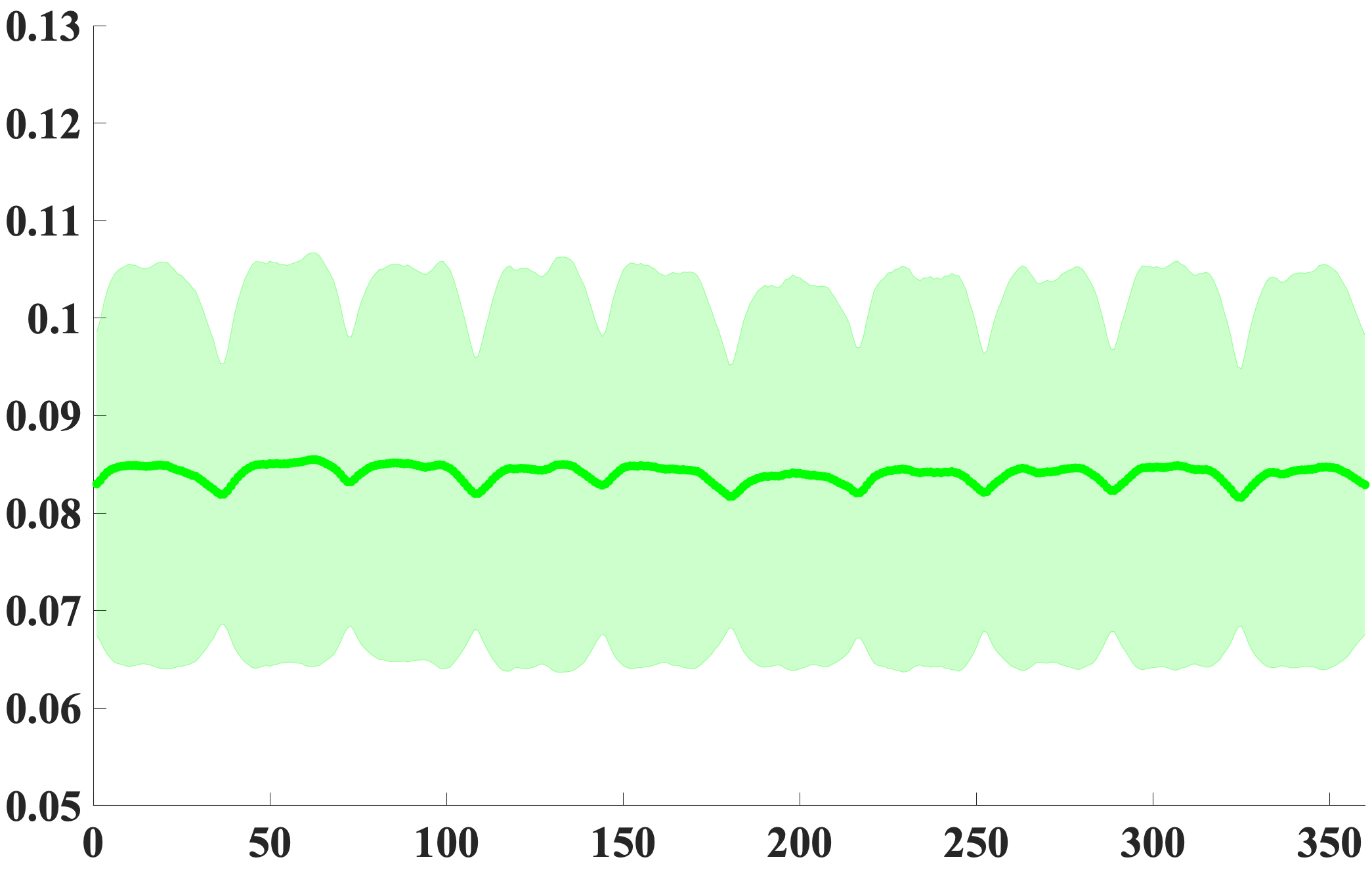}}
        \begin{axis}[
            trim axis left, trim axis right, scale only axis,
        	width = 0.22\textheight,
        	height = 0.16\textheight,
        	ylabel = {Performance [$\frac{iter}{s}$]},
        	xlabel = {Processes ID},
            axis line style={draw=none},
        	ticks=none,
            x label style={font=\footnotesize},
            y label style={font=\footnotesize},
        ]
        \end{axis}
    \end{tikzpicture}
    \caption*{(b) Extra compute-bound workload on a random MPI process every \{$\infty$, 100th, 10th, 5th\} iteration for the subsequent \{2~\mbox{K}, 2~\mbox{K}, 3~\mbox{K}, 3~\mbox{K}\} iterations}
    \end{minipage}\quad
    \begin{minipage}{0.3\textwidth}
    \begin{tikzpicture}
        \put(-0.35,-0.01){\includegraphics[scale=0.115]{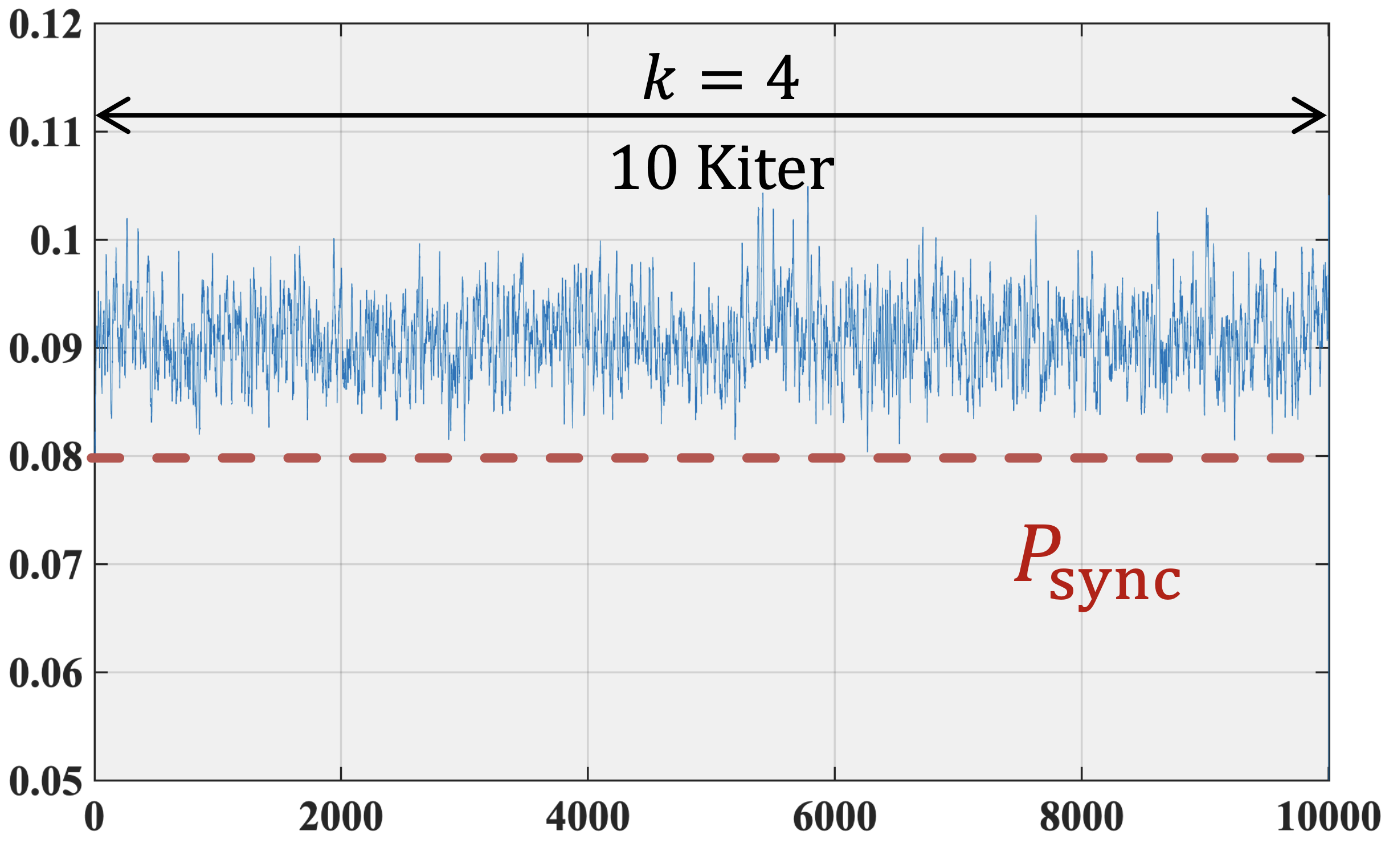}}
        \begin{axis}[
            trim axis left, trim axis right, scale only axis,
        	width = 0.22\textheight,
        	height = 0.16\textheight,
        	ylabel = {Performance [$\frac{iter}{s}$]},
        	xlabel = {Number of iterations},
            axis line style={draw=none},
        	ticks=none,
            x label style={font=\footnotesize},
            y label style={font=\footnotesize},
        ]
        \end{axis}
    \end{tikzpicture}
    
    \begin{tikzpicture}
        \put(-0.35,-0.01){\includegraphics[scale=0.14]{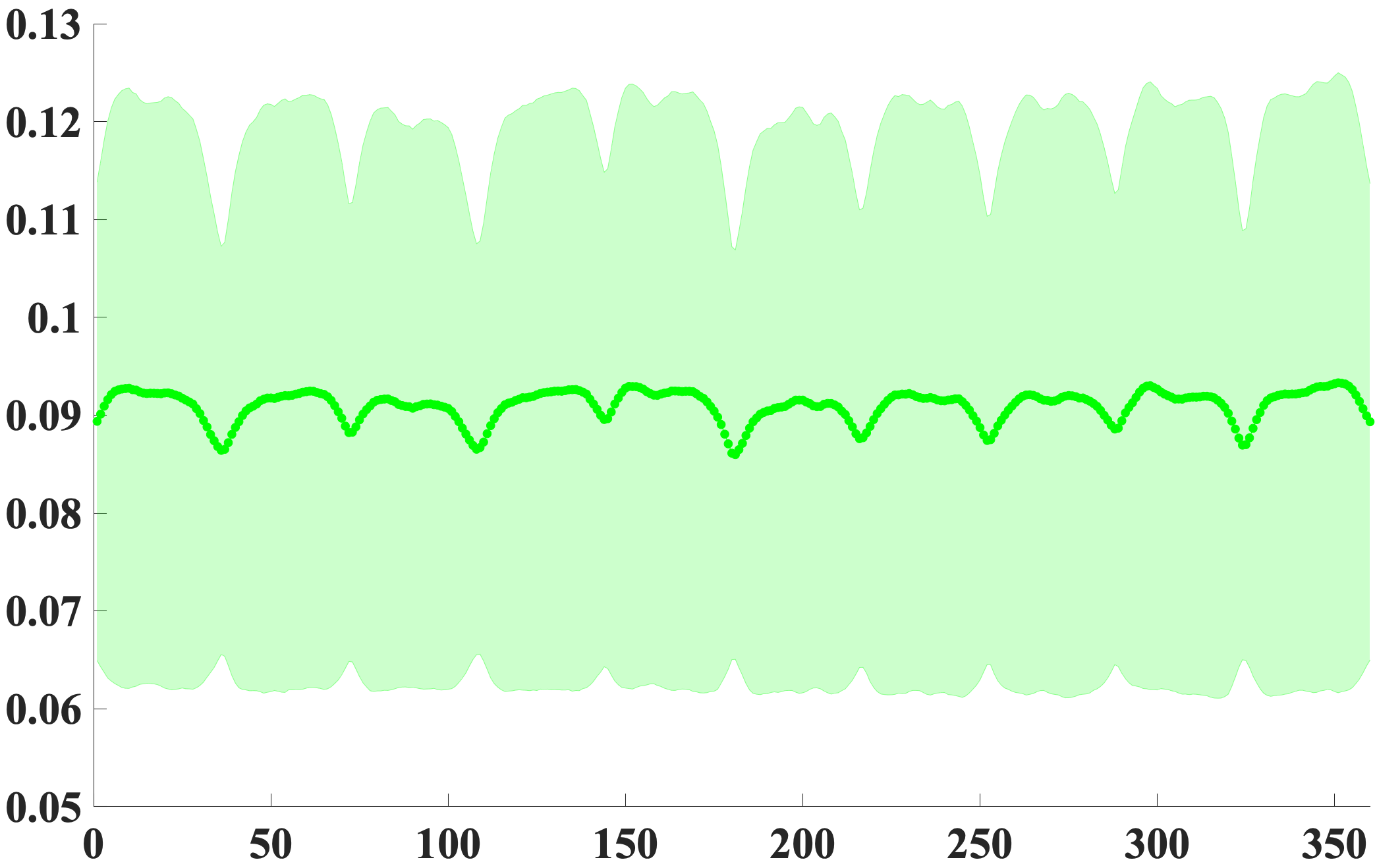}}
        \begin{axis}[
            trim axis left, trim axis right, scale only axis,
        	width = 0.22\textheight,
        	height = 0.16\textheight,
        	ylabel = {Performance [$\frac{iter}{s}$]},
        	xlabel = {Processes ID},
            axis line style={draw=none},
        	ticks=none,
            x label style={font=\footnotesize},
            y label style={font=\footnotesize},
        ]
        \end{axis}    
    \end{tikzpicture}
    \caption*{(c) Extra compute-bound workload on a random MPI process every fourth iteration\\~}
    \end{minipage}
	\caption{Per-process MST benchmark performance on $360$ processes (5 nodes) of Fritz and $10$~\mbox{K} iterations.
    (a) top: mean ($\bar{P}$) and standard deviation ($\sigma$) of performance versus iteration number; bottom: mean and standard deviation of performance versus process rank. 
    (b) Experiment with successively more frequent injections of extra workload on a random process.
    (c) Experiment with constant-frequency delay injection across the whole runtime. 
    The composite performance for synchronized execution is $0.08\,\frac{iter}{s}$ per process.}
	\label{fig:MSTnoise} 
\end{figure*}
\begin{figure*}[tb]
    \begin{minipage}{\textwidth}
    \begin{subfigure}[t]{0.48\textwidth}
    \centering
    \includegraphics[scale=0.42]{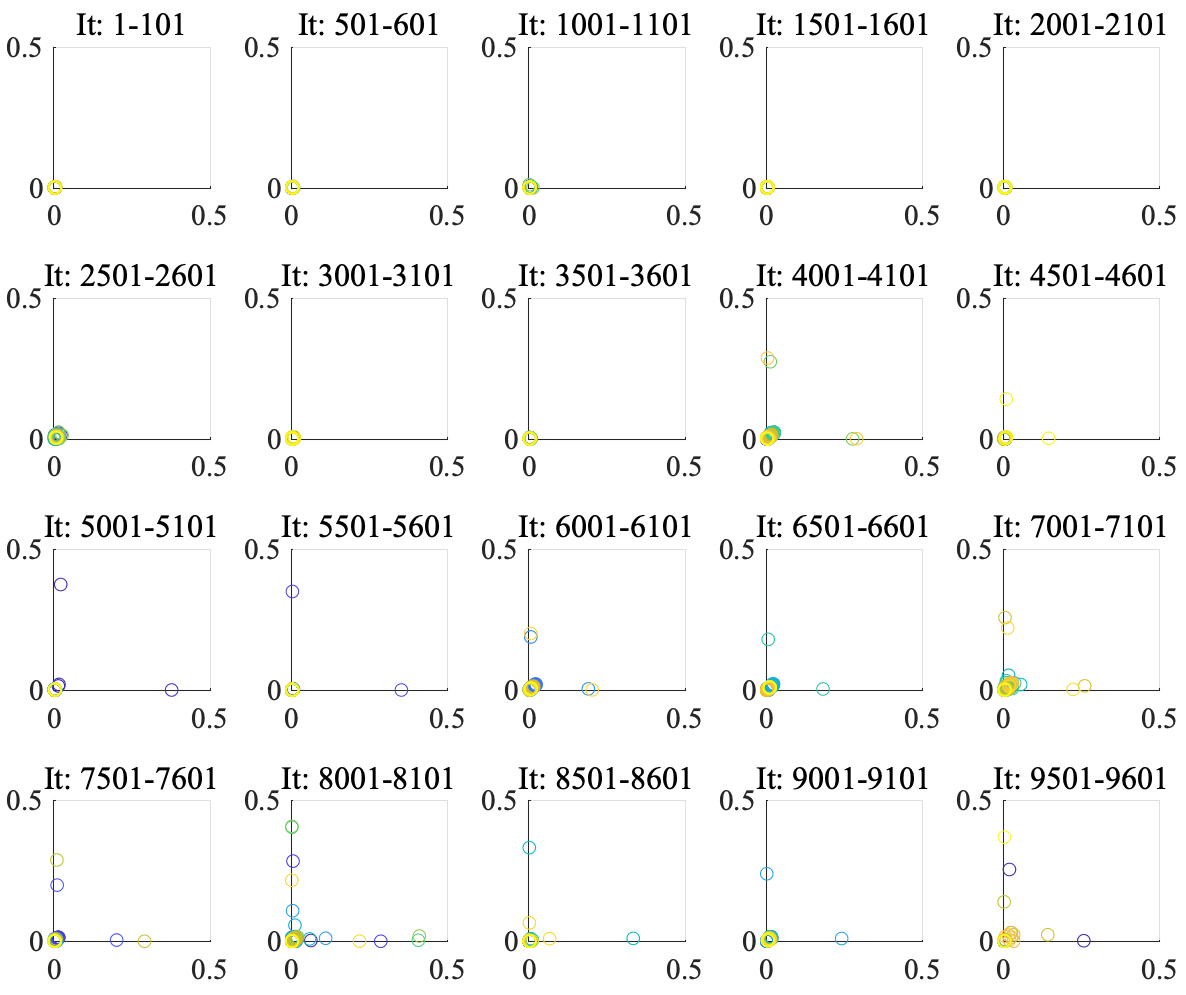}
    \caption{Snippet 100 iterations view of MPI times [s]}
    \end{subfigure} \quad
    \begin{subfigure}[t]{0.48\textwidth}
    \centering
    \includegraphics[scale=0.42]{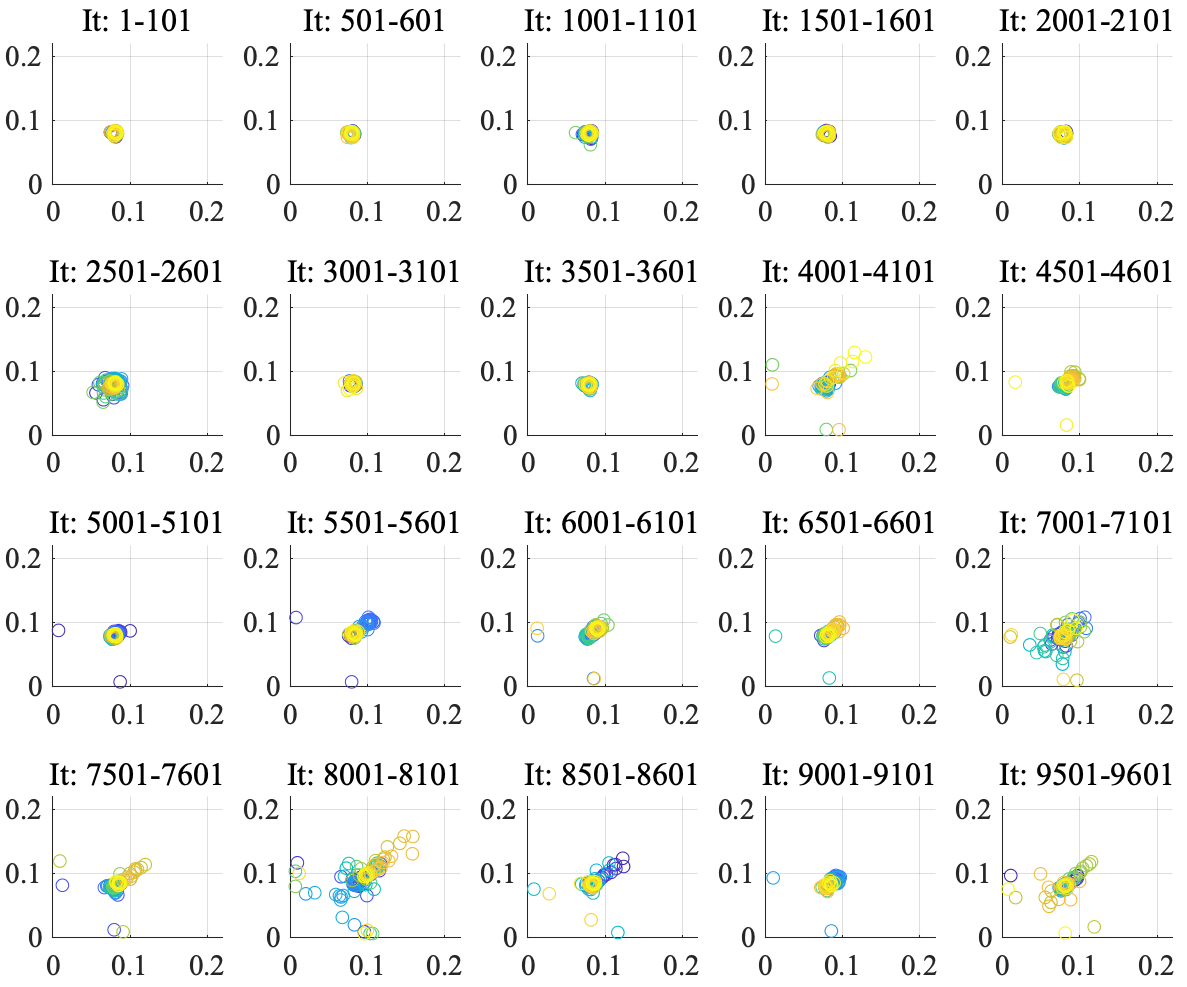}
    \caption{Snippet 100 iterations view of performance [iter/s]}
    \end{subfigure} \quad
    \begin{subfigure}[t]{0.48\textwidth}
    \centering   
    \includegraphics[scale=0.25]{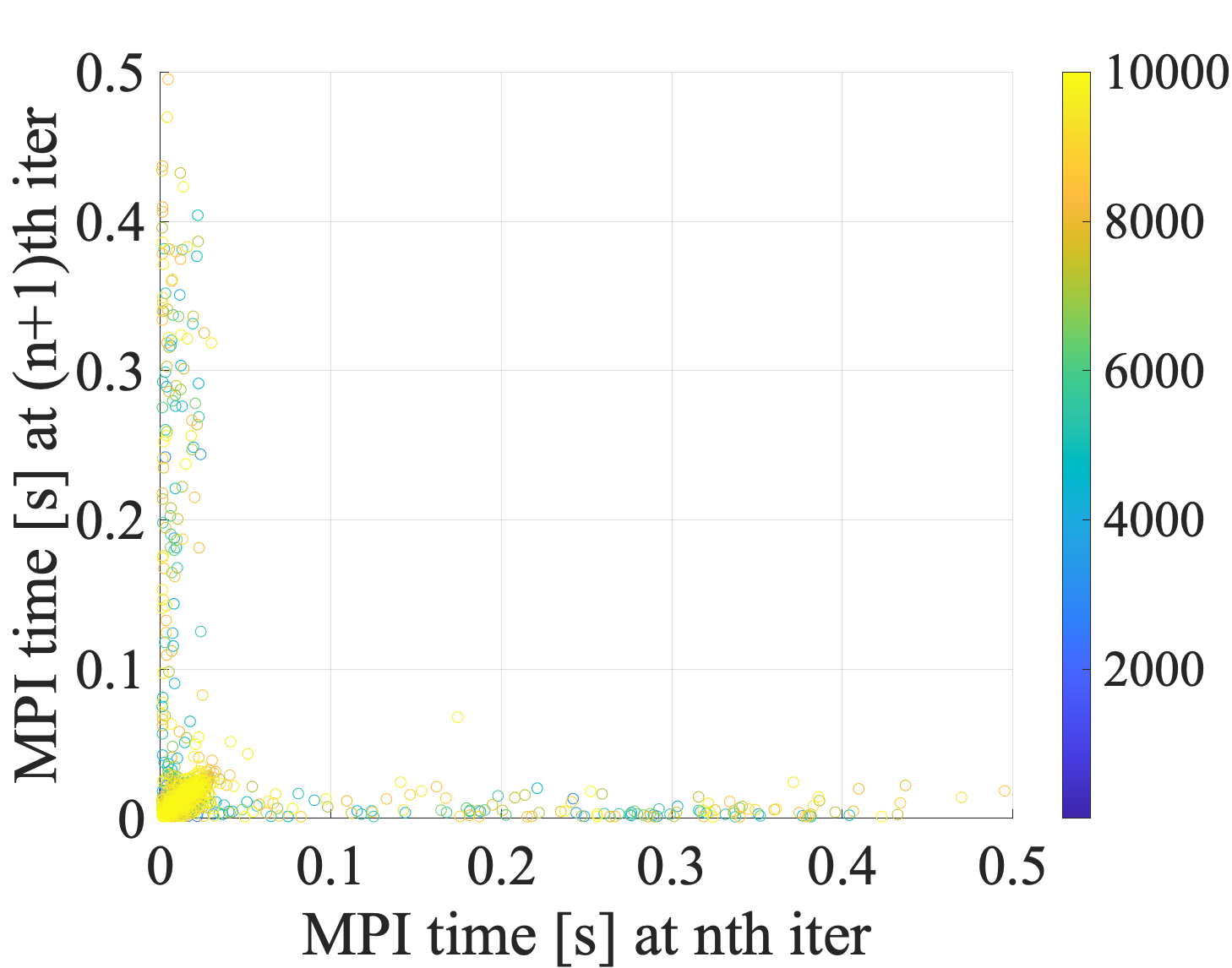}
    \caption{Entire 10K iterations view of MPI times [s]}
    \end{subfigure} \quad
    \begin{subfigure}[t]{0.48\textwidth}
    \centering
    \includegraphics[scale=0.25]{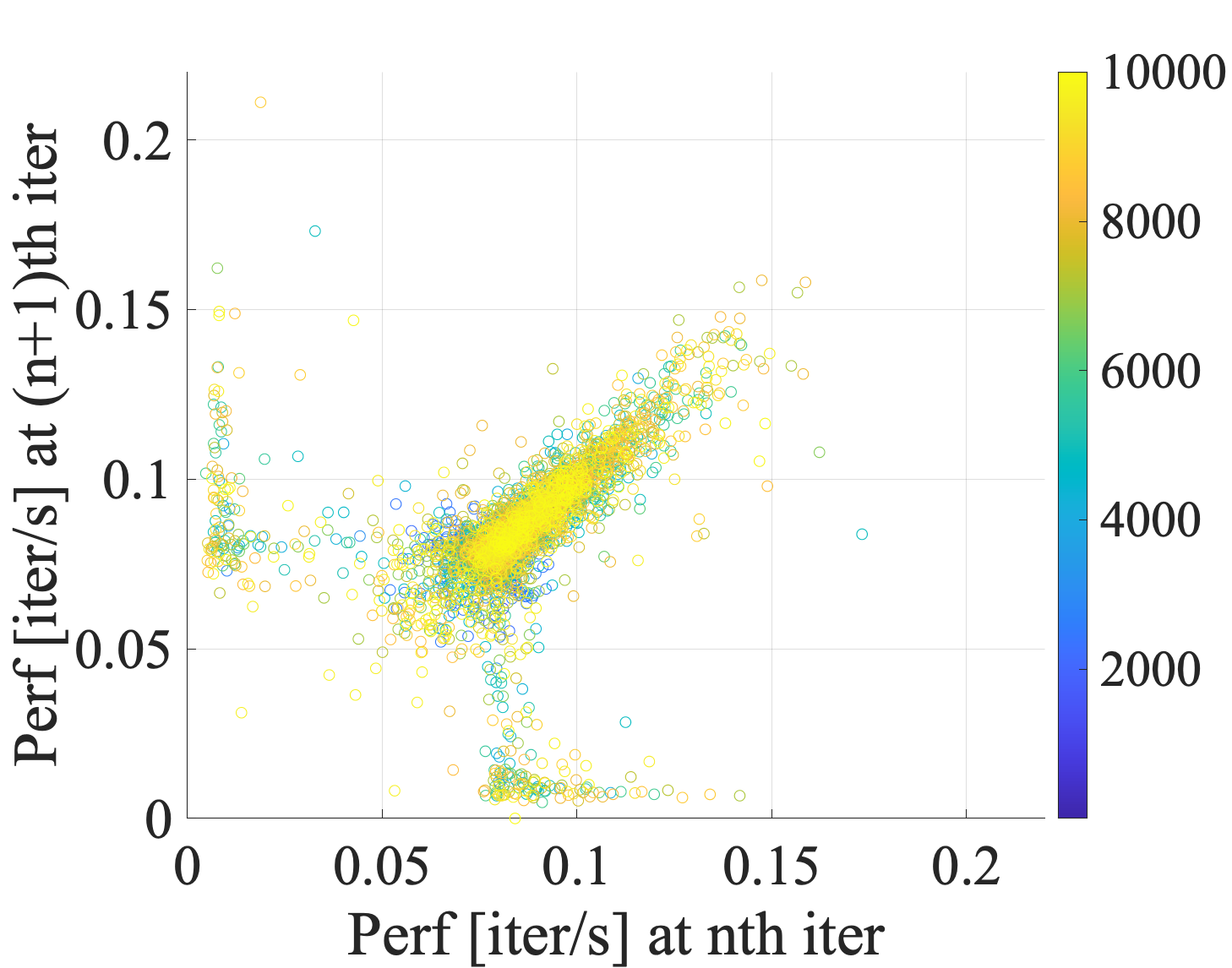}
    \caption{Entire 10K iterations view of performance [iter/s]}
    \end{subfigure}
    \end{minipage}
    \caption{Phase-space analysis of the run in Fig~\ref{fig:MSTnoise}(b) for MPI process rank 36 (on second socket), using MPI times (in seconds, left) and performance (in iterations per second, right).
    Top row: snippet views of 100 time steps each. Bottom row: entire run.
    }
    \label{fig:MSTnoisephasespace}
\end{figure*}
\section{Hardware-software setup and experimental methodology} \label{sec:setup}
Table~\ref{tab:systems} shows the hardware and software environments we used for all experiments.
We chose the following distinct clusters, each with a different interconnect, core count, and memory bandwidth, to ensure the wide applicability of our findings:
\begin{enumerate}
    \item Omni-Path Meggie\footnote{\url{https://hpc.fau.de/systems-services/documentation-instructions/clusters/meggie-cluster}} cluster comprising two Intel Xeon Broadwell CPUs per node with 10 cores each
    \item Omni-Path SuperMUC-NG\footnote{\url{https://doku.lrz.de/display/PUBLIC/SuperMUC-NG}} cluster comprising two Intel Xeon Skylake SP CPUs per node with 24 cores each
    \item Infiniband Fritz\footnote{\url{https://hpc.fau.de/systems-services/documentation-instructions/clusters/fritz-cluster}} cluster comprising two Intel Xeon Ice Lake CPUs per node with 36 cores each
\end{enumerate}
Although hyper-threading is active on the SuperMUC-NG system, in this work we ignore it and only use the consecutive physical cores on a node that are mapped to consecutive MPI processes using the \code{I\_MPI\_PIN\_PROCESSOR\_LIST} environment variable (with \code{mpirun}) or \code{--cpu-bind=rank} (with \code{srun}) in the MPI implementation.
Sub-NUMA Clustering (SNC) is activated on Fritz, which means that the basic scaling unit (i.e., one ccNUMA domain) is half a socket (18 cores).
The clock frequency of all Meggie and Fritz nodes was consistently fixed to the base values of their respective CPUs via the SLURM batch scheduler.
The CPUs on SuperMUC-NG operate by default at an effective clock speed of $2.3$~\GHZ, which is lower than their maximum base clock speed of $3.10$~\GHZ.
The \code{likwid-perfctr} tool was employed to make sure the expected clock frequency was actually set.
We instrumented all programs to collect the timestamps of entering and leaving MPI calls at each iteration of each MPI process over the course of the whole run in order to measure MPI waiting times.
Time stamps were taken using the \CPP{} high-resolution chrono clock (with the \enquote{\code{-std=c++11}} compiler option) for \CPP{} codes, while for C codes we employed \verb.getwalltime()..
\code{ClusterCockpit}~\cite{Clustercockpit:2019} was used for the job monitoring on Meggie and Fritz.
The traces of MPI processes at any point during a run were visualized using the graphical user interface of the Intel Trace Analyzer and Collector (\code{ITAC}) tool (with \enquote{\code{-trace}} compiler option)\footnote{The instrumentation-based \code{ITAC} tool has a single tick of $1$~\NS\ clock resolution and ensures that clocks across MPI processes are synchronized: \url{https://intel.com/content/www/us/en/develop/documentation/itac-user-and-reference-guide/top/intel-trace-collector-reference/time-stamping/clock-synchronization.html}}.
The working set for memory-bound programs was chosen to be at least ten times the size of all \LLC to prevent it from fitting into the available cache.\footnote{The \LLC is made up of the non-inclusive victim L3 plus the L2 caches in the Skylake and Ice lake processors of SuperMUC-NG and Fritz, while it is only the L3 caches in the Broadwell processors of Meggie.}
The ratio of data volume to wall-clock time was used to calculate memory bandwidths.
At least two warm-up time steps, including global synchronization, were run before the actual measurements to give the MPI runtime a chance to settle and get rid of first-call overhead.
To account for variations in runtime, we repeated code executions several times and only significant statistical fluctuations were reported.

In evaluating the analysis results for four applications, automatic overlap of communication and computation non-lockstep execution of MPI processes is highlighted.
The aim is to provide a thorough analysis of the impact of asynchronous execution, rather than a comprehensive analysis of each application.
Any performance improvement caused by MPI asynchronicity must be carefully distinguished from other positive performance effects such as reductions in communication volume.
For example, when attributing performance changes to asynchronicity, the \emph{natural collective cost} measured in a fully synchronized environment is always subtracted from the observed runtime.
This ensures that the trivial positive effect of eliminating the collective is not visible in the data; observed performance effects must come from other sources.

\section{{M}PI-augmented {S}TREAM {T}riad ({MST})}\label{sec:mst}
The McCalpin STREAM Triad \enquote{\code{A(:)=B(:)+s*C(:)}}~\cite{mccalpin1995memory} is often used to measure the attainable memory bandwidth of a processor. 
The MST code adds communication after each full run of the loop (which we call \emph{iteration} in the following) of the loop to mimic a real MPI-parallel, strongly memory-bound program. It is thus a clean setup that can be used to demonstrate the effects of complicated memory-bound applications. The experiments in this section were conducted on the \emph{Fritz} cluster.

\subsection{Implementation}\label{sec:noiseinject}
The fundamental organization of the MST benchmark is shown in Listing.~\ref{fig:mst}.
\begin{lstlisting}[
    style=small,float=ht,
    caption= Pseudo-code implementation of the MPI-augmented STREAM, 
    label=fig:mst
]
for (int iter=0; iter<numIters; iter++) {
    for (int i=0; i<arrayElements; i++) {
      A[i] = B[i]+s*C[i];
    }
    for ( int j = 0; j < 2; j++ ) {
      MPI_Isend(.., &req[j*2]);
      MPI_Irecv(.., &req[1+j*2]);
    }
    MPI_Waitall( 4, &req[0], &stt[0] );
}
\end{lstlisting}

An overall working set of $48$~\GB~($2\times 10^9$ array elements, much larger than the aggregate LLC of the CPUs) is distributed evenly among $360$ MPI processes on 10 nodes of Fritz. 
Each process sends and receives a message of $1$~\MB, which is way beyond the eager limit of the MPI implementation, to each of its two direct neighbors after a full iteration. All processes form a closed chain (periodic boundary conditions). The use of non-blocking point-to-point calls and a final \texttt{MPI\_Waitall} ensures that the communication is bidirectional.
The compiler option \enquote{\code{-qopt-streaming-stores}} compiler option was used to enable the generation of streaming stores, leading to a code balance of 12~\BF\ due to the lack of write-allocate transfers.

\begin{figure*}[ht]
\centering
\includegraphics[scale=1]{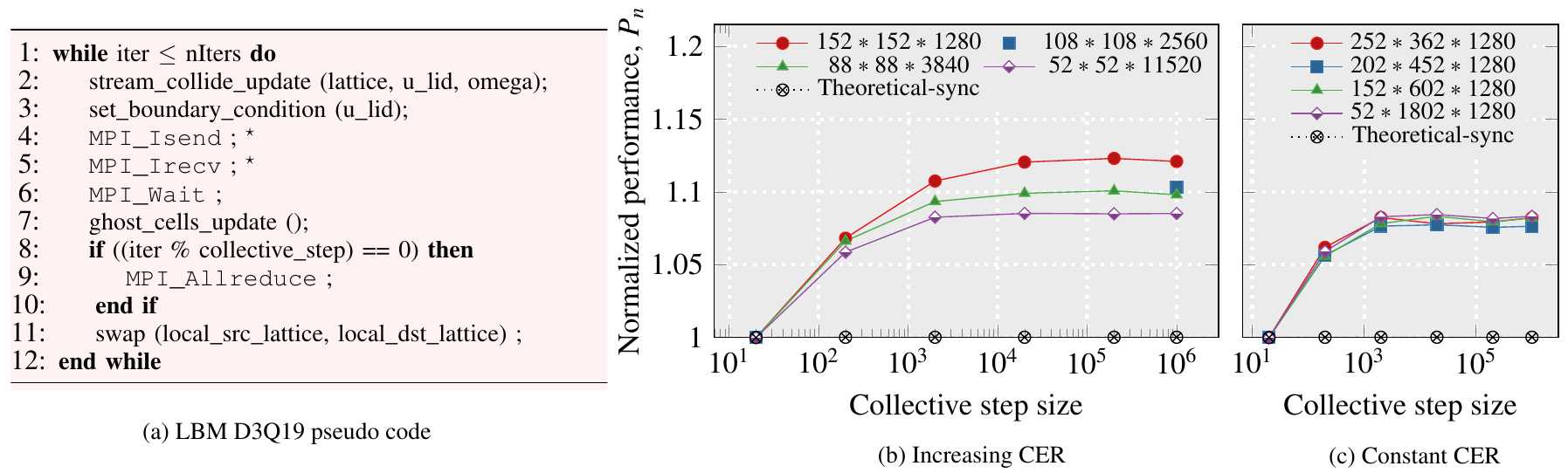}
\caption{{(a)} In the D3Q19 LBM implementation, boundary layers are exchanged using non-blocking MPI routines, including (un)packing in between. (b+c) Influence of the frequency of collectives
			(specifically \texttt{MPI\_Allreduce()}) on LBM performance (executing $10^6$ iterations on $64$ Meggie nodes, i.e., $1280$ MPI processes).
			Graphs are linear on the $y$-axis, and a base-10 log scale is used for the $x$-axis.
            Performance in~\MLUPS\ is normalized to the case of one collective per 20 iterations.
            {(b)} Normalized performance (speedup vs.\ smallest collective stepsize) vs.\ collective step size for constant overall
			problem size ($8.37$~\GB) but different domain shapes, which causes different communication overhead.
			{(c)} Normalized performance for constant \ac{CER} but different shapes of the cross section $n_x \cdot n_y$ at constant overall problem size.
            Finally, the black crosses show the calculated speedup when subtracting the \texttt{MPI\_Allreduce} time from the runtime for the problem size with the best desynchronization speedup in each diagram. This speedup is so close to 1 it cannot be visualized properly.} 
\label{fig:LBMMeggie} 
\end{figure*}

Noise was generated by extending the STREAM Triad computational phases on random processes by burdening them with extra work.
For this extra work, we use the compute-bound loop shown in Listing~\ref{algo:noise}, which is dominated by double-precision floating-point divides. How often this injection takes place is configurable.
\begin{lstlisting}[
    style=small,float=ht,
    caption= Extra workload injection to slow down processes., 
    label=algo:noise
]     
for (std::size_t j = 0; j < 10000000; j++){
    sum = sum + 4.0 / ( 1.0 + j * j);
}
\end{lstlisting}

\subsection{Asynchronicity through noise}
Figure~\ref{fig:MSTnoise}(a) (top) shows a timeline of measured per-process performance vs. iterations on 360 processes (5 Fritz nodes) without deliberate noise injection. The thick line is the average across processes, while the light area shows the standard deviation. The observed average performance of $0.08\,\frac{iter}{s}$ translates into an effective memory bandwidth of 139\,\GBS\ per socket. As the upper limit for the STREAM Triad loop with streaming stores is 162\,\GBS, 14\% of the overall time goes into communication overhead. If this overhead could be overlapped by desynchronization, about 
$0.094\,\frac{iter}{s}$ could be attained. Figure~\ref{fig:MSTnoise}(a) (bottom) shows the performance of each individual process over the entire program runtime as average and standard deviation. 

In the experiments shown in Fig.~\ref{fig:MSTnoise}(b), a random process was picked every $k$ iterations and subjected to the aforementioned extra workload injection. The number $k$ was decreased from $k=\infty$ to $k=5$ in three steps. In addition, Fig.~\ref{fig:MSTnoise}(c) shows a full run with $k=4$ throughout. It can be seen that more frequent injections lead to better performance; if they are not frequent enough (as seen in the $k=100$ phase in Fig.~\ref{fig:MSTnoise}(b) (top)), the system goes back to a synchronized state. With $k=4$, the average performance across processes comes close to the theoretical limit of $0.094\,\frac{iter}{s}$.
These findings give an explanation of why a slightly imbalanced workload is preferable to bulk-synchronous execution in bandwidth-limited parallel programs.
Slowing down processes causes asynchronicity (see the bottom plots of each case in Figure~\ref{fig:MSTnoise}(a-c)) which allows fewer concurrent processes to better utilize available, limited memory bandwidth through communication overlap.
These experiments show that asynchronous execution can be initiated by noise injection, leading to better overall performance. A general theory about how much noise exactly is needed to achieve a certain speedup does not exist yet, however.

\begin{table}[tb]
	\begin{adjustbox}{width=0.48\textwidth}
    \Huge
    \pgfplotstabletypeset[
    every column/.style=,
    color cells={min=0,max=14},
    /pgf/number format/fixed,
    /pgf/number format/precision=2,
    col sep=comma,
    every head row/.style={%
    	before row={\hline
    		& \multicolumn{4}{c}{Domain shape $n_x . n_y . n_z$ at a constant overall problem size} \\
    	},
    	after row=\hline
    },
    @content options for rows={0}{/pgfplots/colormap/redyellow,color cells={min=1.42,max=5.9}},
    @content options for rows={1}{/pgfplots/colormap/redyellow,color cells={min=0.19,max=6.34}},
    @content options for rows={2}{/pgfplots/colormap/redyellow,color cells={min=0.28,max=9.74}},
    @content options for rows={3}{/pgfplots/colormap/redyellow,color cells={min=0.08,max=1.08}},
    colfont={2},
    colfont={3},
    colfont={4},
    colfont={6},
    colfont={5},
    ]{
    	{Program phase},152.152.1280,108.108.2560,88.88.3840,52.52.11520
    	Computation [\MS] ,5.896,4.819,2.857,2.494
    	Communication [\MS] ,6.34,2.283,0.787,0.188
    	Comm. aggregate sum [\GB] ,9.74,1.65,1.07,0.277
    	CER ,1.075305291723202,0.473749740610085,0.275463773188659,0.075380914194066
    }
    \end{adjustbox}
    \caption{In a single LBM iteration, domain shapes and their associated computation and communication run times on the Meggie system.
    The red-to-yellow scale serves only to guide the viewer's attention along the rows and is not applicable to different phases.
    As the cross-section of the domain $n_x \times n_y$ grows, the communication becomes more relevant leaving plenty of room for overlap.
    The messages on each process are communicated via the rendezvous protocol at a large cross-section of the domain until the domain becomes small enough to reach the eager limit.
    }
	\label{tab:LBMdomains}
\end{table}
\mypara{Phase-space analysis}
In order to depict the long-term evolution of a metric such as performance or MPI waiting time, the \emph{phase-space plot} turned out to be useful~\cite{AfzalHW:2022:2}. It is a scatter plot with data points at coordinates $(m_i,m_{i+1})$, where $m_i$ is the value of some metric at time step $i$.
Hence, a data point shows how this metric has evolved from time step $i$ to time step $i+1$. By color coding the data points, one can visualize the long-term evolution of a metric. It turns out that certain behaviors regarding (de)synchronized execution show up as distinctive patterns in the phase space plot. In order to limit clutter, we show data for a single MPI process only, unless noted otherwise. Sometimes it is advantageous to restrict one plot to a certain part (a ``snippet'') of the timeline and show the long-term evolution using multiple plots.

Figure~\ref{fig:MSTnoisephasespace} shows a phase-space analysis of the scenario  depicted in Fig.~\ref{fig:MSTnoise}(b).  In (a) and (c) we use MPI time (in seconds) as a principal metric, while in (b) and (d) we use performance in iter/s. Measurements were taken on process 36 (first process on second socket). In Fig.~\ref{fig:MSTnoisephasespace}(a), each plot shows a snippet of 100 iterations at regular intervals along the iteration space. Up to iteration 4000, execution is mostly synchronous and the MPI waiting time is small. After iteration 4000, the delay injections cause more frequent outliers along the axes; a point near an axis means that this MPI delay is short-lived and disappears in the next step. In addition, as the delays get more frequent, the point cloud near the origin gets diluted, indicating a persistent increase in MPI waiting time. Figure~\ref{fig:MSTnoisephasespace}(c) is less revealing in this scenario since the whole timeline is mapped to the color scale. In Fig.~\ref{fig:MSTnoisephasespace}(b) and (d), performance is used as a metric. The snippet views indicate that performance increases with time (in line with the more frequent delay injections), as can be seen from the dot cloud moving along the diagonal towards higher performance. A point on the diagonal means that a performance level at step $i$ can also be observed in step $i+1$, indicating persistence. One should also add that, even though the average performance across processes increases with injection frequency, so does the performance variation. In Fig.~\ref{fig:MSTnoisephasespace}(d), the whole run is shown again, visualizing this effect more prominently.


\highlight{\emph{Upshot}:
Not all noise is detrimental; in fact, noise can enhance asymptotic performance by pushing the system out of the bulk-synchronous lock-step and thus facilitating spontaneous asynchronous execution and communication overlap. 
Though this is valid only in the absence of high-frequency synchronizing collectives and under the condition that the bottleneck(s) and relevant communication overhead are present.
}

\begin{figure*}[ht]
     \begin{tikzpicture}
        \put(-0.35,-0.01){\includegraphics[scale=0.35]{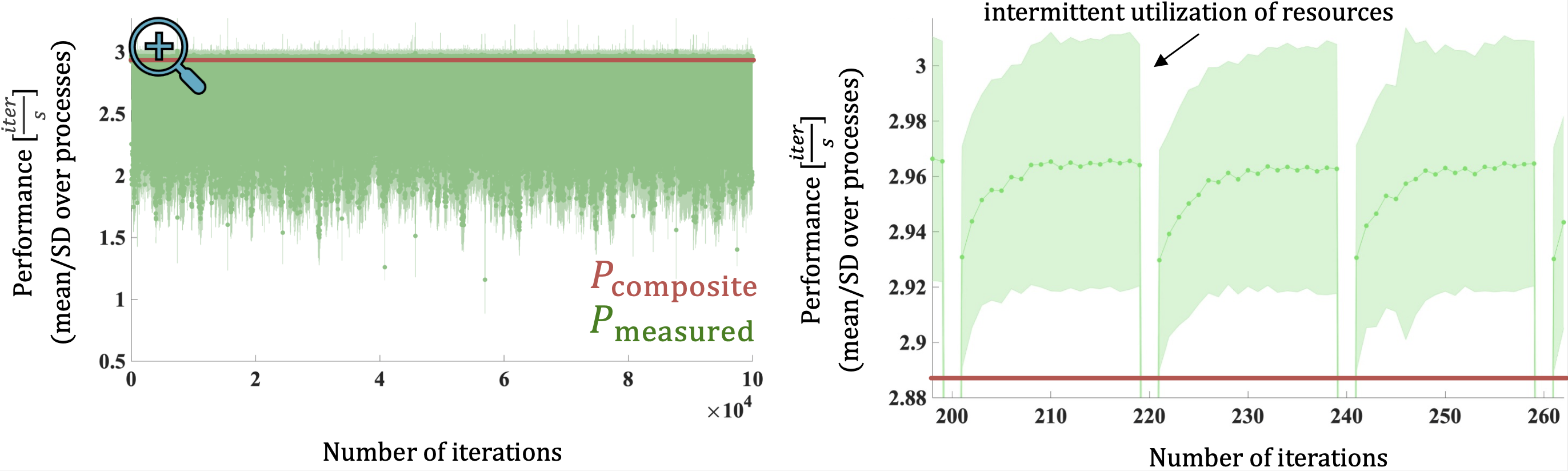}}
        \begin{axis}[
            trim axis left, trim axis right, scale only axis,
        	width = 0.3\textheight,
        	height = 0.18\textheight,
            axis line style={draw=none},
        	ticks=none,
            x label style={font=\footnotesize},
            y label style={font=\footnotesize},
        ]
        \end{axis}
        \node [font=\footnotesize] at (4.5,-0.3){(a) Entire 100~\mbox{K}\ iterations run};
        \node [font=\footnotesize] at (13.5,-0.3){(b) Snippet for 60 iterations};
    \end{tikzpicture}
	\caption{Mean performance (green) with standard deviation (light green) of LBM D3Q19 program run for 100K iterations on 1440 MPI processes of 20 Fritz nodes. The composite performance is being compared with the measured performance of the program run using \code{MPI\_Allreduce} in every 20th iteration.
 }
\label{fig:LBMnonoise} 
\end{figure*}
\begin{figure*}[ht]
    \begin{minipage}{\textwidth}
    \begin{subfigure}[t]{0.76\textwidth}
    \centering
    \includegraphics[scale=0.6]{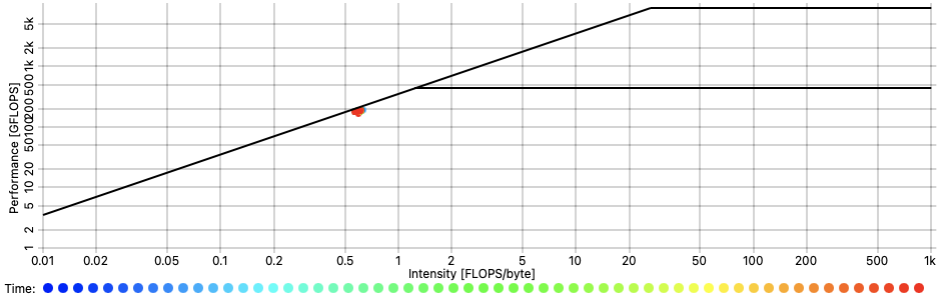}\quad
    \includegraphics[scale=0.6]{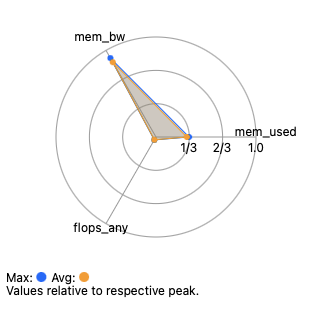}
    \end{subfigure}
    \begin{subfigure}[t]{0.2\textwidth} 
	\begin{tikzpicture}
	\put(-0.4,0) {\includegraphics[width=0.101\textheight,height=0.1\textheight]{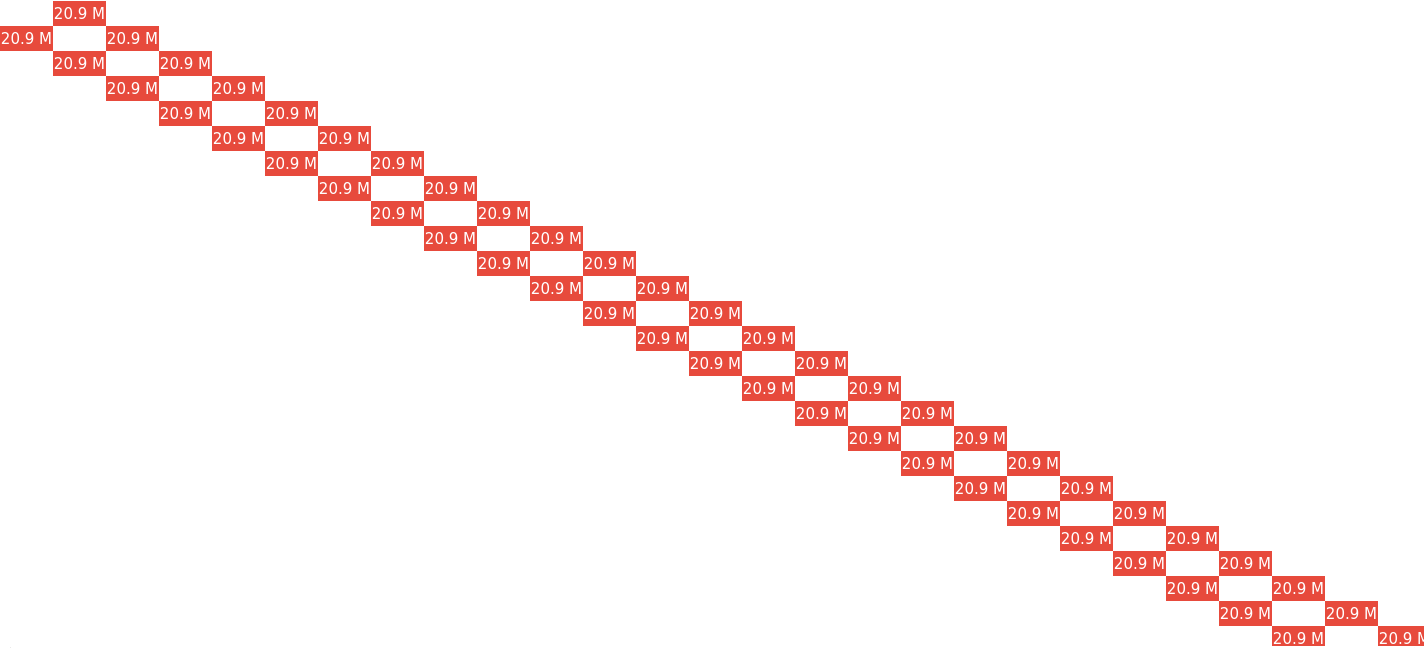}} 
	\begin{axis}[trim axis left, trim axis right, scale only axis,
		width = 0.1\textheight,
		height = 0.1\textheight,
		ylabel = {Sender rank},
		xlabel = {Receiver rank},
		xmin=0, xmax=25,
		ymin=0, ymax=25,
		xtick={0,15,25},
		y dir=reverse,
		ytick={0,5,10,15,20,25},
		axis on top,
		]
	\end{axis}
	\end{tikzpicture}
    \end{subfigure}
    \caption*{(a) Memory-bound D3Q19 implementation}
    \end{minipage}
    \begin{minipage}{\textwidth}
    \begin{subfigure}[t]{0.76\textwidth}
    \centering
    \includegraphics[scale=0.6]{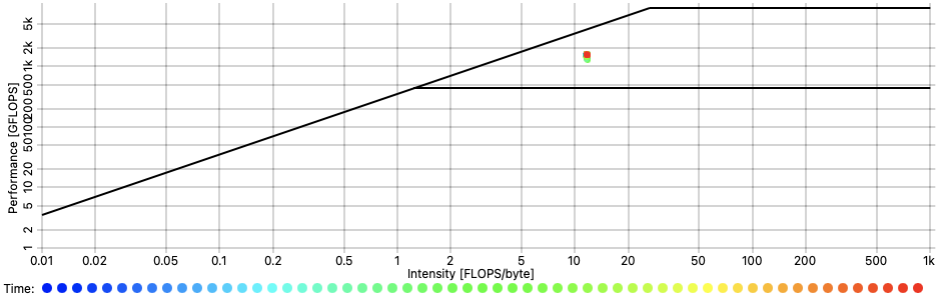}\quad
    \includegraphics[scale=0.6]{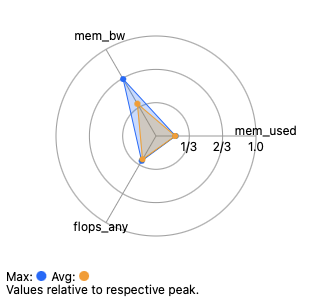}
    \end{subfigure} 
    \begin{subfigure}[t]{0.2\textwidth} 
	\begin{tikzpicture}
	\put(-0.4,0) {\includegraphics[width=0.101\textheight,height=0.1\textheight]{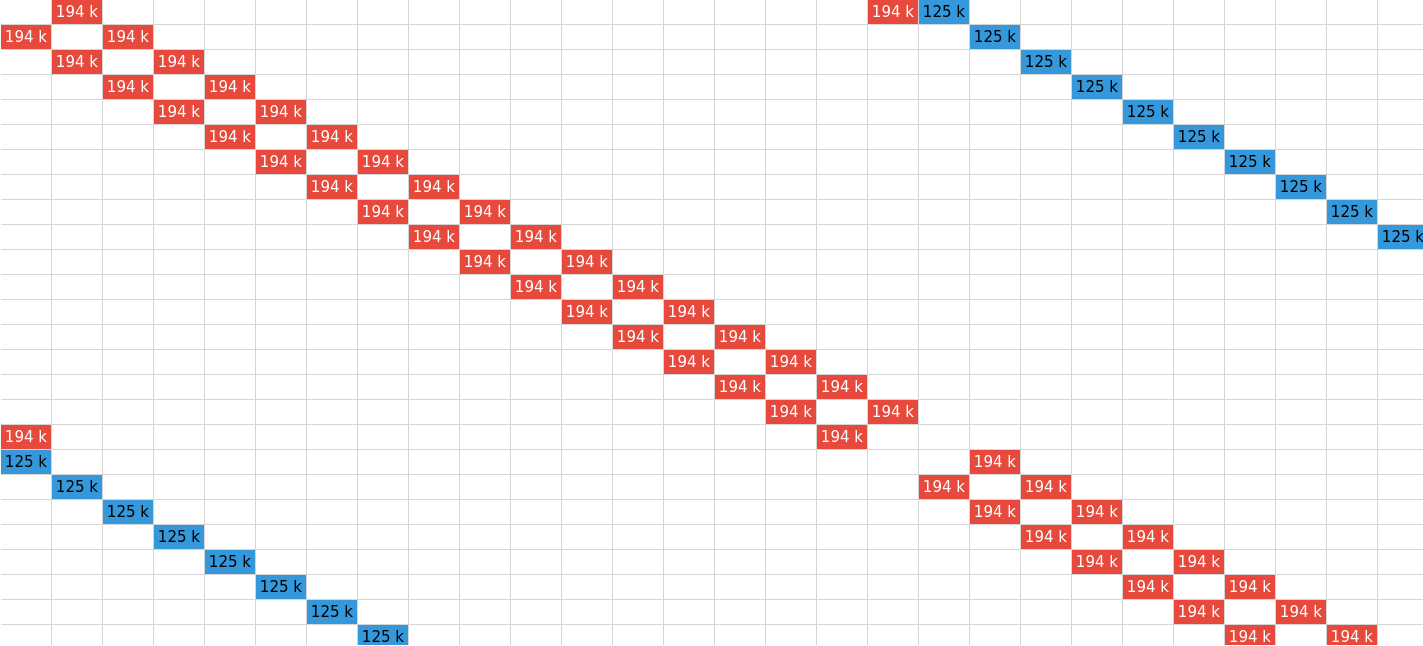}} 
	\begin{axis}[trim axis left, trim axis right, scale only axis,
		width = 0.1\textheight,
		height = 0.1\textheight,
		ylabel = {Sender rank},
		xlabel = {Receiver rank},
		xmin=0, xmax=25,
		ymin=0, ymax=25,
		xtick={0,15,25},
		y dir=reverse,
		ytick={0,5,10,15,20,25},
		axis on top,
		]
	\end{axis}
	\end{tikzpicture}
    \end{subfigure}
    \caption*{(b) Compute-bound D2Q37 SPEChpc 2021 implementation}
    \end{minipage}
    \caption{
    Hardware monitoring data (left, middle) and communication message profile (right) of the LBM benchmarks on 3 nodes (216 cores) of Fritz for (a) D3Q19 and (b) D2Q37. The Roof{}line plots show a color-coded timeline (blue to red) of the job's nodes across the entire run, where each data point is a measurement for a node at a specific point in time.
    The message profile view is zoomed in on the first 26 processes. Each message is (a) 20.9~\MB\ (red) and (b) 194~\KB\ (red) or 125~\KB\ (blue) in size.}
    \label{fig:LBMchar}
\end{figure*}
\begin{figure}[tb]
\centering
	\begin{tikzpicture}
		\pgfplotstableread{figures/LBM/D3Q19-socket.txt}\Ablocksz;
		\pgfplotstableread{figures/LBM/D2Q37-socket.txt}\Bblocksz;
		\begin{axis}[trim axis left, trim axis right, scale only axis,
			width=0.42\textwidth,height=0.13\textheight,
			xlabel = {MPI processes per socket},
			ybar, 
			ybar legend, 
			bar width=0.4mm,
			xmin=0.92,
			ymin=1,
			ymax=300,
			xmax=74,
			xtick pos=left,
			ytick pos=left,
			y label style={at={(-0.08,0.5)},font=\footnotesize},
			x label style={font=\footnotesize},
			x tick label style={font=\footnotesize},
			ylabel = {Memory bandwidth [\GBS]},
			ymajorgrids,
   			y grid style={dotted},
			legend columns = 1, 
			legend style = {
				nodes={inner sep=0.04em},
				draw=none,
				font=\scriptsize,
				cells={align=left},
				anchor=east,
				at={(0.26,0.89)},
				fill=white!95!black,
				/tikz/column 1/.style={column sep=5pt,},
			},
            set layers, 
			]		
			\begin{pgfonlayer}{axis background}
			\fill[shade, left color=blue!2, right color=blue!2]
			(rel axis cs:0,0)--(rel axis cs:0.245,0)--
			(rel axis cs:0.245,1)--(rel axis cs:0,1)--cycle;
			\fill[shade, left color=blue!8, right color=blue!8]
			(rel axis cs:0.245,0)--(rel axis cs:0.826,0)--
			(rel axis cs:0.496,1)--(rel axis cs:0.245,1)--cycle;
			\fill[shade, left color=blue!14, right color=blue!14]
			(rel axis cs:0.496,0)--(rel axis cs:0.805,0)--
			(rel axis cs:0.735,1)--(rel axis cs:0.496,1)--cycle;
			\fill[shade, right color=blue!20, left color=blue!20]
			(rel axis cs:0.735,0)--(rel axis cs:1,0)--
			(rel axis cs:1,1)--(rel axis cs:0.735,1)--cycle;
		\end{pgfonlayer}
			\addplot[ postaction={
				pattern=horizontal lines
			}, fill={red!60!white}, draw=none]
			table
			[
			x expr=\thisrow{Cores}, 
			y expr=\thisrow{BW[GB/s]},
			]{\Bblocksz};
			\addlegendentry{~D2Q37}	  			
		\end{axis}

		\begin{axis}[trim axis left, trim axis right, scale only axis,
			xshift=1.9pt,
			width=0.42\textwidth,height=0.13\textheight,
			ybar, 
			ybar legend, 
			bar width=0.4mm,
			xmin=0.92,
			ymin=1,
			ymax=300,
			xmax=74,
			ytick={},
			ymajorgrids,
			x tick label style={font=\footnotesize},
			axis lines=none, 
			legend columns = 1, 
			legend style = {
				nodes={inner sep=0.04em},
				draw=none,
				font=\scriptsize,
				cells={align=left},
				anchor=east,
				at={(0.25,0.75)},
				fill=white!95!black,
				/tikz/column 1/.style={column sep=5pt,},
			},
			]		
			\addplot[ postaction={
				pattern=north east lines
			}, fill={green!60!white}, draw=none]
			table
			[
			x expr=\thisrow{Cores}, 
			y expr=\thisrow{BW[GB/s]},
			]{\Ablocksz};
			\addlegendentry{~D3Q19}	
		\end{axis}
	\end{tikzpicture}
\caption{Single-node bandwidth scaling of the LBM D3Q19 ($720^3$ domain) and D2Q37 ($10800^2$ domain) implementations over 300 iterations on the Fritz system. The four shaded layers represent four ccNUMA domains of a single Fritz node. The linear scaling beyond the first ccNUMA domain is caused by the compact pinning strategy (filling the node from left to right).}
\label{fig:LBM_socket}
\end{figure}
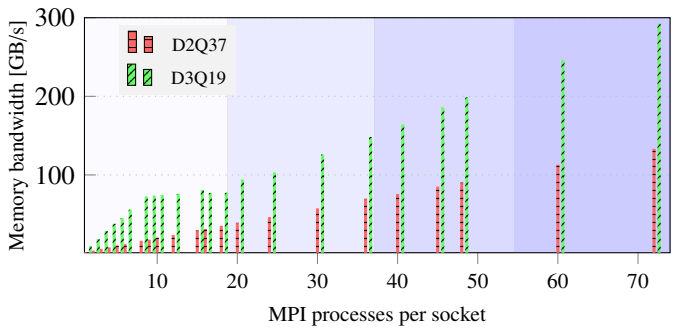
\section{{L}attice {B}oltzmann {M}ethod ({LBM})}\label{sec:lbm}
The \acf{LBM} algorithm is widely used in computational fluid dynamics due to its
ease of implementation and parallelization.
Here employ two \acs{LBM} implementations that allow us to easily tune problem size, domain decomposition, and the \CER. 

\subsection{D3Q19 implementation}
We use a \emph{D3Q19} discretization scheme~\cite{qian1992lattice} and Bhatnagar–Gross–Krook (BGK) collision operator~\cite{bhatnagar1954model} in a two-array implementation with fused stream and collide steps, without streaming stores, and with a stream-optimized \enquote{\acf{SoA}} memory layout~\cite{WELLEIN2006910,WITTMANN2013924} on a lattice of $(n_x+2)\times (n_y+2)\times(n_z+2)$ sites.
The total number of fluid cells in each x, y, and z dimension, $n_x$, $n_y$, and $n_z$, is increased by one boundary wall or lid cell.
The code is MPI parallel with halo exchange and uses double precision.
The domain is cut along the outer ($z$) dimension only so that the communication volume of each process is $2\times 5\times n_x\times n_y\times 8\,\bytes $ per neighbor; ghost elements are exchanged in the $z$ direction via next-neighbor communication using a \code{MPI\_Isend, MPI\_Irecv, MPI\_Wait} sequence. 

Figure~\ref{fig:LBMMeggie} (left) shows the basic structure of the code.
We chose a lid-driven cavity as the application use case. 
The working data set in bytes, including a boundary layer in each direction, is calculated as $19\times 2\times 8 \times (n_x+2)\times (n_y+2)\times (n_z+2)$.  
For a memory-bandwidth-bound data set, the single sweep traffic on the memory bus with write-allocate (also known as \RFO) is $1.5$ times the size of the working data set.
Wittmann et al.~\cite{Wittmann:2016} thoroughly analyzed and modeled socket-level performance.
The code balance of the fused collide/stream sweep is thus $456\,\byte/\lup$ (bytes per lattice site update), which makes the kernel memory bound if implemented efficiently; see Figure~\ref{fig:LBMchar}(a).

\subsubsection{Frequency of collectives}
Our implementation includes an optional correctness check for mass conservation, which employs an \code{MPI\_Allreduce} call (with \code{MPI\_SUM}) after a configurable number of iterations.
The latter is called \emph{collective step size}. We tune it as $2\times 10^n$, where $n$ ranges from $1$ to $6$.
For instance, on the Meggie cluster, the minimum cost for \code{MPI\_Allreduce} is $57\,\mu\seconds$ on 1280 processes (64 nodes), and it grows with the number of processes involved.
Note also that the time that each individual process spends in the routine may significantly deviate from the minimum, depending on desynchronization and load imbalance.
In all cases, the minimum time for \code{MPI\_Allreduce} is negligible compared to the duration of the minimum collective step size (20 LBM sweeps). 
Note that a direct measurement of the cost of a collective for large step sizes would be problematic because, in asynchronous execution, the time spent within the MPI call fluctuates significantly across processes.
In our experiments, we always compare the measured average performance for $10^6$ LBM iterations at a given collective step size with the run at the minimum collective step size, i.e., with a \code{MPI\_Allreduce} after every 20th sweep.

\begin{figure*}[tb]
    \begin{minipage}{\textwidth}
    \begin{subfigure}[t]{0.48\textwidth}
    \centering
    \includegraphics[scale=0.42]{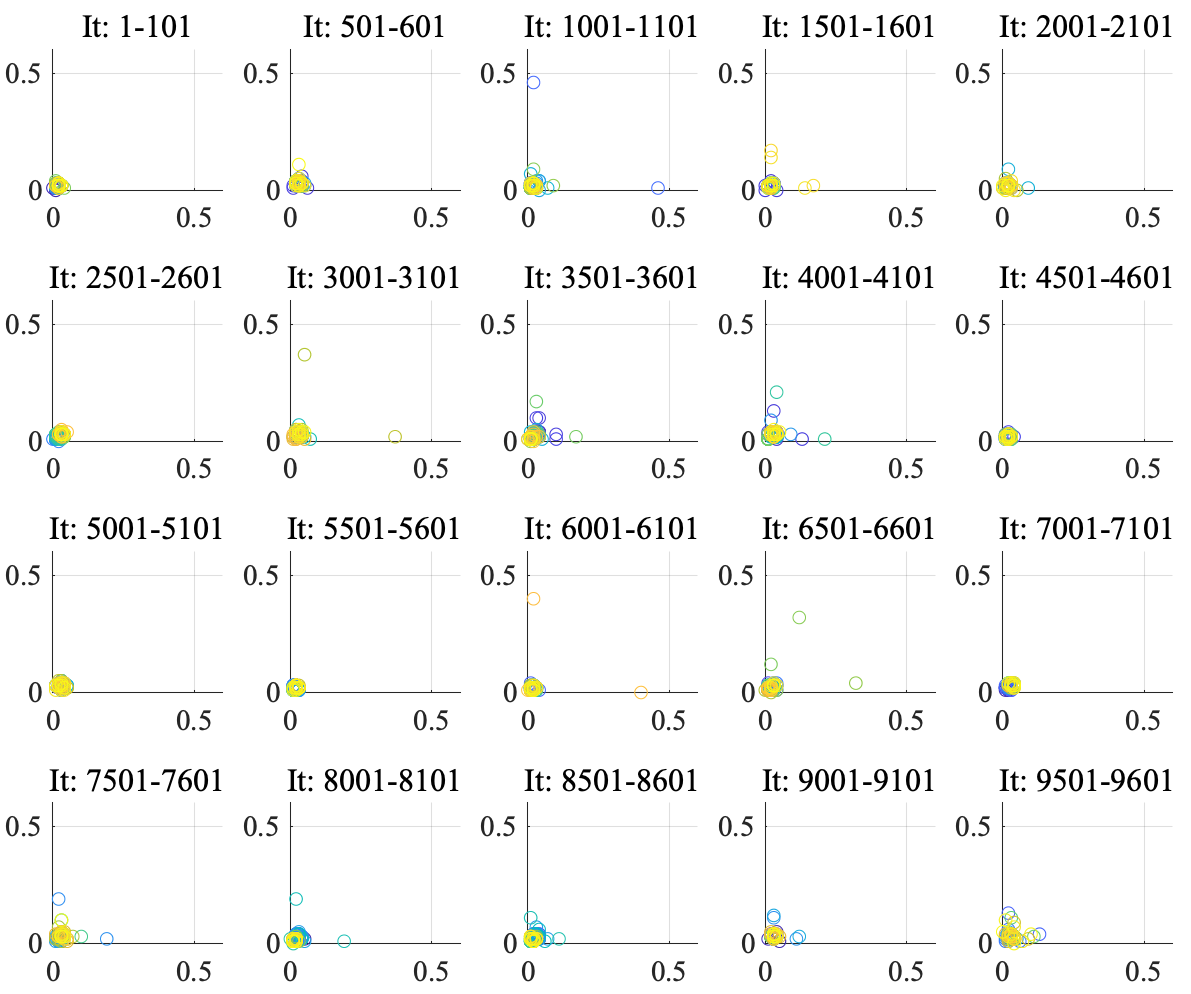}
    \caption{Snippet 100 iterations view of MPI times [s]}
    \end{subfigure} \quad
    \begin{subfigure}[t]{0.48\textwidth}
    \centering
    \includegraphics[scale=0.42]{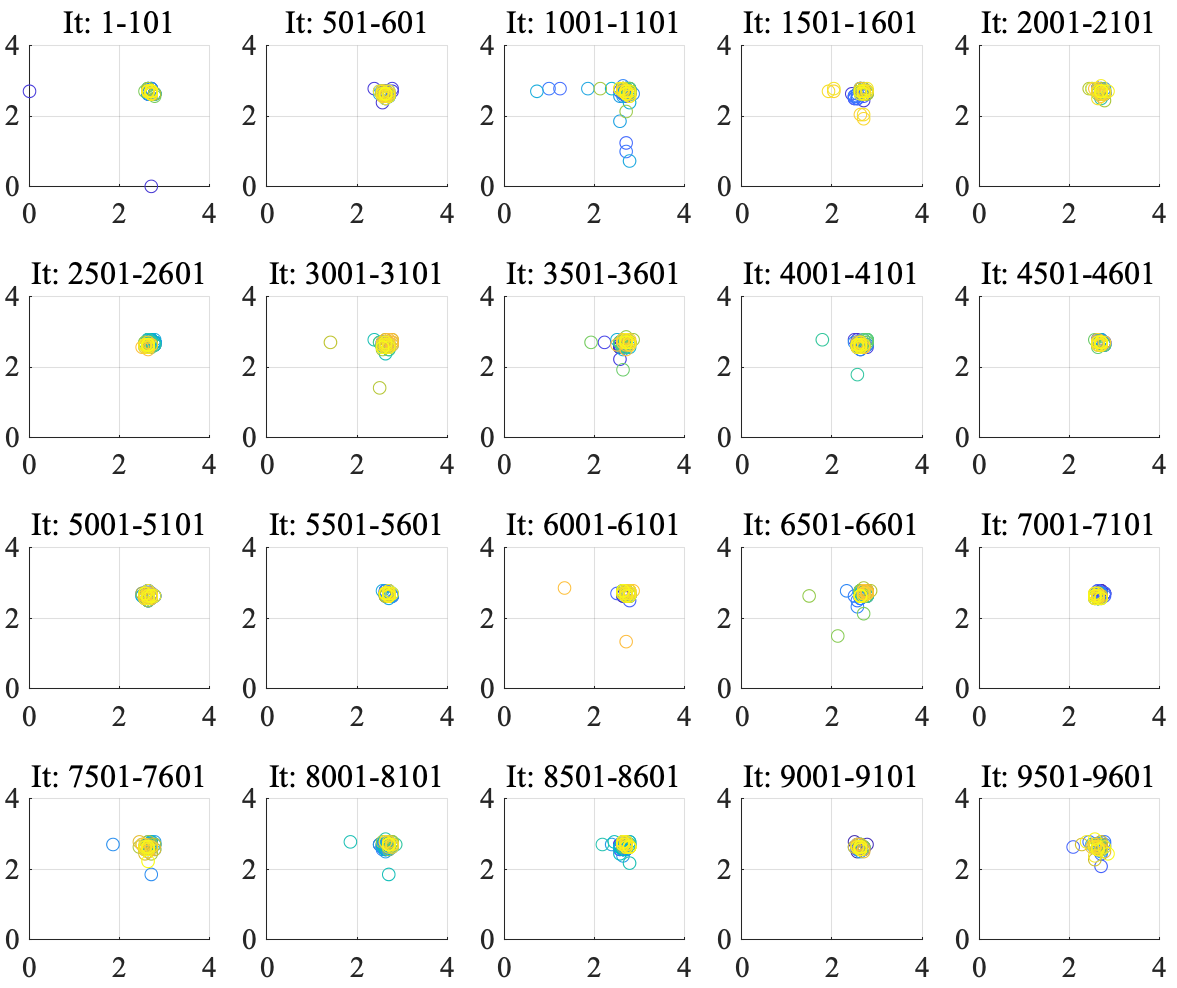}
    \caption{Snippet 100 iterations view of performance [iter/s]}
    \end{subfigure} \quad
    \begin{subfigure}[t]{0.48\textwidth}
    \centering
    \includegraphics[scale=0.3]{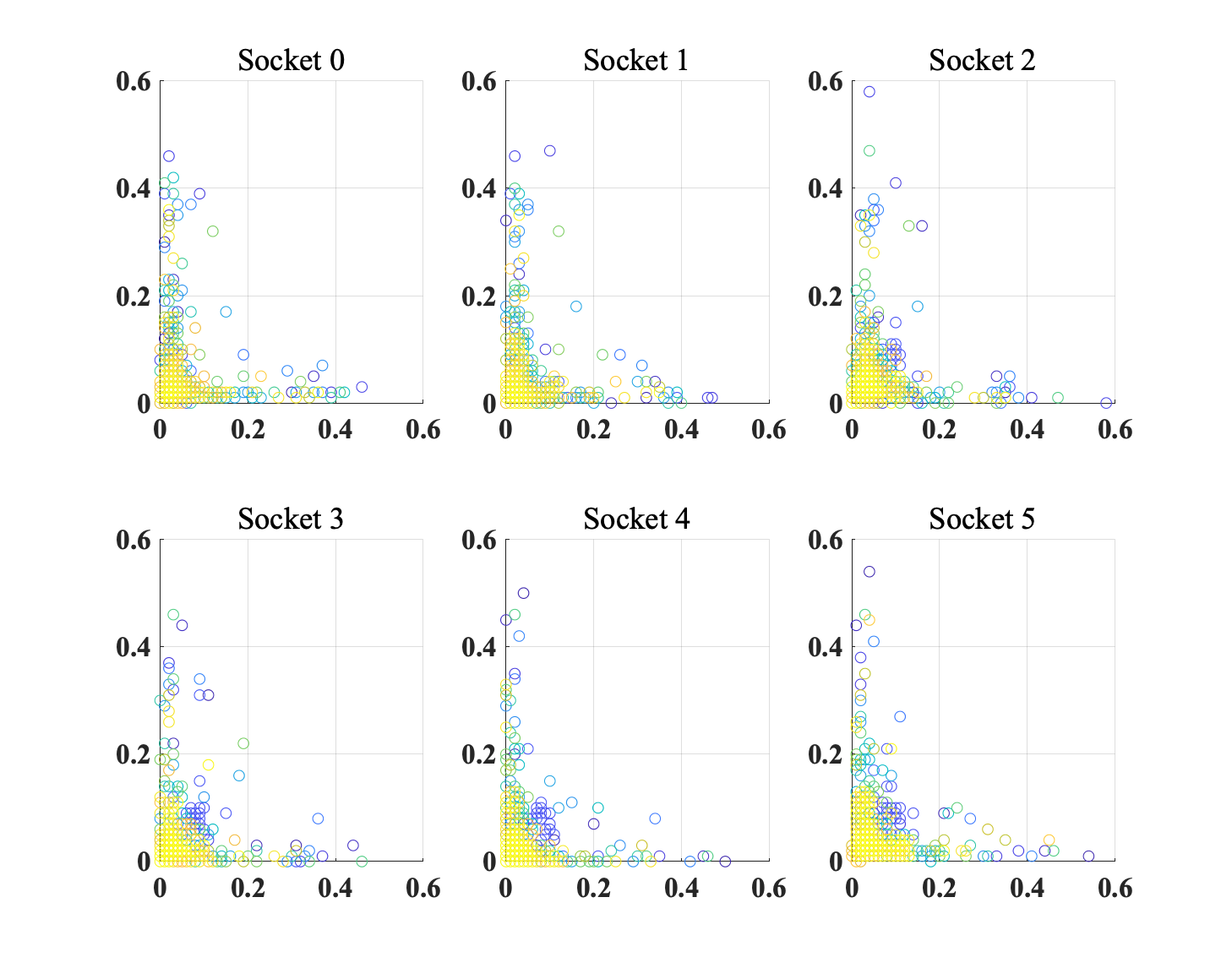}
    \includegraphics[scale=0.28]{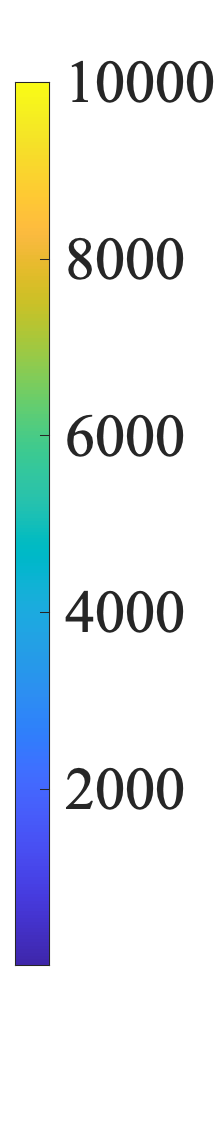}
    \caption{Entire 10K iterations view of MPI times [s]}
    \end{subfigure} \quad
    \begin{subfigure}[t]{0.48\textwidth}
    \centering
    \includegraphics[scale=0.3]{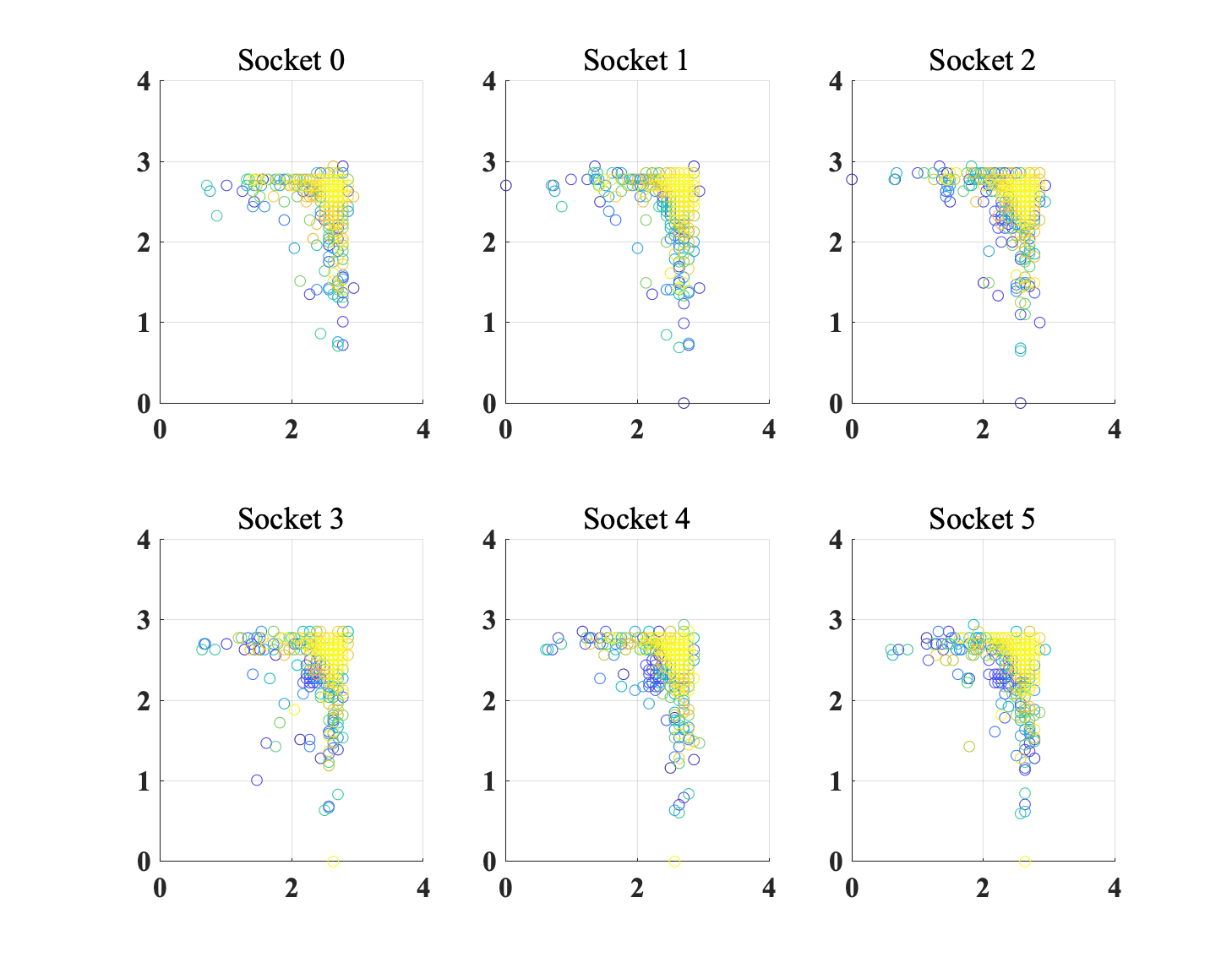}
    \includegraphics[scale=0.28]{figures/LBM/color-coding.png}
    \caption{Entire 10K iterations view of performance [iter/s]}
    \end{subfigure}
    \end{minipage}
    \caption{Phase-space analysis of SPEC D2Q37 implementation with displaying MPI times (in seconds) and performance (in iterations per seconds) in parallel for (a, b) 37th MPI process on second Fritz socket (c, d) first process on each Fritz socket.}
    \label{fig:2DQ37phasespace}
\end{figure*}
\begin{figure*}[tb]
    \begin{minipage}{\textwidth}
    \begin{subfigure}[t]{0.48\textwidth}
    \centering
    \includegraphics[scale=0.42]{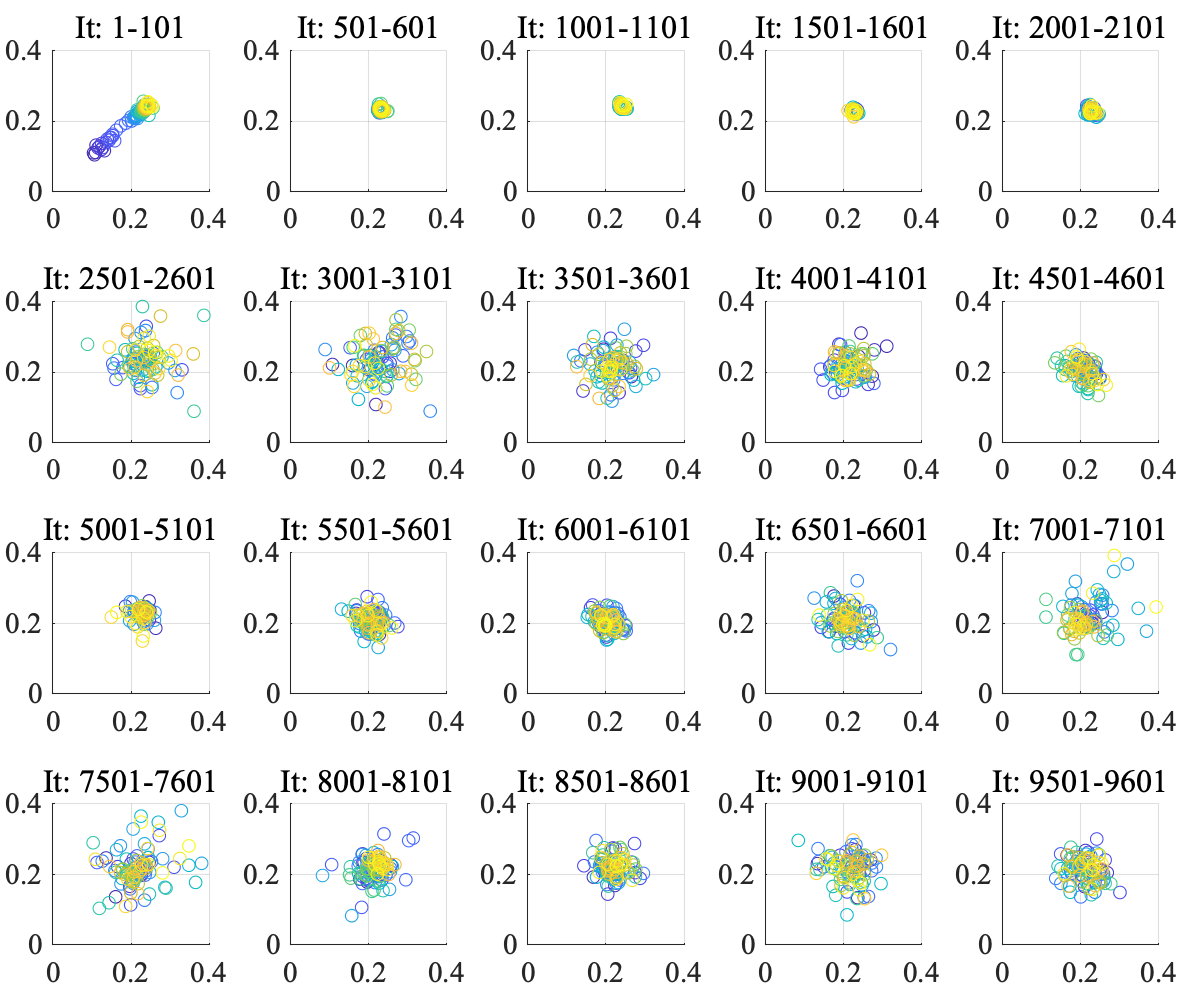}
    \caption{Snippet 100 iterations view of MPI times [s]}
    \end{subfigure} \quad
    \begin{subfigure}[t]{0.48\textwidth}
    \centering
    \includegraphics[scale=0.42]{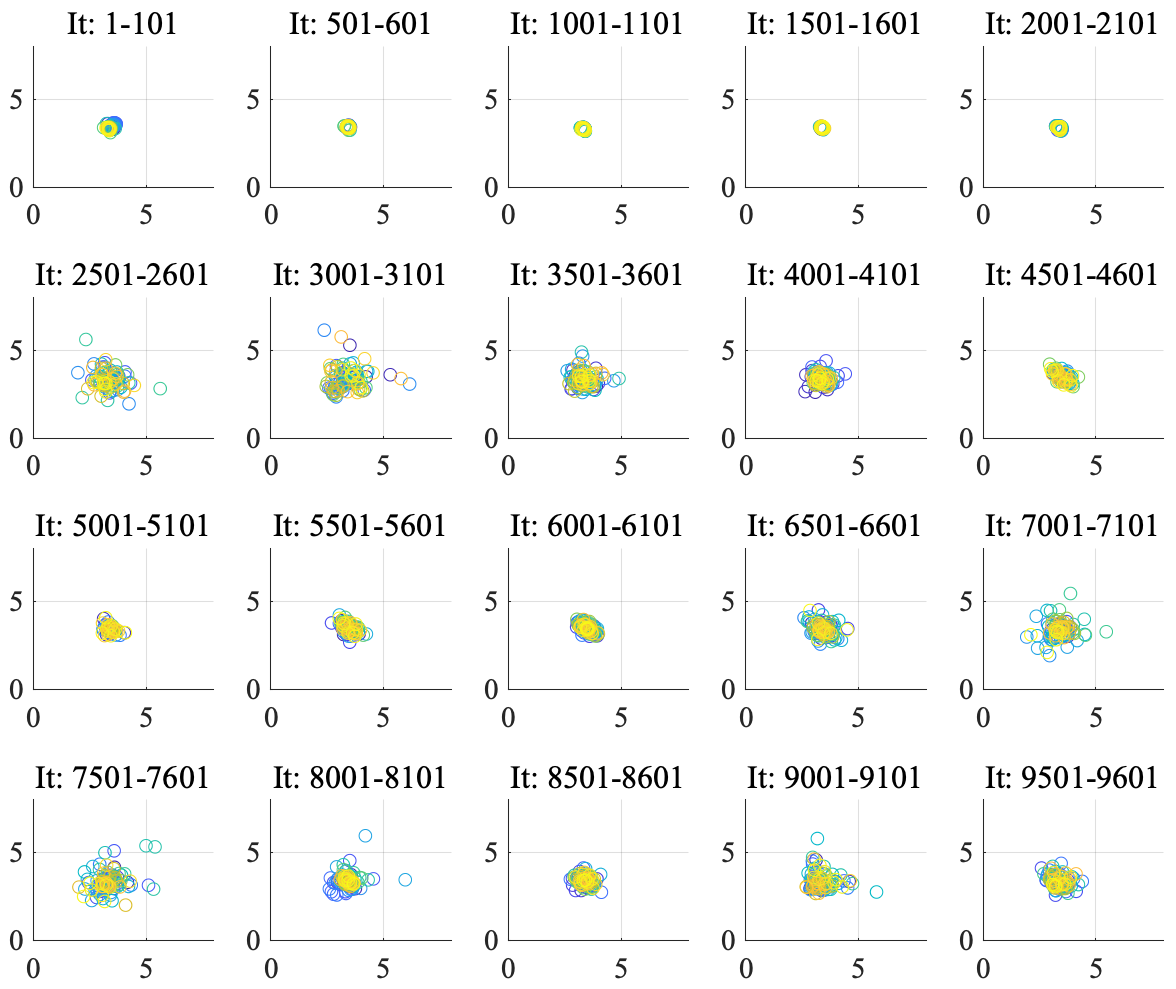}
    \caption{Snippet 100 iterations view of performance [iter/s]}
    \end{subfigure} \quad
    \begin{subfigure}[t]{0.48\textwidth}
    \centering
    \includegraphics[scale=0.3]{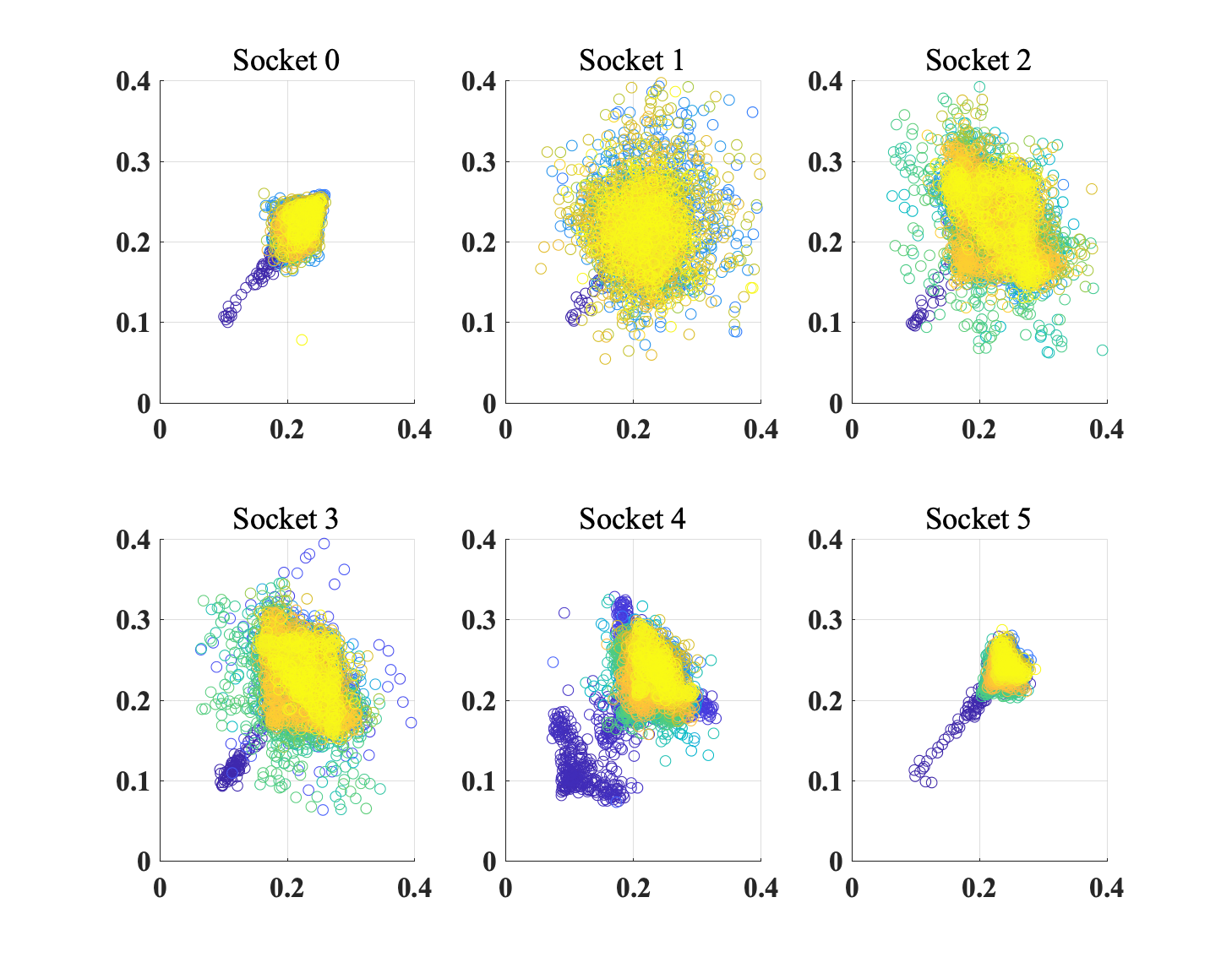}
    \includegraphics[scale=0.28]{figures/LBM/color-coding.png}
    \caption{Entire 10K iterations view of MPI times [s]}
    \end{subfigure} \quad
    \begin{subfigure}[t]{0.48\textwidth}
    \centering
    \includegraphics[scale=0.3]{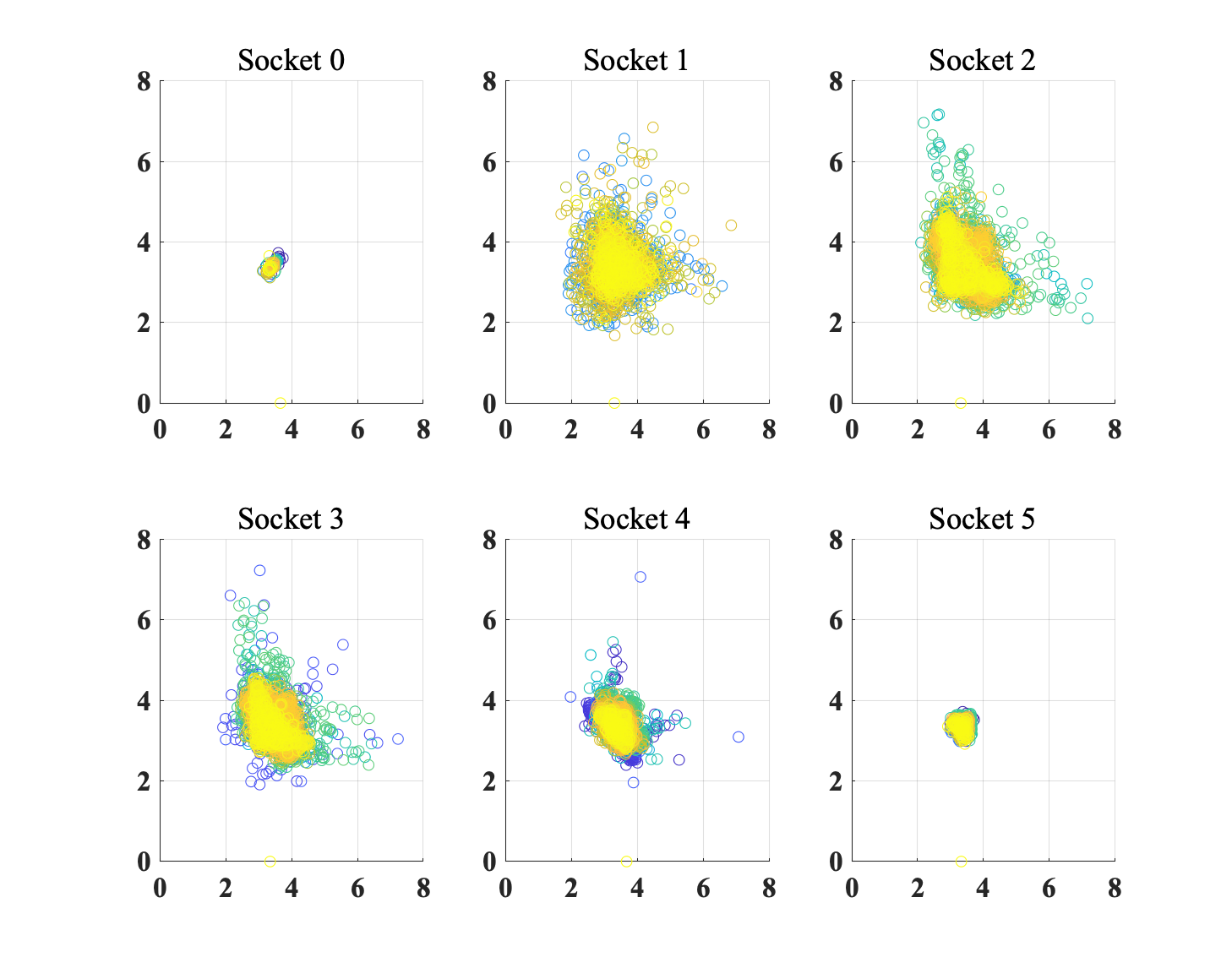}
    \includegraphics[scale=0.28]{figures/LBM/color-coding.png}
    \caption{Entire 10K iterations view of performance [iter/s]}
    \end{subfigure}
    \end{minipage}
    \caption{Phase-space analysis of D3Q19 with displaying MPI times (in seconds) and performance (in iterations per seconds) in parallel for (a, b) 37th MPI process on second Fritz socket (c, d) first process on each Fritz socket.
    } 
    \label{fig:3DQ19phasespace}
\end{figure*}
\subsubsection[Section title sans citation]{Asynchronicity through fewer collectives and CER $\to$ 1}
In Fig.~\ref{fig:LBMMeggie}(b+c), we show relative performance $P_n$ versus collective step size with respect to the minimum collective step size baseline (20 iterations) on the Meggie system, running the DqQ19 code on 1280 processes.
The value $P_n$ measures the relative speedup brought on by asynchronous execution and bottleneck avoidance; a value of 1 denotes no performance improvement. A greater $P_n$ value suggests a more effective communication overlap and better scalability, and a value of 2 signifies a doubling of the performance.
Unless specified otherwise, we choose a working set of approximately $29.5\times 10^6$ lattice sites ($8.97$~\GB) arranged in various cuboid geometries.
In the first series of experiments, we modified the computation-to-execution ratio by changing the overall geometry from $152\times 152\times 1280$ to $52\times 52\times 11520$, i.e., from strongly communication dominated to execution dominated (as shown by the CER row in Table~\ref{tab:LBMdomains}).
Note that even with automatic overlap, the impact of communication overhead is still significant.
The maximum performance of the Meggie system according to the Roof{}line model is 14~\GLUPS.
The best observed performance of 8.5~\GLUPS\ is significantly lower, which shows that there is considerable residual communication overhead.
          
On 64 Meggie nodes, the data for different domain shapes in Fig.~\ref{fig:LBMMeggie}(b) shows how, depending on the \ac{CER}, maximum speedups between 7~\% and 13~\% can be achieved.
The highest speedup can be observed at $152\times 152\times 1280$ (red circles), which has a CER that is  close to 1.
This is to be expected because the performance potential for overlapping communication with computation is highest at this point.
Deviating from this optimum CER in either direction reduces the speedup.
For instance, lowering the CER by reducing both $n_x$ and $n_y$ from 152 to 108, 88, and 52 caused the performance boost to decrease from 10.8 \% to 9.3 \%, 9.2 \% or 8 \%, respectively.
We also include the calculated speedup when subtracting the minimum \code{MPI\_Allreduce} time from the runtime for the best-performing case (black crosses).
It shows that the actual time for the call is insignificant.
In the experiment shown in Fig.~\ref{fig:LBMMeggie}(c), decreasing $n_x$ from 252 to 52, increasing $n_y$ from 362 to 1802, and keeping $n_z = 1280$ led to a nearly constant overall problem size.
We kept the CER constant and modified the shape of the $n_x\times n_y$ cross-section to check if it affected the speedup, which was not the case as expected.
    
Figure~\ref{fig:LBMnonoise} illustrates the impact of collectives using a timeline view and a run with 1440 processes (20 nodes) on the Fritz system.
The problem size was chosen to be $720\times720\times5760$ (907~\GB), with a reduction occurring in every 20th iteration.
The performance (as measured by taking the mean and standard deviation across processes) drops sharply right after the collectives and then quickly settles to a steady state. This shows clearly that gradual desynchronization is the cause for performance getting better with fewer collectives.\medskip

\highlight{\emph{Upshot}: For memory-bound LBM, if collectives are unavoidable but their frequency can be adapted, it is advantageous to make them less frequent to spend more time in an asynchronous execution even if the absolute overhead of the collective is negligible. The highest performance boost can be expected if the communication to computation ratio is close to 1.}

\subsection{D2Q37 implementation}
The double-precision vectorized \emph{D2Q37} LBM code from SPEChpc 2021\footnote{Version 1.1, \url{https://spec.org/hpc2021}} uses LBM to simulate the evolution of the Rayleigh-Taylor instability using 37 velocity components.
The \ac{SPEC} benchmarks concentrate on compute-intensive parallel performance and call for minimum main memory requirements of $\{$0.06, 0.48, 4, 14.5$\}$~\TB\ for the workloads that fall under the categories of $\{$tiny, small, medium, large$\}$, respectively.
Here, we employ the tiny workload for the \code{505.lbm\_t} benchmark, which can use a maximum of 60~\GB\ of memory and 256 processes.
The code supports 1D and 2D domain decomposition; we use 2D here in a $12\times 18$ process grid (three Fritz nodes). 
To communicate with its four neighbors, each MPI process uses non-blocking point-to-point \texttt{MPI\_Isend} and \texttt{MPI\_Irecv} calls.
By default, an \texttt{MPI\_Barrier} is used at the end of each iteration to maintain their synchronization, which we removed for the benchmarking.

The code has an arithmetic intensity of about 11~\FB\ for the stream-collide sweep, which makes the code compute bound; the monitoring data supports this assumption (see Figure~\ref{fig:LBMchar}(b)).

\subsection{Asynchronicity through resource bottleneck}
Figure~\ref{fig:LBM_socket} displays the socket-level performance in terms of memory bandwidth for both LBM variants. 
In the D2Q37 implementation ($10800^2$ domain), each message is 194~\KB\ when communicating with a direct neighbor and 125~\KB\ when communicating with a distant process, as opposed to 20.9~\MB\ when communicating in the D3Q19 implementation ($720^3$ domain); see Figure~\ref{fig:LBMchar}.
The D2Q37 implementation, in contrast to the D3Q19 implementation, has a low \ac{CER} ratio, additional long-distance bidirectional communication with the eighteenth process, and no bandwidth bottleneck.
Hence, it is not expected that performance can be gained by automatic communication overlap and avoiding bottlenecks. D3Q19 implementation, on the other hand, should fit the bill. These differences should show up in their corresponding phase-space plots.

Figure~\ref{fig:2DQ37phasespace} shows phase-space plots of MPI waiting times (in seconds) and performance (in iter/s) for D2Q37 as snippet views on one process (top) and as full end-to-end views on one process per socket (bottom).
Because of the absence of any contention on the memory interface or on the network, all MPI processes are self synchronizing. 
This is reflected by the low MPI times clustered around zero; the very few outliers on the axes are random noise that has no permanent effect, else these would move towards the diagonal (see Figure~\ref{fig:2DQ37phasespace}(a)).
Consequently, the code performance is temporarily affected by the random noise but there is no permanent positive effect as can be seen from the performance snippet view (Fig.~\ref{fig:2DQ37phasespace}(b)), where most points are clustered in a ``lump'' on the diagonal, with outliers in parallel to the axes.
As expected, the processes across sockets are correlated in absence of contention~\cite{AfzalHW:2022:2}, as illustrated by the socket-wise view in~\ref{fig:2DQ37phasespace}(c).

In contrast, for the D3Q19 implementation with contention on the memory interface, the MPI times grow over time already at the beginning of the run as illustrated in Fig.~\ref{fig:3DQ19phasespace}(a, c).
Since the progress is steady, there are no dot clouds along either axis.
Given that the initial state was synchronized, noise does generally not result in performance slowdown. 
The entire view on the MPI time and performance phase spaces (Figures~\ref{fig:3DQ19phasespace}(c,d)) shows that the observed process on two out of the six sockets (socket 0 and 5) has very little performance variation and almost constant MPI time, which indicates that these sockets are still (almost) in sync. Deliberate, random noise injections might thus boost the overall performance further. 

\medskip

\highlight{\emph{Upshot}: If the memory bandwidth cannot be saturated, LBM is not a candidate for performance improvement through spontaneous communication overlap.}

	\begin{figure*}[t]
	\begin{minipage}{\textwidth}
		\centering
		\includegraphics[scale=0.45]{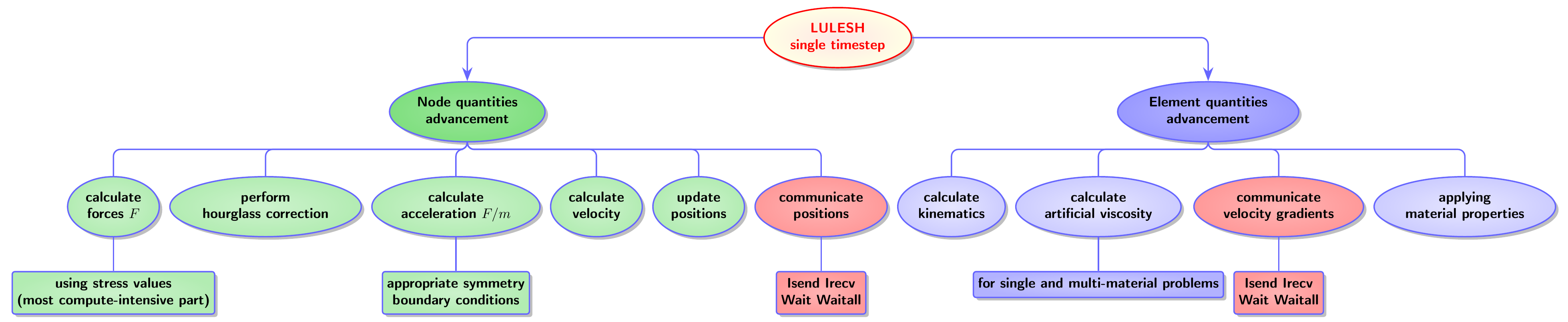}
		\caption{
			Algorithm in a single LULESH time step: communication (red), nodal updates (green) and element updates (blue).
            Most of the algorithm's time is spent in the initial stages of updating elemental and nodal quantities, i.e., updating the calculations for the elemental kinematics (compute-bound kernel), nodal stress integration (memory-bound kernel), and nodal hourglass corrections (memory-bound kernel).
		}  
		\label{fig:Fig_LULESHAlgo}
	\end{minipage}
	\end{figure*}
\begin{figure*}[t]
		\centering
\begin{minipage}{\textwidth}
\begin{subfigure}[t]{0.18\textwidth} 
	\begin{tikzpicture}
	\put(-0.4,0) {\includegraphics[width=0.101\textheight,height=0.14\textheight]{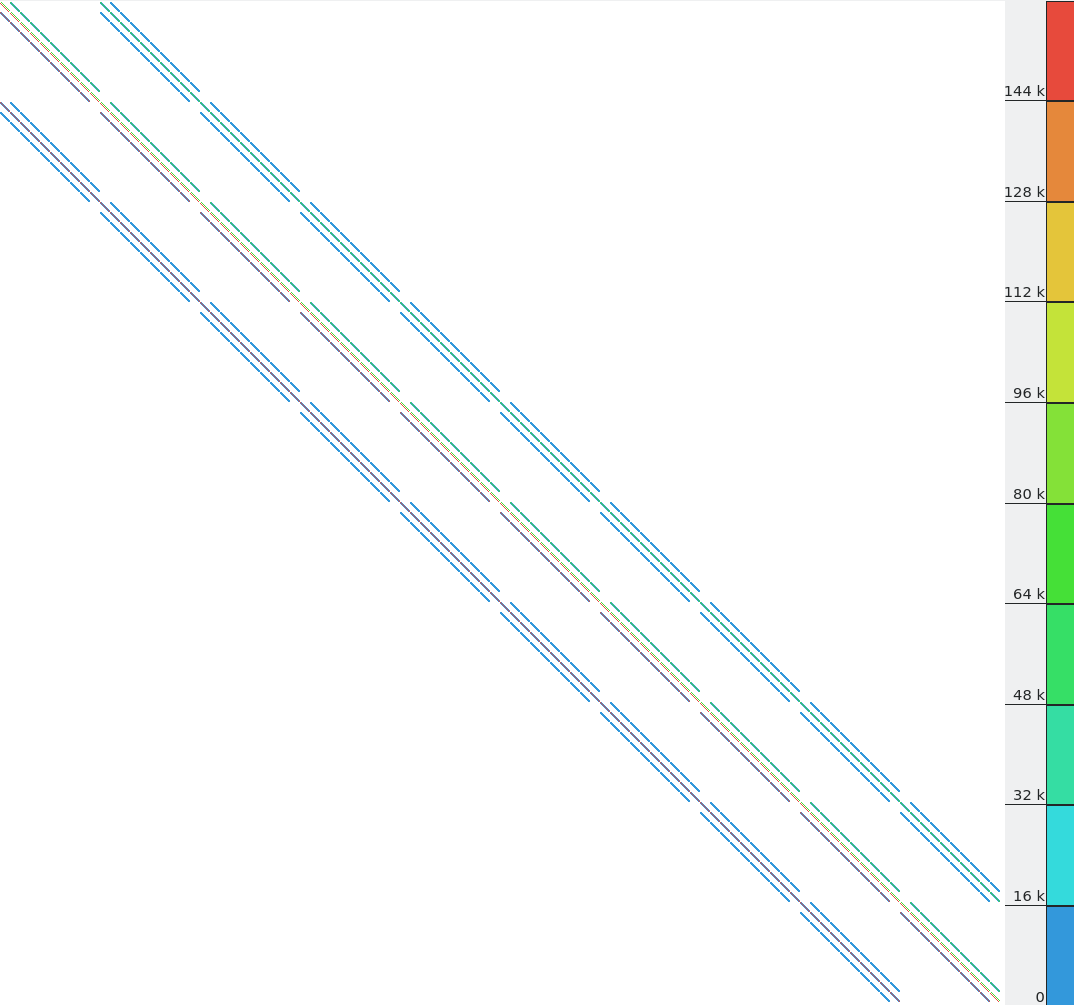}} 
	\begin{axis}[trim axis left, trim axis right, scale only axis,
		width = 0.1\textheight,
		height = 0.14\textheight,
		ylabel = {Sender rank},
		xlabel = {Receiver rank},
		xmin=0, xmax=1004,
		ymin=0, ymax=999,
		xtick={0,500,999},
		y dir=reverse,
		ytick={0,200,400,600,800,999},
		axis on top,
		]
	\end{axis}
	\node [] at (0.98,-1.4){\textcolor{black}{(a)} Message profile};
	\end{tikzpicture}
\end{subfigure}
\hspace{1em}
\begin{subfigure}[t]{0.24\textwidth}
\begin{tikzpicture}
	\pgfplotstableread{figures/LULESH/LULESH_Meggie_b0_c0.txt}\chronoA;	
	\begin{axis}[trim axis left, trim axis right, scale only axis,
		axis background/.style={fill=white!92!black},
		width = 0.85\textwidth,
		height = 0.14\textheight,
		xlabel = {Cubic domain size per process},
		ylabel = {Elements solved [z/\second]},
		ymajorgrids,
		y grid style={very thick, dotted},
		xmajorgrids,
		x grid style={very thick, white, dotted},
		axis on top,
		legend style = {
			nodes={inner sep=0.04em}, 
			font=\footnotesize, 
			anchor=north, 
			at={(0.655,0.99)},
			fill=white!92!black,
			draw=none,
		},
		legend columns = 1, 
		xmax=160 
		]
		
		\addplot+[]
		table
		[
		x expr=\thisrow{Size}, 
		y expr=\thisrow{Fixedit_FOM}
		]{\chronoA};
		\addlegendentry{~no reduction}
		
		\addplot+[]
		table
		[
		x expr=\thisrow{Size}, 
		y expr=\thisrow{FOM}
		]{\chronoA};
		\addlegendentry{~reduction}
	\end{axis}		
	\node [] at (2.2,-1.4){\textcolor{black}{(b)} Performance (load balanced)};
\end{tikzpicture}
\end{subfigure}
\hspace{1.6em}
\begin{subfigure}[t]{0.522\textwidth}
	\pgfplotsset{
		cycle list/Set1-5,
		cycle multiindex* list={
			mark list*\nextlist
			Set1-5\nextlist
		},
	}
	\begin{tikzpicture}
		\pgfplotstableread{figures/LULESH/LULESH_Meggie_b_c.txt}\chronoA;	
		\begin{axis}[trim axis left, trim axis right, scale only axis,
		axis background/.style={fill=white!92!black},
		width = 0.85\textwidth,
		height = 0.14\textheight,
		xlabel = {Imbalance \texttt{-b}},
		ylabel = {Elements solved [z/\second]},
		ymajorgrids,
		y grid style={very thick, dotted},
		x grid style={very thick, white, dotted},
		axis on top,
		legend style = {
			nodes={inner sep=0.04em},
			font=\footnotesize, 
			anchor=north, 
			at={(0.64,1)},
			fill=white!92!black,
			draw=none,
		},
		legend columns = 2,
		xmin=0, 
		xmax=25,
		ymin=280000,
		ymax=1060000,
		xtick={1,5,6,10,11,15,16,20,21,25},
		xticklabels={0,4,0,4,0,4,0,4,0,4},
		set layers, 
		]		
		\begin{pgfonlayer}{axis background}
			\fill[shade, left color=red!2, right color=red!2]
			(rel axis cs:0,0)--(rel axis cs:0.2,0)--
			(rel axis cs:0.2,1)--(rel axis cs:0,1)--cycle;
			\fill[shade, left color=red!8, right color=red!8]
			(rel axis cs:0.2,0)--(rel axis cs:0.4,0)--
			(rel axis cs:0.4,1)--(rel axis cs:0.2,1)--cycle;
			\fill[shade, left color=red!14, right color=red!14]
			(rel axis cs:0.4,0)--(rel axis cs:0.6,0)--
			(rel axis cs:0.6,1)--(rel axis cs:0.4,1)--cycle;
			\fill[shade, right color=red!20, left color=red!20]
			(rel axis cs:0.6,0)--(rel axis cs:0.8,0)--
			(rel axis cs:0.8,1)--(rel axis cs:0.6,1)--cycle;
			\fill[shade, right color=red!26, left color=red!26]
			(rel axis cs:0.8,0)--(rel axis cs:1,0)--
			(rel axis cs:1,1)--(rel axis cs:0.8,1)--cycle;
		\end{pgfonlayer} 
	
		\addplot+[mark=square*, blue]
		table
		[
		x expr=\thisrow{cases}, 
		y expr=\thisrow{S40_FOM[z/s]}
		]{\chronoA};
		\addlegendentry{~$40^3$, reduction}
		
		\addplot+[mark=*, blue]
		table
		[
		x expr=\thisrow{cases}, 
		y expr=\thisrow{S40_FIXEDdt8e-10_FOM[z/s]}
		]{\chronoA};
		\addlegendentry{~$40^3$,  no reduction}
		
		\addplot+[mark=square*, teal]
		table
		[
		x expr=\thisrow{cases}, 
		y expr=\thisrow{S60_FOM[z/s]}
		]{\chronoA};
		\addlegendentry{~$60^3$, reduction}
		
		\addplot+[mark=*, teal]
		table
		[
		x expr=\thisrow{cases}, 
		y expr=\thisrow{S60_FIXEDdt8e-10_FOM[z/s]}
		]{\chronoA};
		\addlegendentry{~$60^3$,  no reduction}
		
		\addplot+[mark=square*, red]
		table
		[
		x expr=\thisrow{cases}, 
		y expr=\thisrow{S90_FOM[z/s]}
		]{\chronoA};
		\addlegendentry{~$90^3$, reduction}
		
		\addplot+[mark=*, red]
		table
		[
		x expr=\thisrow{cases}, 
		y expr=\thisrow{S90_FIXEDdt8e-10_FOM[z/s]}
		]{\chronoA};
		\addlegendentry{~$90^3$,  no reduction}		
		\end{axis}		
		\node [] at (4,-1.4){\textcolor{black}{(c)} Performance (trigger load imbalance)};
		\node [] at (1,0.15){\textcolor{black}{$c=0$}};
		\node [] at (2.5,0.15){\textcolor{black}{$c=1$}};
		\node [] at (4.1,0.15){\textcolor{black}{$c=2$}};
		\node [] at (5.8,0.15){\textcolor{black}{$c=3$}};
		\node [] at (7.4,0.15){\textcolor{black}{$c=4$}};
		\discontarrow(0,0)(0,0)(0,0.15)(0,0.15);
	\end{tikzpicture}
\end{subfigure}
\end{minipage}%
		\caption{ {\textcolor{black}{(a)}} LULESH communication
                  topology matrix. Maximum message sizes are
                  \{$144$~\KB, $630$~\KB, $1.44$~\MB, $1.26$~\MB\}
                  with domain sizes of
                  $\{40, 60, 90, 120\}^3$ per process.
                  {\textcolor{black}{(b)}} Performance in elements solved
                  per second versus domain size without load imbalance, comparing
                  runs with reductions (squares) with runs without (circles)
                  on $50$ Meggie nodes ($1000$ processes).
                  {\textcolor{black}{(c)}}
                  Impact of load
                  imbalance on performance for $10^3$ MPI
                  processes on $50$ Meggie nodes. The imbalance is
                  triggered among domains via the \texttt{-b} flag and
                  among regions within a domain via the \texttt{-c}
                  flag.}
		\label{fig:LULESHimbalance} 

\end{figure*}
	\begin{figure}[t]
		\centering
		\centering
\begin{minipage}{\textwidth}
\begin{subfigure}[t]{0.49\textwidth}
	\pgfplotsset{
		cycle list/Set1-5,
		cycle multiindex* list={
			mark list*\nextlist
			Set1-5\nextlist
		},
	}
	\begin{tikzpicture}
		\pgfplotstableread{figures/LULESH/LULESH_SuperMUC-NG_b_c.txt}\chronoA;	
		\begin{axis}[trim axis left, trim axis right, scale only axis,
		axis background/.style={fill=white!92!black},
		width = 0.85\textwidth,
		height = 0.14\textheight,
		xlabel = {Imbalance \texttt{-b}},
		ylabel = {Elements solved [z/\second]},
		ymajorgrids,
		y grid style={very thick, dotted},
		x grid style={very thick, white, dotted},
		axis on top,
		legend style = {
			nodes={inner sep=0.04em},
			font=\footnotesize, 
			anchor=north, 
			at={(0.6,1)},
			fill=white!92!black,
			draw=none,
		},
		legend columns = 2,
		xmin=0, 
		xmax=25,
		ymin=280000,
		ymax=1060000,
		xtick={1,5,6,10,11,15,16,20,21,25},
		xticklabels={0,4,0,4,0,4,0,4,0,4},
		set layers, 
		]		
		\begin{pgfonlayer}{axis background}
			\fill[shade, left color=green!2, right color=green!2]
			(rel axis cs:0,0)--(rel axis cs:0.2,0)--
			(rel axis cs:0.2,1)--(rel axis cs:0,1)--cycle;
			\fill[shade, left color=green!8, right color=green!8]
			(rel axis cs:0.2,0)--(rel axis cs:0.4,0)--
			(rel axis cs:0.4,1)--(rel axis cs:0.2,1)--cycle;
			\fill[shade, left color=green!14, right color=green!14]
			(rel axis cs:0.4,0)--(rel axis cs:0.6,0)--
			(rel axis cs:0.6,1)--(rel axis cs:0.4,1)--cycle;
			\fill[shade, right color=green!20, left color=green!20]
			(rel axis cs:0.6,0)--(rel axis cs:0.8,0)--
			(rel axis cs:0.8,1)--(rel axis cs:0.6,1)--cycle;
			\fill[shade, right color=green!26, left color=green!26]
			(rel axis cs:0.8,0)--(rel axis cs:1,0)--
			(rel axis cs:1,1)--(rel axis cs:0.8,1)--cycle;
		\end{pgfonlayer} 
	
		\addplot+[mark=square*, blue]
		table
		[
		x expr=\thisrow{cases}, 
		y expr=\thisrow{S40_FOM[z/s]}
		]{\chronoA};
		\addlegendentry{~$40^3$, reduction}
		
		\addplot+[mark=*, blue]
		table
		[
		x expr=\thisrow{cases}, 
		y expr=\thisrow{S40_FIXEDdt8e-10_FOM[z/s]}
		]{\chronoA};
		\addlegendentry{~$40^3$,  no reduction}
		
		\addplot+[mark=square*, teal]
		table
		[
		x expr=\thisrow{cases}, 
		y expr=\thisrow{S60_FOM[z/s]}
		]{\chronoA};
		\addlegendentry{~$60^3$, reduction}
		
		\addplot+[mark=*, teal]
		table
		[
		x expr=\thisrow{cases}, 
		y expr=\thisrow{S60_FIXEDdt8e-10_FOM[z/s]}
		]{\chronoA};
		\addlegendentry{~$60^3$,  no reduction}
		
		\addplot+[mark=square*, red]
		table
		[
		x expr=\thisrow{cases}, 
		y expr=\thisrow{S90_FOM[z/s]}
		]{\chronoA};
		\addlegendentry{~$90^3$, reduction}
		
		\addplot+[mark=*, red]
		table
		[
		x expr=\thisrow{cases}, 
		y expr=\thisrow{S90_FIXEDdt8e-10_FOM[z/s]}
		]{\chronoA};
		\addlegendentry{~$90^3$,  no reduction}		
		\end{axis}		
		\node [] at (0.8,0.15){\textcolor{blue}{$c=0$}};
		\node [] at (2.3,0.15){\textcolor{blue}{$c=1$}};
		\node [] at (3.9,0.15){\textcolor{blue}{$c=2$}};
		\node [] at (5.6,0.15){\textcolor{blue}{$c=3$}};
		\node [] at (7.1,0.15){\textcolor{blue}{$c=4$}};
		\discontarrow(0,0)(0,0)(0,0.15)(0,0.15);
	\end{tikzpicture}
\end{subfigure}
\end{minipage}%
		\caption{
			Load imbalance impact (trigging cost \texttt{-c} and imbalance \texttt{-b} flag, varying between zero to four) on performance for $10^3$ MPI processes on $21$ SuperMUC-NG nodes.
		}
		\label{fig:LULESHimbalance_superMUC} 
	\end{figure}
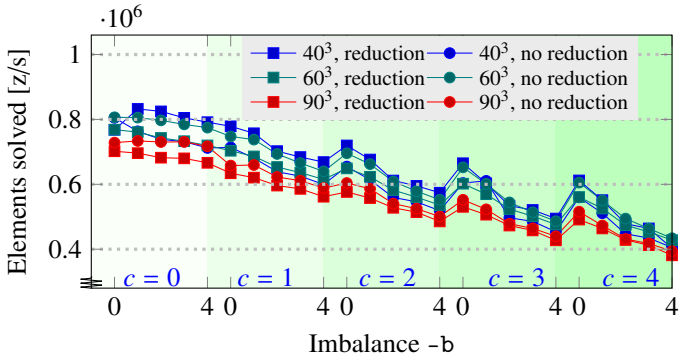
\section{{L}ivermore {U}nstructured {L}agrangian {E}xplicit {S}hock {H}ydrodynamics ({LULESH})}\label{sec:lulesh}
LULESH\footnote{\Ac{LULESH} application, version $2.0$: \url{http://asc.llnl.gov/codes/proxy-apps/lulesh}} is a MPI-parallel proxy application for shock hydrodynamic simulation.
Although it uses a Cartesian mesh, it employs an unstructured data layout and indirection arrays to mimic the unstructured complex hexahedral mesh geometry in the full application code. 
LULESH approximates the hydrodynamic equations discretely by partitioning the spatial problem domain into a collection of volumetric elements of the mesh.
Routines are performed on a region-by-region basis to make the memory access patterns non-unit stride and to easily introduce artificial load imbalances.
The mapping between materials and regions is important, since the hydrodynamics codes typically simulate problems containing multiple materials, which are then mapped onto regions (subsets of the mesh).
LULESH solves the single-material \emph{Sedov} blast wave problem; to mimic multiple materials, it uses multiple regions with varying sizes, each modeling the same ideal gas material.
Materials' relative motion as a result of forces is described by hydrodynamic modeling.
Load imbalance can be introduced by differently-sized regions as well as the amount of computation per grid point. 
The code intensity (in~\FB) becomes low for large domain sizes using few process counts.
The execution and data transfer characteristics make the code memory bound on modern architectures and thus, in principle, a candidate for desynchronization dynamics.

    \begin{table*}[t]
		\centering
		\caption{ {\textcolor{black}{(Left)}} Parallel HPCG algorithm.
                  {\textcolor{black}{(Right)}}
                  Domain sizes and corresponding runtime breakdown for
                  execution and communication in one iteration of HPCG
                  (including three reductions) and an undisturbed
                  fully synchronized state (first iteration after an \texttt{MPI\_Barrier}) on (a) 1280 processes on Meggie and (b) 1296 processes on SuperMUC-NG.
                  Communication plays less
                  role for large problems (small CER) and
                  gets higher for small problems. All runs used the default \texttt{MPI\_Allreduce}
                  implementation.}
		\label{tab:HPCG0}
		\begin{minipage}[c]{0.3\textwidth}
	\resizebox{\columnwidth}{!}{%
		\includegraphics[scale=0.065]{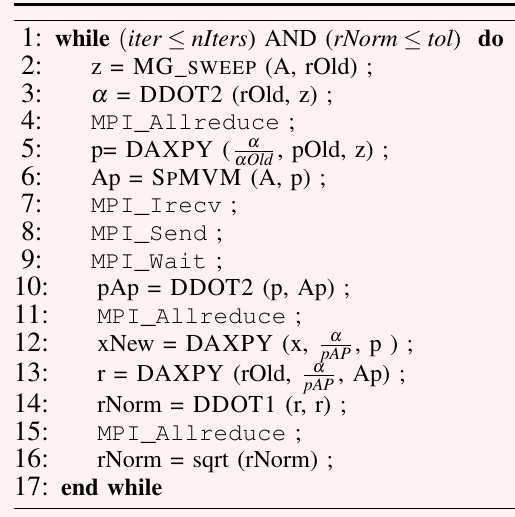}
	}
\end{minipage}\hfill
\begin{minipage}[c]{0.65\textwidth}
	\hspace{-5em}
	\begin{adjustbox}{width=1.15\textwidth}
		\pgfplotstabletypeset[
		/pgfplots/colormap/cool,
		every column/.style=,
		after row=,
		color cells={min=50000,max=122850,textcolor=black},
		/pgf/number format/fixed,
		/pgf/number format/precision=3,
		col sep=comma,
		]{
			{(a) M: Subdomain (full domain)},{Exec [ms]},{Comm [ms]},{Comm [GB]},{Allreduce min [us]},{Allreduce mean [ms]},{CER}
			{$32^3$ (256x512x320) },18.77,2.626,0.303,690,2.913,0.139904102
			{$48^3$ (384x768x480) },53.27,1.335,0.665,202,0.639,0.02506101
			{$64^3$ (512x1024x640) },135.3,2.27,1.17,2988,7.3,0.016777531
			{$96^3$ (768x1536x960) },493,17.766,2.6,116,37.36,0.036036511
			{$128^3$ (1024x2048x1280) },1102.1,20.852,4.59,119,102.93,0.018920243
			{$144^3$ (1152x2304x1440) },1573.19,6.348,5.8,116,153,0.004035113
		}
		\end{adjustbox}
		\begin{adjustbox}{width=1.015\textwidth}
		\hspace{-7em}
		\pgfplotstabletypeset[
		/pgfplots/colormap/cool,
		every column/.style=,
		after row=,
		color cells={min=50000,max=122850,textcolor=black},
		/pgf/number format/fixed,
		/pgf/number format/precision=3,
		col sep=comma,
		]{
			{(b) S: Subdomain (full domain)},{Exec [ms]},{Comm [ms]},{Comm [GB]},{Allreduce min [us]},{Allreduce mean [ms]},{CER}
			{$32^3$ (384x384x288) },16.5782,2.686,0.308,1200,4.03,0.162
			{$48^3$ (576x576x432) },61.722,1.545786,0.758,400,1.12241,0.025
			{$64^3$ (768x768x576) },147.708,5.48298,1.27,169,3.76779,0.037
			{$96^3$ (1152x1152x864) },516.171,7.487,2.64,1500,4.09406,0.0145
		}
	\end{adjustbox}
\end{minipage}
	\end{table*}
\subsection{Implementation}
The algorithm is represented visually in Fig.~\ref{fig:Fig_LULESHAlgo}.
The updates in physical quantities are done in two steps: at the corners of the hexahedra (more specifically, cubes) (node) and at the center of each hexahedron (element).
\emph{Nodes} store kinematic values (positions and velocities).
The steps involve the most compute-intensive calculation of the nodal forces from the elemental contribution of stresses and volume force.
After a diagnostic check for negative volumes, the hourglass contribution is applied element-wise to each nodal force.
Then it computes the accelerations via $F = ma$ with appropriate symmetry boundary conditions to calculate new nodal velocities and positions.
\emph{Elements} store thermodynamic variables (energy and pressure).
The steps involve the calculation of elemental kinematic values and new elemental and regional artificial viscosities.
Then, material properties are applied to each element and the \acf{EOS} is evaluated. 
The implementation uses an outer loop over the regions and an inner loop over the elements in a region.
The communication of ghost fields happens twice: First, the exchange of positions 
ensures the same nodal values of the neighboring elements.
Second, after their computation, viscosity gradients 
are exchanged.

\subsection{Load imbalance, problem sizes, and cost}
We use the latest code version v$2.0$, which by default calculates
time constraints (Courant and hydro) and then determines the minimum required time step across domains.
This dynamic time step calculation incurs additional $(n_\mathrm{iter}-1)$ reductions (\code{MPI\_Allreduce}) using the \code{MPI\_MIN} operation.
Independent of time step size, we used a fixed number of steps $n_\mathrm{iter}=5000$, and $10^3$ MPI processes on the Meggie system.
The results are shown in the default setting.
The message profile (Fig.~\ref{fig:LULESHimbalance}\textcolor{black}{(a)}) lets us expect fast-traveling idle waves due to long-distance point-to-point communication (\code{MPI\_Isend, MPI\_Irecv, MPI\_Wait, MPI\_Waitall} sequence). 
	
Initially (see Fig.~\ref{fig:LULESHimbalance}\textcolor{black}{(b)}), we
switched off the load imbalance by setting the cost \code{-c} and balance \code{-b} command line options to be equal to zero.
Performance in the number of elements solved per second is shown with and without reduction operations.
To avoid reductions, we run with a fixed time step by defining a sufficiently small step size of $8\times 10^{-10}$~\second\ (\code{dtfixed}) 
in the setup code (\code{lulesh-init.cc}). 
In this case, the steps to solution differ, since the initial time step previously with reduction scales to an arbitrary size based on an analytical equation.

In Fig.~\ref{fig:LULESHimbalance}\textcolor{black}{(c)} and Fig.~\ref{fig:LULESHimbalance_superMUC}, we scan different degrees of load imbalance to explore its performance impact on Meggie and SuperMUC-NG, the rationale being that load imbalance serves as a natural source of idle waves and thus as a trigger for automatic communication overlap.
Both the number of regions and their load imbalance are very problem-dependent.
The minimum number of regions must equal the number of distinct materials.
Hence, we used $11$ distinct regions, as defined by default.
Load imbalance is caused by adding the additional computation cost to some regions in evaluating material properties for the various Equations of State (EOS).
The cost (\code{-c \#}) value sets the relative imbalance between regions within a domain.
It increases the cost of about 45\% of the regions by the indicated value times the time spent in \code{EvalEOSForElems()} and the cost of 5\% of the regions by ten times the indicated value.
Consequently, half the regions have no extra cost.
The load balance (\code{-b \#}) value can artificially add imbalance between domains (due to a different mix of regions in each domain).
It changes the relative weight of regions within a domain.  
This weight imbalance has a limit in that the time spent in \code{EvalEOSForElems()} remains less than half of the overall runtime.
The imbalance imposed via the \code{-b \#} and \code{-c \#} flags is representative of the varying cost of evaluating material properties of various equations of state or strength models.
Comparing no imbalance (\code{-c 0 -b 0}) with maximum (\code{-c 4 -b 4}), the performance gain when eliminating the reduction reduced from $4.8$\% to $4$\%.
Speedup by desynchronization is only relevant at a very slight inter-domain load imbalance (\code{-c 1 -b 0} on Meggie; on SuperMUC-NG, imbalance has no positive effect.
However, the impact becomes better when using vectorization flags for strong bandwidth saturation.
Since LULESH has a low CER, long-distance communication
, and workloads are not balanced even when no load imbalance (\code{-c 0 -b 0}) setting is used, therefore, increasing asynchronicity by an algorithm variant of reduction or injecting load imbalance will result in performance loss rather than any benefit.\medskip

\highlight{\emph{Upshot}: Speedup with asynchronicity is only relevant with a slight load imbalance for LULESH, if at all. In strong load imbalance scenarios, the overlap effect is swamped by dominating laggers.}

\section{{H}igh {P}erformance {C}onjugate {G}radient ({HPCG})}\label{sec:hpcg}
HPCG\footnote{\Ac{HPCG} benchmark, version $3.1$: \url{http://hpcg-benchmark.org}} complements the LINPACK benchmark when ranking supercomputer systems; both together provide a better measure for real-world application performance.
In HPCG, a linear system of equations is solved whose coefficient matrix emerges from a 27-point stencil at each grid point in a 3-D domain.
On many systems, its performance is determined by memory bandwidth at large problem sizes because of its low operational intensity.
Hence, it shows the typical saturating performance pattern when scaling across the cores of a contention domain.

	\begin{figure*}[t]
		\centering
		\includegraphics[scale=0.84]{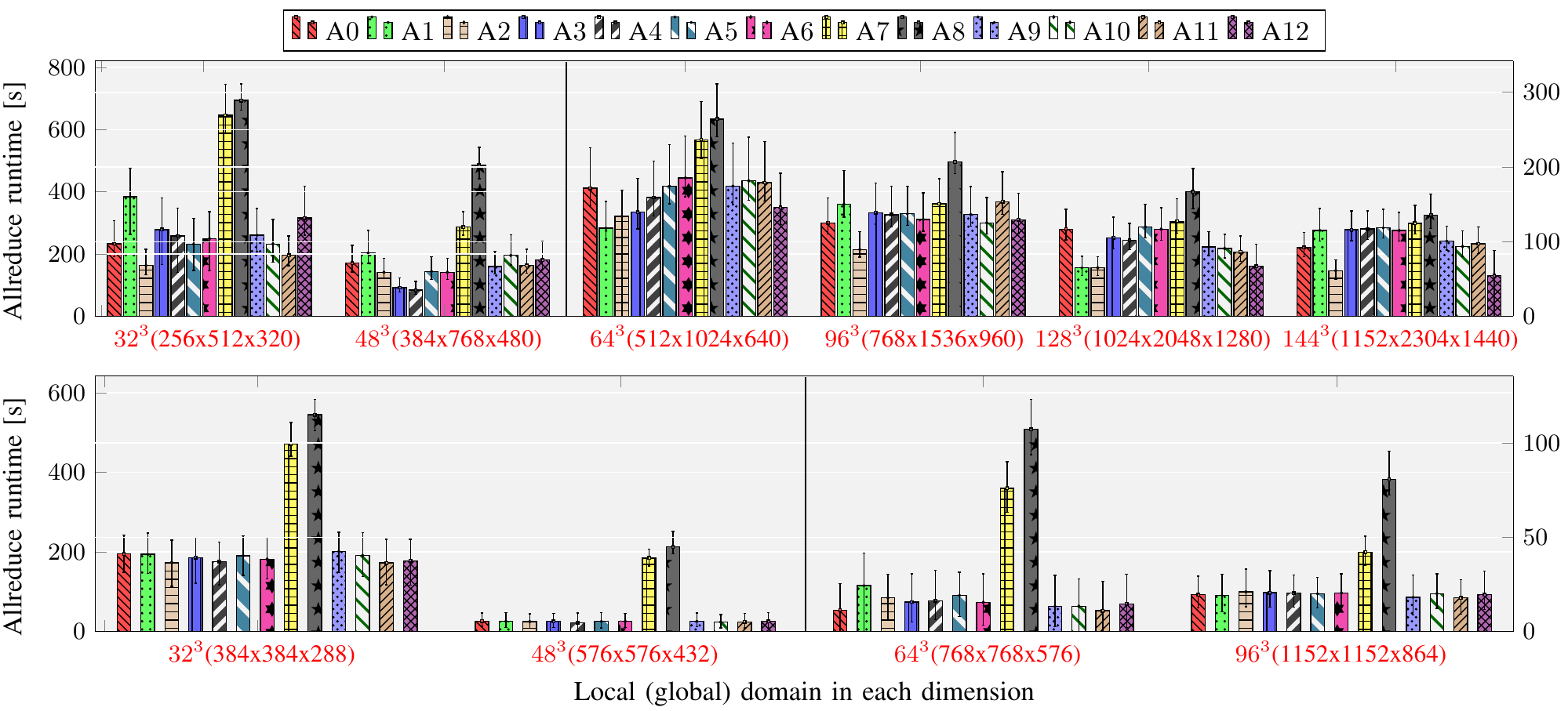}
		\caption{Average, minimum, and maximum aggregated
                  time spent in \texttt{MPI\_Allreduce} versus implementation
                  variants and local domain size for 
                  {\textcolor{black}{(top)}} $1280$ processes
                  distributed on $64$ Meggie nodes and
                  {\textcolor{black}{(bottom)}} $1296$ processes
                  distributed on $27$ SuperMUC-NG nodes.
                  The total runtime was about $1800$~\seconds\ in all cases.
                  Note the different $y$ axis scaling in the left and right parts of these plots (separated by a solid line).}
		\label{fig:allreduce}
	\end{figure*}
\subsection{Implementation}
The MPI-parallel reference implementation of HPCG comprises a \acf{MG} preconditioner and seven compute kernels (see Table~\ref{tab:HPCG0}(left)); six of the kernels are of \code{BLAS-1} type, i.e., one \code{DDOT1} (4~\BF), two \code{DDOT2} (8~\BF) and three \code{DAXPY} (16~\BF\ without write-allocate), and one is a sparse matrix-vector multiplication \code{SPMV} (optimal 6~\BF).
The \code{MG} preconditioner (optimal 6~\BF) comprises five kernels: \code{RESTRICTION}, \code{PROLONGATION}, \code{SPMV}, and two symmetric Gauss-Seidel (\code{SYMGS}) kernels serving as pre- and post-smoothers (each with forward and backward sweeps) for coarsening and refinement, respectively.
MPI parallelization is performed on a grid of $n_{px} n_{py} n_{pz} = n_p$ processes, where $n_{px}$ is the inner dimension.
Domain sizes are always given per process (weak scaling).
There are two types of MPI communication:
Three \code{MPI\_Allreduce} collectives are required for the dot products.
Within SpMV and SymGS, point-to-point communication is used to handle halo exchanges for each subdomain with \code{MPI\_Irecv}/\code{MPI\_Send}/\code{MPI\_Wait} sequences.
Communication is symmetric throughout, excluding boundaries, and the number of communication partners per process can vary between 7 (corners) and 26 (interior).
        
In all our \acs{HPCG} experiments, desynchronization across MPI processes occurs automatically, i.e., it is not provoked~\cite{AfzalHWcpe22}.
The propagation speed of idle waves within back-to-back \ac{SpMVM} operations using the HPCG matrix was modeled and analyzed in~\cite{AfzalHW2021}.

    \begin{figure*}[t]
		\centering
		\includegraphics[scale=0.84]{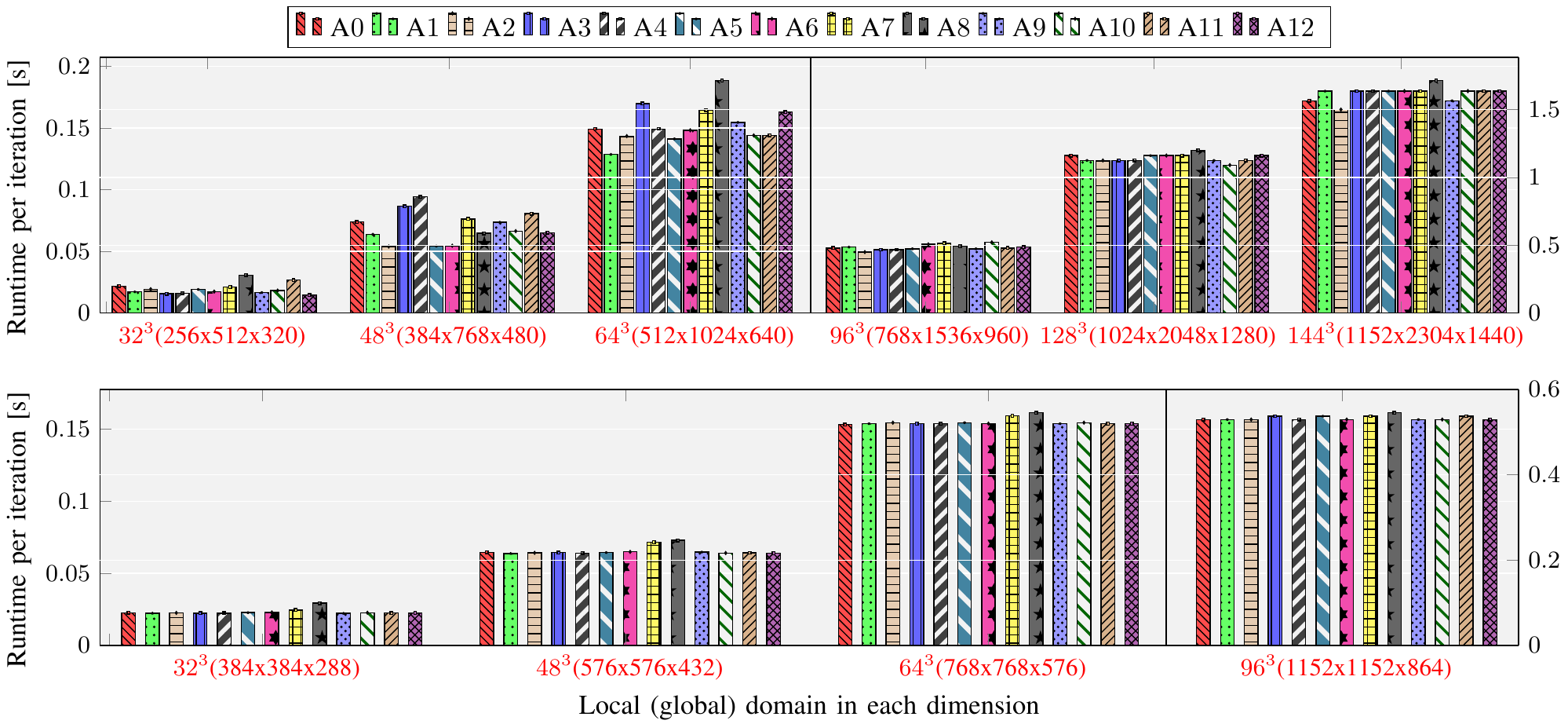}
		\caption{
		  Average runtime per HPCG iteration at fixed run time of $1800$~\seconds\ for MPI-parallelized
          {\textcolor{black}{(top)}} $1280$ processes distributed on $64$ Meggie nodes and {\textcolor{black}{(bottom)}} $1296$ processes distributed on $27$ SuperMUC-NG nodes.
		  The $x$-axis denotes local (global) domain dimensions in  $x$, $y$, and $z$ directions.
		  The legend represents diverse \texttt{MPI\_Allreduce} algorithms.
          Note the different $y$ axis scalings left and right of the solid line.
		}  
		\label{fig:HPCG1}
	\end{figure*}
\subsection{Reduction algorithms}
We chose HPCG to analyze design alternatives for the \code{MPI\_Allreduce} collective, especially with regard to desynchronization.
In HPCG, the collectives have a different effect than in LBM (perfectly balanced load, gradual desync with natural noise) and in LULESH (heavily load imbalanced).
The \code{MPI\_Allreduce} is a significant and widely used collective that aggregates the elements to compute their minimum, maximum, sum, or other values in numerous distributed applications.
The implementation details of \code{MPI\_Allreduce} have a decisive influence on the desynchronization and overlapping behavior of the algorithm.
Less-synchronizing implementations allow for better overlapping with preceding (here \code{DDOT}) and subsequent memory-bound kernels (here \code{DAXPY}) and will generally also lead to more overlap within the \code{SPMV}/\code{SYMGS} operations.
All reductions in HPCG occur on single 8-byte \code{double} values per process.
In each iteration, since \code{MPI\_Allreduce} occurs three times, each process transfers an aggregated 61.44~\KB\ and 62.16~\KB\ in both directions for 1280 and 1295 processes, respectively. 

Even under the necessary condition that no process can leave \code{MPI\_Allreduce} before all processes have entered it, there are considerable variations in the interaction details of this collective routine with idle waves and asynchronous execution.\footnote{In~\cite{AfzalHW2021}, we studied these variations using \code{MPI\_Reduce} in the Intel MPI library. Naturally, \code{MPI\_Reduce} is more permeable and thus allows for higher desynchronization, which is influenced by both the specifics of the reduction algorithm and the idle wave velocity.} 
In fact, a less-optimal but also less-synchronizing implementation can lead to better performance for HPCG.
Still, an asynchronous execution is more affected by the details of the reduction algorithm than by the idle wave velocity.
Ideally, the asynchronous performance boost would be unchanged by the collective. 

In Table~\ref{tab:HPCG0}\textcolor{black}{(right)} we show a breakdown of computation and communication times, communication volumes, \code{MPI\_Allreduce} timings (default implementation with the Intel MPI version used), and the CER for 1280-process runs at different subdomain sizes, from small (communication heavy) to large (computation dominant).
These measurements, although care was taken to maintain an undisturbed, fully synchronized execution, already exhibit considerable variation in the \code{MPI\_Allreduce} timings across processes. 

To study the effects of \code{MPI\_Allreduce} implementations on asynchronous execution, we explore the configuration space provided by Intel MPI's \code{I\_MPI\_ADJUST\_<opname>} variable.\footnote{Implementation details of \code{MPI\_Allreduce} algorithmic variants can be found at \url{https://software.intel.com/content/www/us/en/develop/documentation/mpi-developer-reference-windows/top/environment-variable-reference/i-mpi-adjust-family-environment-variables.html}.}
We label the twelve available implementations with {A$0$} to {A$11$} (default, recursive doubling, Rabenseifner's, reduce{-}plus{-}broadcast, topology\code{-}aware reduce-plus\code{-}broadcast, binomial gather{-}plus{-}scatter, topology\code{-}aware binominal gather{-}plus{-}scatter, Shumilin's ring, ring, Knomial, topology{-}aware SHM{-}based flat and topology{-}aware SHM-based Knomial).
The {A$12$} label denotes the setting \code{I\_MPI\_COLL\_INTRANODE=pt2pt}, which specifies that intranode collectives should use point-to-point calls instead of shared-memory transfers.
It was used in combination with the default \code{MPI\_Allreduce} variant.

\subsection{Asynchronicity through slower reduction algorithms}
Figure~\ref{fig:allreduce} shows average, minimum, and maximum \code{MPI\_Allreduce} times across 1280 processes for all variants.
The \enquote{best} or \enquote{worst} algorithm in terms of pure \code{MPI\_Allreduce} overhead is heavily dependent on the message sizes and the number of processes per node~\cite{supalov2014optimizing}.
However, this overhead is not always the most relevant metric when desynchronization is present.

Runtime in seconds per iteration for the full HPCG algorithm is shown in Fig.~\ref{fig:HPCG1} (Tables~\ref{tab:HPCG1}, \ref{tab:HPCG2} and \ref{tab:HPCG3} in~\ref{appx:hpcg} show the corresponding higher-is-better metrics like \GFS\ and total number of iterations in 1800\,s; note that \GFS\ numbers are as printed by the benchmark, so the ranking is not the same on both scales).
As expected, smaller problem sizes show a larger impact from changes in \code{MPI\_Allreduce} because communication is more relevant (higher CER).
The ring algorithm A$8$ (and then Shumilin's ring A$7$) is the most synchronizing implementation, while recursive doubling A$1$ is the least synchronizing one (maximum desync but not the most efficient design, however).

The largest differences in runtime per iteration across \code{MPI\_Allreduce} implementations can be observed with small domains per process because the CER is larger while the point-to-point communication overhead is not dominant.
In these scenarios, desynchronization can lead to performance improvements and will survive collective operations, implementations permitting.
Our results show that the ring algorithms are generally the worst choice for HPCG performance, not only based on their efficiency but also due to their strongly synchronizing implementation.
The global synchronizations happen twice in the ring algorithm, i.e., at the end of both \emph{scatter-reduce} and \emph{allgather} phases~\cite{iakymchuk2021efficient}.
Recursive doubling (A$1$) or Rabenseifner's (A$2$) are often among the best alternatives. 
        
Naively, one assumes that the fastest \code{MPI\_Allreduce} should give you the best overall performance, but this is not always the case. 
This can be easily visualized by comparing the 
time spent in \code{MPI\_Allreduce} (Fig. \ref{fig:allreduce}) with the overall performance of HPCG (Fig.~\ref{fig:HPCG1}).
For instance, A$2$/A$11$ for the $32^3$ domain and A$3$/A$4$ for the $48^3$ domain show the least time spent in the reduction, but the overall HPCG performance is on the low end across all variants.  
Therefore, in the presence of significant communication overhead, a large amount of time spent in the MPI library is not necessarily harmful if the processes are desynchronized, thereby overlapping useful work with waiting times.\medskip
        
\highlight{\emph{Upshot}: Faster collectives are not always the best; in fact, there is a trade-off between collective overhead and overlap benefit, depending on how often collectives occur.
For better HPCG performance, we recommend employing a less-synchronizing \code{MPI\_Allreduce} variant that is permeable to asynchronous execution, even if it has higher overhead than the most efficient implementation.}

\section{Outlook and future work}\label{sec:conclusion}
Parallel programs on HPC systems exhibit intriguing dynamics.
We demonstrated using the MPI-augmented STREAM Triad (MST) microbenchmark that deliberately injected noise can accelerate the onset of an asynchronous state from an initial bulk-synchronous state and thereby mitigate some of the communication overhead by hiding it behind useful computation.
Using a D3Q19 Lattice-Boltzmann (LBM) implementation, we showed why it is advisable to keep the frequency of collectives as low as possible, not only to reduce the communication overhead but mainly to avoid resynchronization and thus allow asynchronous execution across processes. Comparing this memory-bound code with a scalable D2Q37 implementation from SPEChpc 2021, we also demonstrated that lack of contention on bottlenecks (memory bandwidth in our case) and low communication-to-computation ratio do not yield positive performance effects from desynchronization. For MST and LBM we employed the \emph{phase space plot} as a new option for visualizing the temporal behavior of parallel programs in terms of MPI waiting time and performance; synchronized and desynchronized programs exhibit distinctly different patterns in this visualization.

The LULESH proxy application, although it does fulfill the condition of memory boundedness, cannot profit from automatic communication overlap, because its load imbalance (once configured) is too strong and nullifies any positive effect from desynchronization. 

Using the HPCG benchmark we demonstrated that, for certain problem sizes, the collective's synchronizing quality is more significant than its bare overhead: ``slower'' collectives seem to be more transparent to idle waves and allow for at least part of the desynchronization to survive.

One general conclusion from this work is that the presence of a relevant communication overhead and (groups of) processes being subject to a common hardware bottleneck are prerequisites for automatic overlap.
%
Future research will aim to provide an analytical description of the effects of asynchronous execution of MPI processes on performance, including overlap effects not only between computation and communication but also between parts of an application with different behavior towards bottlenecks.
In order to achieve this, we will employ a parallel simulator that is currently under development.
The simulator can model the dynamics of large-scale applications in a controlled setting, allowing for more in-depth investigation while taking the contention into account and without running programs on actual systems.

\section*{Acknowledgments}

The authors gratefully acknowledge the scientific support and HPC resources provided by the Erlangen National High Performance Computing Center (NHR@FAU) of the Friedrich-Alexander-Universität Erlangen-Nürnberg (FAU) and LRZ Gar\-ching. NHR funding is provided by federal and Bavarian state authorities. NHR@FAU hardware is partially funded by the German Research Foundation (DFG) -- 440719683. This work was partly supported by the Competence Network for Scientific High-Performance Computing in Bavaria (KONWIHR) under project ``OMI4Papps.''

\appendix

    



\section{HPCG performance metrics}\label{appx:hpcg}
Additional performance metrics resulting from the HPCG program are addressed in Tables~\ref{tab:HPCG1}, \ref{tab:HPCG2} and \ref{tab:HPCG3}. 
	\begin{table*}[hb]
		\centering
		\caption{
			Number of HPCG iteration at fixed run time of $1800$~\seconds\ for MPI-parallelized $1280$ ($1296$) processes distributed on $64$ Meggie ($27$ SuperMUC-NG) nodes.
			First column donates local (global) domain dimensions in all three x, y and z directions.
			First row represents diverse \texttt{MPI\_Allreduce} algorithms.}
		\label{tab:HPCG1}
		\hspace{-3em}
\begin{adjustbox}{width=\textwidth}
	\centering
	\huge
	\pgfplotstabletypeset[
	/pgfplots/colormap/greenyellow,
	every column/.style=,
	after row=,
	color cells={min=58350,max=122850},
	col sep=comma,
	@content options for rows={0}{/pgfplots/colormap/greenyellow,color cells={min=58350,max=122850}},
	@content options for rows={1}{/pgfplots/colormap/greenyellow,color cells={min=19100,max=33350}},
	@content options for rows={2}{/pgfplots/colormap/greenyellow,color cells={min=9550,max=14000}},
	@content options for rows={3}{/pgfplots/colormap/greenyellow,color cells={min=3450,max=4000}},
	@content options for rows={4}{/pgfplots/colormap/greenyellow,color cells={min=1500,max=1600}},
	@content options for rows={5}{/pgfplots/colormap/greenyellow,color cells={min=1050,max=1200}},
	@content options for rows={6}{/pgfplots/colormap/greenyellow,color cells={min=61550,max=81150}},
	@content options for rows={7}{/pgfplots/colormap/greenyellow,color cells={min=24700,max=28250}},
	@content options for rows={8}{/pgfplots/colormap/greenyellow,color cells={min=11150,max=11750}},
	@content options for rows={9}{/pgfplots/colormap/greenyellow,color cells={min=3300,max=3400}},
	every row 0 column 13/.style={
		postproc cell content/.append style={
			/pgfplots/table/@cell content/.add={\bfseries\boldmath\color{green!45!black}}{}
		},
	},
	every row 1 column 3/.style={
		postproc cell content/.append style={
			/pgfplots/table/@cell content/.add={\bfseries\boldmath\color{green!45!black}}{}
		},
	},
	every row 2 column 2/.style={
		postproc cell content/.append style={
			/pgfplots/table/@cell content/.add={\bfseries\boldmath\color{green!45!black}}{}
		},
	},
	every row 3 column 3/.style={
		postproc cell content/.append style={
			/pgfplots/table/@cell content/.add={\bfseries\boldmath\color{green!45!black}}{}
		},
	},
	every row 4 column 2/.style={
		postproc cell content/.append style={
			/pgfplots/table/@cell content/.add={\bfseries\boldmath\color{green!45!black}}{}
		},
	},
	every row 5 column 3/.style={
		postproc cell content/.append style={
			/pgfplots/table/@cell content/.add={\bfseries\boldmath\color{green!45!black}}{}
		},
	},
	every row 6 column 10/.style={
		postproc cell content/.append style={
			/pgfplots/table/@cell content/.add={\bfseries\boldmath\color{green!45!black}}{}
		},
	},
	every row 7 column 2/.style={
		postproc cell content/.append style={
			/pgfplots/table/@cell content/.add={\bfseries\boldmath\color{green!45!black}}{}
		},
	},
	every row 8 column 1/.style={
		postproc cell content/.append style={
			/pgfplots/table/@cell content/.add={\bfseries\boldmath\color{green!45!black}}{}
		},
	},
	every row 9 column 2/.style={
		postproc cell content/.append style={
			/pgfplots/table/@cell content/.add={\bfseries\boldmath\color{green!45!black}}{}
		},
	},
	colfont={10},
	]{
		{Domain vs. allreduce},A0,A1,A2,A3,A4,A5,A6,A7,A8,A9,A10,A11,A12
		$32^3$ (256x512x320) (M),82450,104500,92750,115000,111250,93950,103150,85150,58350,108150,98550,67350,122850
		$48^3$ (384x768x480) (M),24350,28200,33350,20800,27850,33150,32950,23600,19100,24450,27100,22350,27600
		$64^3$ (512x1024x640) (M),12050,14000,12550,10600,12050,12750,12150,10950,9550,11650,12500,12500,11050
		$96^3$ (768x1536x960) (M),3750,3700,4000,3850,3850,3800,3550,3500,3450,3800,3650,3750,3700
		$128^3$ (1024x2048x1280) (M),1550,1600,1600,1600,1600,1550,1550,1550,1500,1600,1650,1600,1550
		$144^3$ (1152x2304x1440) (M),1150,1100,1200,1100,1100,1100,1100,1100,1050,1150,1100,1100,1100
		$32^3$ (384x384x288) (S),79600,80850,79950,80600,80600,79350,78800,73000,61550,81150,79650,80500,79900
		$48^3$ (576x576x432) (S),27900,28250,28050,27950,28100,27950,27750,25150,24700,27800,28100,28050,28150
		$64^3$ (768x768x576) (S),11750,11700,11650,11700,11700,11650,11700,11300,11150,11700,11650,11700,11700
		$96^3$ (1152x1152x864) (S),3400,3400,3400,3350,3400,3350,3400,3350,3300,3400,3400,3350,3400
	}
\end{adjustbox}
	\end{table*}
	
	\begin{table*}[ht]
		\centering
		\caption{HPCG performance in~\GFS\ (convergence and optimization phase overhead included) at fixed run time of $1800$~\seconds\ for MPI-parallelized $1280$ processes distributed on $64$ Meggie nodes.
			First column donates local (global) domain dimensions in all three x, y and z directions.
			First row represents diverse \texttt{MPI\_Allreduce} algorithms.}
		\label{tab:HPCG2}
		\hspace{-3.5em}
\begin{adjustbox}{width=\textwidth}
	\centering
	\scriptsize
	\pgfplotstabletypeset[
	/pgfplots/colormap/greenyellow,
	every column/.style=,
	after row=,
	color cells={min=500,max=900},
	/pgf/number format/fixed,
	/pgf/number format/precision=0,
	col sep=comma,
	@content options for rows={0}{/pgfplots/colormap/greenyellow,color cells={min=506.1,max=844.6}},
	@content options for rows={1}{/pgfplots/colormap/greenyellow,color cells={min=669.1,max=846.7}},
	@content options for rows={2}{/pgfplots/colormap/greenyellow,color cells={min=722.9,max=852.3}},
	@content options for rows={3}{/pgfplots/colormap/greenyellow,color cells={min=781.7,max=855.8}},
	@content options for rows={4}{/pgfplots/colormap/greenyellow,color cells={min=787.1,max=860}},
	@content options for rows={5}{/pgfplots/colormap/greenyellow,color cells={min=808.3,max=865.1}},
	@content options for rows={6}{/pgfplots/colormap/greenyellow,color cells={min=537.4,max=689.4}},
	@content options for rows={7}{/pgfplots/colormap/greenyellow,color cells={min=717.6,max=815.3}},
	@content options for rows={8}{/pgfplots/colormap/greenyellow,color cells={min=761,max=806.3}},
	@content options for rows={9}{/pgfplots/colormap/greenyellow,color cells={min=757,max=786.8}},
	every row 0 column 3/.style={
		postproc cell content/.append style={
			/pgfplots/table/@cell content/.add={\bfseries\boldmath\color{green!45!black}}{}
		},
	},
	every row 1 column 3/.style={
		postproc cell content/.append style={
			/pgfplots/table/@cell content/.add={\bfseries\boldmath\color{green!45!black}}{}
		},
	},
	every row 2 column 2/.style={
		postproc cell content/.append style={
			/pgfplots/table/@cell content/.add={\bfseries\boldmath\color{green!45!black}}{}
		},
	},
	every row 3 column 3/.style={
		postproc cell content/.append style={
			/pgfplots/table/@cell content/.add={\bfseries\boldmath\color{green!45!black}}{}
		},
	},
	every row 4 column 2/.style={
		postproc cell content/.append style={
			/pgfplots/table/@cell content/.add={\bfseries\boldmath\color{green!45!black}}{}
		},
	},
	every row 5 column 3/.style={
		postproc cell content/.append style={
			/pgfplots/table/@cell content/.add={\bfseries\boldmath\color{green!45!black}}{}
		},
	},
	every row 6 column 3/.style={
		postproc cell content/.append style={
			/pgfplots/table/@cell content/.add={\bfseries\boldmath\color{green!45!black}}{}
		},
	},
	every row 7 column 2/.style={
		postproc cell content/.append style={
			/pgfplots/table/@cell content/.add={\bfseries\boldmath\color{green!45!black}}{}
		},
	},
	every row 8 column 3/.style={
		postproc cell content/.append style={
			/pgfplots/table/@cell content/.add={\bfseries\boldmath\color{green!45!black}}{}
		},
	},
	every row 9 column 2/.style={
		postproc cell content/.append style={
			/pgfplots/table/@cell content/.add={\bfseries\boldmath\color{green!45!black}}{}
		},
	},
	colfont={10},
	]{
		{Domain vs. allreduce},A0,A1,A2,A3,A4,A5,A6,A7,A8,A9,A10,A11,A12
		$32^3$ (256x512x320) (M),764.991,765.387,844.599,761.357,757.96,763.934,760.654,612.044,506.116,795.927,781.922,749.23,785.701
		$48^3$ (384x768x480) (M),782.264,782.239,846.707,846.409,845.024,847.268,844.083,735.986,669.128,798.434,780.8,782.292,796.467
		$64^3$ (512x1024x640) (M),800.493,852.268,849.286,804.576,803.238,803.852,789.087,756.952,722.938,789.599,794.731,790.88,804.182
		$96^3$ (768x1536x960) (M),826.623,806.449,855.806,819.014,820.022,818.057,817.25,804.861,781.728,818.349,816.867,807.024,820.012
		$128^3$ (1024x2048x1280) (M),821.926,860.037,859.453,830.126,834.007,818.419,820.883,814.758,787.086,840.445,841.796,842.965,808.792
		$144^3$ (1152x2304x1440) (M),842.999,822.141,865.075,822.309,821.771,821.64,822.521,819.236,809.313,837.501,837.219,834.642,813.035
		$32^3$ (384x384x288) (S),690.017,688.413,689.382,688.887,696.275,687.147,686.01,621.871,537.357,687.773,689.699,695.124,689.024
		$48^3$ (576x576x432) (S),815.23,815.346,813.156,813.565,814.766,811.619,813.158,737.109,717.608,815.672,814.895,814.607,814.832
		$64^3$ (768x768x576) (S),809.567,806.08,806.289,807.541,807.863,806.156,807.285,779.575,760.956,807.255,807.337,809.614,807.827
		$96^3$ (1152x1152x864) (S),787.786,786.797,785.793,785.784,785.254,785.348,786.752,776.247,757.015,786.901,786.427,786.803,786.667
	}
\end{adjustbox}
	\end{table*}
	
	\begin{table*}[!hb]
		\centering
		\caption{HPCG B/W in~\GBS\ at fixed run time of $1800$~\seconds\ for MPI-parallelized $1280$ processes distributed on $64$ Meggie nodes.
			First column donates local (global) domain dimensions in all three x, y and z directions.
			First row represents diverse \texttt{MPI\_Allreduce} algorithms.
		}
		\label{tab:HPCG3}
		\hspace{-3em}
\begin{adjustbox}{width=\textwidth}
	\centering
	\pgfplotstabletypeset[
	/pgfplots/colormap/greenyellow,
	every column/.style=,
	after row=,
	color cells={min=4000,max=7000},
	/pgf/number format/fixed,
	/pgf/number format/precision=0,
	col sep=comma,
	@content options for rows={0}{/pgfplots/colormap/greenyellow,color cells={min=4008.1,max=6679.6}},
	@content options for rows={1}{/pgfplots/colormap/greenyellow,color cells={min=5260.3,max=6718.9}},
	@content options for rows={2}{/pgfplots/colormap/greenyellow,color cells={min=5722.7,max=6755.1}},
	@content options for rows={3}{/pgfplots/colormap/greenyellow,color cells={min=6178.8,max=6790.1}},
	@content options for rows={4}{/pgfplots/colormap/greenyellow,color cells={min=6237.5,max=6836.4}},
	@content options for rows={5}{/pgfplots/colormap/greenyellow,color cells={min=6423.6,max=6882.4}},
	@content options for rows={6}{/pgfplots/colormap/greenyellow,color cells={min=4075.1,max=5228}},
	@content options for rows={7}{/pgfplots/colormap/greenyellow,color cells={min=5441.8,max=6166.3}},
	@content options for rows={8}{/pgfplots/colormap/greenyellow,color cells={min=5770.3,max=6114.1}},
	@content options for rows={9}{/pgfplots/colormap/greenyellow,color cells={min=5886.1,max=5958.5}},
	every row 0 column 3/.style={
		postproc cell content/.append style={
			/pgfplots/table/@cell content/.add={\bfseries\boldmath\color{green!45!black}}{}
		},
	},
	every row 1 column 3/.style={
		postproc cell content/.append style={
			/pgfplots/table/@cell content/.add={\bfseries\boldmath\color{green!45!black}}{}
		},
	},
	every row 2 column 2/.style={
		postproc cell content/.append style={
			/pgfplots/table/@cell content/.add={\bfseries\boldmath\color{green!45!black}}{}
		},
	},
	every row 3 column 3/.style={
		postproc cell content/.append style={
			/pgfplots/table/@cell content/.add={\bfseries\boldmath\color{green!45!black}}{}
		},
	},
	every row 4 column 2/.style={
		postproc cell content/.append style={
			/pgfplots/table/@cell content/.add={\bfseries\boldmath\color{green!45!black}}{}
		},
	},
	every row 5 column 3/.style={
		postproc cell content/.append style={
			/pgfplots/table/@cell content/.add={\bfseries\boldmath\color{green!45!black}}{}
		},
	},
	every row 6 column 3/.style={
		postproc cell content/.append style={
			/pgfplots/table/@cell content/.add={\bfseries\boldmath\color{green!45!black}}{}
		},
	},
	every row 7 column 2/.style={
		postproc cell content/.append style={
			/pgfplots/table/@cell content/.add={\bfseries\boldmath\color{green!45!black}}{}
		},
	},
	every row 8 column 3/.style={
		postproc cell content/.append style={
			/pgfplots/table/@cell content/.add={\bfseries\boldmath\color{green!45!black}}{}
		},
	},
	every row 9 column 2/.style={
		postproc cell content/.append style={
			/pgfplots/table/@cell content/.add={\bfseries\boldmath\color{green!45!black}}{}
		},
	},
	colfont={10},
	]{
		{Domain vs. allreduce},A0,A1,A2,A3,A4,A5,A6,A7,A8,A9,A10,A11,A12
		$32^3$ (256x512x320) (M),6013.1,5996.4,6679.63,5971.03,5979.96,5998.04,5965.92,4786.06,4008.14,6237.6,6214.32,5984.1,6172.33
		$48^3$ (384x768x480) (M),6196.03,6148.56,6718.88,6674.34,6663.43,6679.34,6693.88,5784.74,5260.29,6283.55,6166.82,6160.23,6264.07
		$64^3$ (512x1024x640) (M),6328.27,6767.46,6755.14,6384.19,6384.73,6368.92,6263.33,5995.51,5722.74,6274.52,6302.81,6273.37,6382.71
		$96^3$ (768x1536x960) (M),6538.31,6378.87,6790.1,6483.7,6493.49,6479.76,6466.69,6378.56,6178.75,6478.51,6471.24,6383.42,6490.87
		$128^3$ (1024x2048x1280) (M),6517.99,6835.82,6836.23,6590.42,6622.69,6498.13,6515.57,6466.33,6237.5,6675.17,6689.64,6700.82,6431.92
		$144^3$ (1152x2304x1440) (M),6694.97,6533.94,6882.42,6526.65,6528.66,6522.42,6531.43,6502.44,6423.58,6655.62,6655.05,6633.13,6474.36
		$32^3$ (384x384x288) (S),5232.85,5220.69,5228.03,5224.27,5280.3,5211.09,5202.46,4716.05,4075.13,5215.83,5230.43,5271.58,5225.32
		$48^3$ (576x576x432) (S),6182.06,6182.94,6166.34,6169.44,6178.54,6154.69,6166.35,5589.66,5441.78,6185.42,6179.53,6177.35,6179.05
		$64^3$ (768x768x576) (S),6138.95,6112.5,6114.09,6123.58,6126.03,6113.08,6121.65,5911.52,5770.33,6121.42,6122.04,6139.31,6125.75
		$96^3$ (1152x1152x864) (S),5973.61,5966.12,5958.5,5958.43,5954.41,5955.13,5965.77,5886.12,5886.12,5966.9,5963.31,5966.16,5965.12
	}
\end{adjustbox}
	\end{table*}

 \bibliographystyle{elsarticle-harv} 
 \bibliography{references}

\begin{thebibliography}{30}
\expandafter\ifx\csname natexlab\endcsname\relax\def\natexlab#1{#1}\fi
\providecommand{\url}[1]{\texttt{#1}}
\providecommand{\href}[2]{#2}
\providecommand{\path}[1]{#1}
\providecommand{\DOIprefix}{doi:}
\providecommand{\ArXivprefix}{arXiv:}
\providecommand{\URLprefix}{URL: }
\providecommand{\Pubmedprefix}{pmid:}
\providecommand{\doi}[1]{\href{http://dx.doi.org/#1}{\path{#1}}}
\providecommand{\Pubmed}[1]{\href{pmid:#1}{\path{#1}}}
\providecommand{\bibinfo}[2]{#2}
\ifx\xfnm\relax \def\xfnm[#1]{\unskip,\space#1}\fi
\bibitem[{Afzal et~al.(2019a)Afzal, Hager and Wellein}]{AfzalEuroMPI19Poster}
\bibinfo{author}{Afzal, A.}, \bibinfo{author}{Hager, G.},
  \bibinfo{author}{Wellein, G.}, \bibinfo{year}{2019}a.
\newblock \bibinfo{title}{Delay flow mechanisms on clusters}.
\newblock \URLprefix
  \url{https://hpc.fau.de/files/2019/09/EuroMPI2019_AHW-Poster.pdf}.
  \bibinfo{note}{poster at the EuroMPI 2019, September 10--13, 2019, Zurich,
  Switzerland}.
\bibitem[{Afzal et~al.(2019b)Afzal, Hager and Wellein}]{AfzalHW19}
\bibinfo{author}{Afzal, A.}, \bibinfo{author}{Hager, G.},
  \bibinfo{author}{Wellein, G.}, \bibinfo{year}{2019}b.
\newblock \bibinfo{title}{{Propagation} and {Decay} of {Injected} {One}-{Off}
  {Delays} on {Clusters}: {A} {Case} {Study}}, in:
  \bibinfo{booktitle}{Proceedings of the 2019 {IEEE} International Conference
  on Cluster Computing}, \bibinfo{publisher}{Institute of Electrical and
  Electronics Engineers Inc.}. pp. \bibinfo{pages}{1--10}.
\newblock \DOIprefix\doi{10.1109/CLUSTER.2019.8890995}.
\bibitem[{Afzal et~al.(2020)Afzal, Hager and Wellein}]{AfzalHW20}
\bibinfo{author}{Afzal, A.}, \bibinfo{author}{Hager, G.},
  \bibinfo{author}{Wellein, G.}, \bibinfo{year}{2020}.
\newblock \bibinfo{title}{{Desynchronization} and {Wave} {Pattern} {Formation}
  in {MPI}-{Parallel} and {Hybrid} {Memory}-{Bound} {Programs}}, in:
  \bibinfo{editor}{Sadayappan, P.}, \bibinfo{editor}{Chamberlain, B.L.},
  \bibinfo{editor}{Juckeland, G.}, \bibinfo{editor}{Ltaief, H.} (Eds.),
  \bibinfo{booktitle}{Lecture Notes in Computer Science (including subseries
  Lecture Notes in Artificial Intelligence and Lecture Notes in
  Bioinformatics)}, \bibinfo{publisher}{Springer International Publishing},
  \bibinfo{address}{Cham}. pp. \bibinfo{pages}{391--411}.
\newblock \DOIprefix\doi{10.1007/978-3-030-50743-5_20}.
\bibitem[{Afzal et~al.(2021)Afzal, Hager and Wellein}]{AfzalHW2021}
\bibinfo{author}{Afzal, A.}, \bibinfo{author}{Hager, G.},
  \bibinfo{author}{Wellein, G.}, \bibinfo{year}{2021}.
\newblock \bibinfo{title}{{Analytic} {Modeling} of {Idle} {Waves} in {Parallel}
  {Programs}: {Communication}, {Cluster} {Topology}, and {Noise} {Impact}}, in:
  \bibinfo{editor}{Chamberlain, B.L.}, \bibinfo{editor}{Varbanescu, A.L.},
  \bibinfo{editor}{Ltaief, H.}, \bibinfo{editor}{Luszczek, P.} (Eds.),
  \bibinfo{booktitle}{Lecture Notes in Computer Science (including subseries
  Lecture Notes in Artificial Intelligence and Lecture Notes in
  Bioinformatics)}, \bibinfo{publisher}{Springer Science and Business Media
  Deutschland GmbH}. pp. \bibinfo{pages}{351--371}.
\newblock \DOIprefix\doi{10.1007/978-3-030-78713-4_19}.
\bibitem[{Afzal et~al.(2022a)Afzal, Hager and Wellein}]{AfzalHWcpe22}
\bibinfo{author}{Afzal, A.}, \bibinfo{author}{Hager, G.},
  \bibinfo{author}{Wellein, G.}, \bibinfo{year}{2022}a.
\newblock \bibinfo{title}{{Analytic} performance model for parallel overlapping
  {memory}-{bound} kernels}.
\newblock \bibinfo{journal}{Concurrency and Computation: Practice and
  Experience} \bibinfo{volume}{34}, \bibinfo{pages}{e6816}.
\newblock \DOIprefix\doi{10.1002/cpe.6816}.
\bibitem[{Afzal et~al.(2022b)Afzal, Hager and Wellein}]{AfzalHW:2022:4}
\bibinfo{author}{Afzal, A.}, \bibinfo{author}{Hager, G.},
  \bibinfo{author}{Wellein, G.}, \bibinfo{year}{2022}b.
\newblock \bibinfo{title}{The role of idle waves, desynchronization, and
  bottleneck evasion in the performance of parallel programs}.
\newblock \bibinfo{journal}{IEEE Transactions on Parallel and Distributed
  Systems, TPDS} \DOIprefix\doi{10.1109/TPDS.2022.3221085}.
\bibitem[{Afzal et~al.(2022c)Afzal, Hager, Wellein and
  Marakidis}]{AfzalHW:2022:2}
\bibinfo{author}{Afzal, A.}, \bibinfo{author}{Hager, G.},
  \bibinfo{author}{Wellein, G.}, \bibinfo{author}{Marakidis, S.},
  \bibinfo{year}{2022}c.
\newblock \bibinfo{title}{Exploring techniques for the analysis of spontaneous
  asynchronicity in {MPI}-{Parallel} applications}, in:
  \bibinfo{booktitle}{Proceedings of the International Conference on Parallel
  Processing and Applied Mathematics}.
\newblock \DOIprefix\doi{10.48550/arXiv.2205.13963}.
\bibitem[{Afzal et~al.(2022d)Afzal, Wellein and Hager}]{AfzalHW:2022:3}
\bibinfo{author}{Afzal, A.}, \bibinfo{author}{Wellein, G.},
  \bibinfo{author}{Hager, G.}, \bibinfo{year}{2022}d.
\newblock \bibinfo{title}{Addressing {White}-{Box} modeling and simulation
  challenges in parallel computing}, in: \bibinfo{booktitle}{Proceedings of the
  2022 ACM SIGSIM Conference on Principles of Advanced Discrete Simulation},
  \bibinfo{publisher}{Association for Computing Machinery},
  \bibinfo{address}{New York, NY, USA}. p. \bibinfo{pages}{25–26}.
\newblock \DOIprefix\doi{10.1145/3518997.3534986}.
\bibitem[{{Bhatele} et~al.(2013){Bhatele}, {Mohror}, {Langer} and
  {Isaacs}}]{Bhatele2013}
\bibinfo{author}{{Bhatele}, A.}, \bibinfo{author}{{Mohror}, K.},
  \bibinfo{author}{{Langer}, S.H.}, \bibinfo{author}{{Isaacs}, K.E.},
  \bibinfo{year}{2013}.
\newblock \bibinfo{title}{There goes the neighborhood: Performance degradation
  due to nearby jobs}, in: \bibinfo{booktitle}{SC '13: Proceedings of the
  International Conference on High Performance Computing, Networking, Storage
  and Analysis}, pp. \bibinfo{pages}{1--12}.
\newblock \DOIprefix\doi{10.1145/2503210.2503247}.
\bibitem[{Bhatnagar et~al.(1954)Bhatnagar, Gross and
  Krook}]{bhatnagar1954model}
\bibinfo{author}{Bhatnagar, P.L.}, \bibinfo{author}{Gross, E.P.},
  \bibinfo{author}{Krook, M.}, \bibinfo{year}{1954}.
\newblock \bibinfo{title}{A model for collision processes in gases. i. small
  amplitude processes in charged and neutral one-component systems}.
\newblock \bibinfo{journal}{Physical review} \bibinfo{volume}{94},
  \bibinfo{pages}{511}.
\bibitem[{B{\"o}hme et~al.(2016)B{\"o}hme, Geimer, Arnold, Voigtlaender and
  Wolf}]{Boehme:2016}
\bibinfo{author}{B{\"o}hme, D.}, \bibinfo{author}{Geimer, M.},
  \bibinfo{author}{Arnold, L.}, \bibinfo{author}{Voigtlaender, F.},
  \bibinfo{author}{Wolf, F.}, \bibinfo{year}{2016}.
\newblock \bibinfo{title}{Identifying the root causes of wait states in
  large-scale parallel applications}.
\newblock \bibinfo{journal}{ACM Trans. Parallel Comput.} \bibinfo{volume}{3},
  \bibinfo{pages}{11:1--11:24}.
\newblock \DOIprefix\doi{10.1145/2934661}.
\bibitem[{Eitzinger et~al.(2019)Eitzinger, Gruber, Afzal, Zeiser and
  Wellein}]{Clustercockpit:2019}
\bibinfo{author}{Eitzinger, J.}, \bibinfo{author}{Gruber, T.},
  \bibinfo{author}{Afzal, A.}, \bibinfo{author}{Zeiser, T.},
  \bibinfo{author}{Wellein, G.}, \bibinfo{year}{2019}.
\newblock \bibinfo{title}{{ClusterCockpit}-{A} web application for job-specific
  performance monitoring}, in: \bibinfo{booktitle}{Proceedings of the 2019
  {IEEE} International Conference on Cluster Computing},
  \bibinfo{organization}{IEEE}. pp. \bibinfo{pages}{1--7}.
\newblock \DOIprefix\doi{10.1109/CLUSTER.2019.8891017}.
\bibitem[{Ferreira et~al.(2008)Ferreira, Bridges and
  Brightwell}]{ferreira2008characterizing}
\bibinfo{author}{Ferreira, K.B.}, \bibinfo{author}{Bridges, P.},
  \bibinfo{author}{Brightwell, R.}, \bibinfo{year}{2008}.
\newblock \bibinfo{title}{Characterizing application sensitivity to {OS}
  interference using kernel-level noise injection}, in:
  \bibinfo{booktitle}{Proceedings of the 2008 ACM/IEEE conference on
  Supercomputing}, \bibinfo{organization}{IEEE Press}. p.~\bibinfo{pages}{19}.
\bibitem[{{Gamell} et~al.(2015){Gamell}, {Teranishi}, {Heroux}, {Mayo},
  {Kolla}, {Chen} and {Parashar}}]{Gamell:2015}
\bibinfo{author}{{Gamell}, M.}, \bibinfo{author}{{Teranishi}, K.},
  \bibinfo{author}{{Heroux}, M.A.}, \bibinfo{author}{{Mayo}, J.},
  \bibinfo{author}{{Kolla}, H.}, \bibinfo{author}{{Chen}, J.},
  \bibinfo{author}{{Parashar}, M.}, \bibinfo{year}{2015}.
\newblock \bibinfo{title}{Local recovery and failure masking for stencil-based
  applications at extreme scales}, in: \bibinfo{booktitle}{SC '15: Proceedings
  of the International Conference for High Performance Computing, Networking,
  Storage and Analysis}, pp. \bibinfo{pages}{1--12}.
\newblock \DOIprefix\doi{10.1145/2807591.2807672}.
\bibitem[{Iakymchuk et~al.(2021)Iakymchuk, Faustino, Emerson, Barreto, Bartsch,
  Rodrigues and Monteiro}]{iakymchuk2021efficient}
\bibinfo{author}{Iakymchuk, R.}, \bibinfo{author}{Faustino, A.},
  \bibinfo{author}{Emerson, A.}, \bibinfo{author}{Barreto, J.},
  \bibinfo{author}{Bartsch, V.}, \bibinfo{author}{Rodrigues, R.},
  \bibinfo{author}{Monteiro, J.C.}, \bibinfo{year}{2021}.
\newblock \bibinfo{title}{Efficient and eventually consistent collective
  operations}, in: \bibinfo{booktitle}{2021 IEEE International Parallel and
  Distributed Processing Symposium Workshops (IPDPSW)},
  \bibinfo{organization}{IEEE}. pp. \bibinfo{pages}{621--630}.
\bibitem[{Jolliffe and Cadima(2016)}]{jolliffe2016principal}
\bibinfo{author}{Jolliffe, I.T.}, \bibinfo{author}{Cadima, J.},
  \bibinfo{year}{2016}.
\newblock \bibinfo{title}{Principal component analysis: a review and recent
  developments}.
\newblock \bibinfo{journal}{Philosophical Transactions of the Royal Society A:
  Mathematical, Physical and Engineering Sciences} \bibinfo{volume}{374},
  \bibinfo{pages}{20150202}.
\newblock \DOIprefix\doi{10.1098/rsta.2015.0202}.
\bibitem[{Kaufman and Rousseeuw(2009)}]{kaufman2009finding}
\bibinfo{author}{Kaufman, L.}, \bibinfo{author}{Rousseeuw, P.J.},
  \bibinfo{year}{2009}.
\newblock \bibinfo{title}{Finding groups in data: an introduction to cluster
  analysis}.
\newblock \bibinfo{publisher}{John Wiley \& Sons}.
\newblock \DOIprefix\doi{10.1002/9780470316801}.
\bibitem[{{Le{\'o}n} et~al.(2016){Le{\'o}n}, {Karlin} and {Moody}}]{Leon2016}
\bibinfo{author}{{Le{\'o}n}, E.A.}, \bibinfo{author}{{Karlin}, I.},
  \bibinfo{author}{{Moody}, A.T.}, \bibinfo{year}{2016}.
\newblock \bibinfo{title}{System noise revisited: Enabling application
  scalability and reproducibility with {SMT}}, in: \bibinfo{booktitle}{2016
  IEEE International Parallel and Distributed Processing Symposium (IPDPS)},
  pp. \bibinfo{pages}{596--607}.
\newblock \DOIprefix\doi{10.1109/IPDPS.2016.48}.
\bibitem[{Markidis et~al.(2015)Markidis, Vencels, Peng, Akhmetova, Laure and
  Henri}]{markidis2015idle}
\bibinfo{author}{Markidis, S.}, \bibinfo{author}{Vencels, J.},
  \bibinfo{author}{Peng, I.B.}, \bibinfo{author}{Akhmetova, D.},
  \bibinfo{author}{Laure, E.}, \bibinfo{author}{Henri, P.},
  \bibinfo{year}{2015}.
\newblock \bibinfo{title}{Idle waves in high-performance computing}.
\newblock \bibinfo{journal}{Physical Review E} \bibinfo{volume}{91},
  \bibinfo{pages}{013306}.
\newblock \DOIprefix\doi{10.1103/PhysRevE.91.013306}.
\bibitem[{McCalpin et~al.(1995)}]{mccalpin1995memory}
\bibinfo{author}{McCalpin, J.D.}, et~al., \bibinfo{year}{1995}.
\newblock \bibinfo{title}{Memory bandwidth and machine balance in current high
  performance computers}.
\newblock \bibinfo{journal}{IEEE computer society technical committee on
  computer architecture (TCCA) newsletter} \bibinfo{volume}{2}.
\bibitem[{Nataraj et~al.(2007)Nataraj, Morris, Malony, Sottile and
  Beckman}]{nataraj2007ghost}
\bibinfo{author}{Nataraj, A.}, \bibinfo{author}{Morris, A.},
  \bibinfo{author}{Malony, A.D.}, \bibinfo{author}{Sottile, M.},
  \bibinfo{author}{Beckman, P.}, \bibinfo{year}{2007}.
\newblock \bibinfo{title}{The ghost in the machine: observing the effects of
  kernel operation on parallel application performance}, in:
  \bibinfo{booktitle}{Proceedings of the 2007 ACM/IEEE conference on
  Supercomputing}, pp. \bibinfo{pages}{1--12}.
\bibitem[{Petrini et~al.(2003)Petrini, Kerbyson and Pakin}]{petrini2003case}
\bibinfo{author}{Petrini, F.}, \bibinfo{author}{Kerbyson, D.J.},
  \bibinfo{author}{Pakin, S.}, \bibinfo{year}{2003}.
\newblock \bibinfo{title}{The case of the missing supercomputer performance:
  {A}chieving optimal performance on the 8,192 processors of {ASCI} {Q}}, in:
  \bibinfo{booktitle}{Supercomputing, 2003 ACM/IEEE Conference},
  \bibinfo{organization}{IEEE}. pp. \bibinfo{pages}{55--55}.
\newblock \DOIprefix\doi{10.1145/1048935.1050204}.
\bibitem[{Qian et~al.(1992)Qian, d'Humi{\`e}res and
  Lallemand}]{qian1992lattice}
\bibinfo{author}{Qian, Y.H.}, \bibinfo{author}{d'Humi{\`e}res, D.},
  \bibinfo{author}{Lallemand, P.}, \bibinfo{year}{1992}.
\newblock \bibinfo{title}{Lattice bgk models for navier-stokes equation}.
\newblock \bibinfo{journal}{EPL (Europhysics Letters)} \bibinfo{volume}{17},
  \bibinfo{pages}{479}.
\bibitem[{Supalov et~al.(2014)Supalov, Semin, Klemm and
  Dahnken}]{supalov2014optimizing}
\bibinfo{author}{Supalov, A.}, \bibinfo{author}{Semin, A.},
  \bibinfo{author}{Klemm, M.}, \bibinfo{author}{Dahnken, C.},
  \bibinfo{year}{2014}.
\newblock \bibinfo{title}{Optimizing HPC applications with intel cluster
  tools}.
\newblock \bibinfo{publisher}{Springer Nature}.
\bibitem[{Vassilvitskii and Arthur(2006)}]{vassilvitskii2006k}
\bibinfo{author}{Vassilvitskii, S.}, \bibinfo{author}{Arthur, D.},
  \bibinfo{year}{2006}.
\newblock \bibinfo{title}{{k-means++}: The advantages of careful seeding}, in:
  \bibinfo{booktitle}{Proceedings of the eighteenth annual ACM-SIAM symposium
  on Discrete algorithms}, pp. \bibinfo{pages}{1027--1035}.
\newblock \URLprefix \url{https://dl.acm.org/doi/10.5555/1283383.1283494}.
\bibitem[{Vetterling et~al.(1992)Vetterling, Vetterling, Press, Press,
  Teukolsky, Flannery and Flannery}]{vetterling1992numerical}
\bibinfo{author}{Vetterling, W.T.}, \bibinfo{author}{Vetterling, W.T.},
  \bibinfo{author}{Press, W.H.}, \bibinfo{author}{Press, W.H.},
  \bibinfo{author}{Teukolsky, S.A.}, \bibinfo{author}{Flannery, B.P.},
  \bibinfo{author}{Flannery, B.P.}, \bibinfo{year}{1992}.
\newblock \bibinfo{title}{Numerical Recipes: Example book C}.
\newblock \bibinfo{publisher}{Cambridge University Press}.
\bibitem[{Weisbach et~al.(2018)Weisbach, Gerofi, Kocoloski, H{\"a}rtig and
  Ishikawa}]{Weisbach:2018}
\bibinfo{author}{Weisbach, H.}, \bibinfo{author}{Gerofi, B.},
  \bibinfo{author}{Kocoloski, B.}, \bibinfo{author}{H{\"a}rtig, H.},
  \bibinfo{author}{Ishikawa, Y.}, \bibinfo{year}{2018}.
\newblock \bibinfo{title}{Hardware performance variation: A comparative study
  using lightweight kernels}, in: \bibinfo{editor}{Yokota, R.},
  \bibinfo{editor}{Weiland, M.}, \bibinfo{editor}{Keyes, D.},
  \bibinfo{editor}{Trinitis, C.} (Eds.), \bibinfo{booktitle}{High Performance
  Computing}, \bibinfo{publisher}{Springer International Publishing},
  \bibinfo{address}{Cham}. pp. \bibinfo{pages}{246--265}.
\newblock \DOIprefix\doi{10.1007/978-3-319-92040-5_13}.
\bibitem[{Wellein et~al.(2006)Wellein, Zeiser, Hager and
  Donath}]{WELLEIN2006910}
\bibinfo{author}{Wellein, G.}, \bibinfo{author}{Zeiser, T.},
  \bibinfo{author}{Hager, G.}, \bibinfo{author}{Donath, S.},
  \bibinfo{year}{2006}.
\newblock \bibinfo{title}{On the single processor performance of simple lattice
  {Boltzmann} kernels}.
\newblock \bibinfo{journal}{Computers \& Fluids} \bibinfo{volume}{35},
  \bibinfo{pages}{910--919}.
\newblock \DOIprefix\doi{10.1016/j.compfluid.2005.02.008}.
  \bibinfo{note}{proceedings of the First International Conference for
  Mesoscopic Methods in Engineering and Science}.
\bibitem[{Wittmann et~al.(2016)Wittmann, Hager, Zeiser, Treibig and
  Wellein}]{Wittmann:2016}
\bibinfo{author}{Wittmann, M.}, \bibinfo{author}{Hager, G.},
  \bibinfo{author}{Zeiser, T.}, \bibinfo{author}{Treibig, J.},
  \bibinfo{author}{Wellein, G.}, \bibinfo{year}{2016}.
\newblock \bibinfo{title}{Chip-level and multi-node analysis of
  energy-optimized lattice {Boltzmann} {CFD} simulations}.
\newblock \bibinfo{journal}{Concurrency and Computation: Practice and
  Experience} \bibinfo{volume}{28}, \bibinfo{pages}{2295--2315}.
\newblock \DOIprefix\doi{10.1002/cpe.3489}.
\bibitem[{Wittmann et~al.(2013)Wittmann, Zeiser, Hager and
  Wellein}]{WITTMANN2013924}
\bibinfo{author}{Wittmann, M.}, \bibinfo{author}{Zeiser, T.},
  \bibinfo{author}{Hager, G.}, \bibinfo{author}{Wellein, G.},
  \bibinfo{year}{2013}.
\newblock \bibinfo{title}{Comparison of different propagation steps for lattice
  {Boltzmann} methods}.
\newblock \bibinfo{journal}{Computers \& Mathematics with Applications}
  \bibinfo{volume}{65}, \bibinfo{pages}{924--935}.
\newblock \DOIprefix\doi{10.1016/j.camwa.2012.05.002}.
  \bibinfo{note}{mesoscopic Methods in Engineering and Science}.

\end{thebibliography}
 




\end{document}